\documentclass[useAMS,usenatbib]{mnras}
\usepackage{multicol}
\usepackage{multirow}
\usepackage{natbib}

\usepackage{amsmath} % usage of some command in math mode (e.g. \substack)
\usepackage{graphicx}
\usepackage{txfonts}
\usepackage{psfig}
\usepackage{natbib}
\usepackage{soul}

\usepackage{xcolor}

\title[Tree-based solvers for AMR code FLASH - II]
{Tree-based solvers for adaptive mesh refinement code FLASH - II:\\
radiation transport module TreeRay.}
\author[W\"unsch et al.]{
Richard W\"unsch$^{1}$ \thanks{E-mail: richard@wunsch.cz},
Stefanie Walch$^{2}$, 
Franti\v{s}ek Dinnbier$^{2,1,4}$,
Daniel Seifried$^2$, 
Sebastian Haid$^2$, \newauthor
Andre Klepitko$^2$, 
Anthony P. Whitworth$^{3}$,
Jan Palou\v{s}$^{1}$ \\
$^{1}$Astronomical Institute, Academy of Sciences of the Czech Republic, Bocni II 1401, 141 31 Prague, Czech Republic\\
$^{2}$I. Physikalisches Institut, Universit{\"a}t zu K{\"o}ln, Z{\"u}lpicher Str. 77, 50937 K{\"o}ln, Germany\\
$^{3}$School of Physics \& Astronomy, Cardiff University, The Parade, Cardiff CF24 3AQ, Wales, UK\\
$^{4}$Charles University in Prague, Faculty of Mathematics and Physics, Astronomical Institute, V Hole\v{s}ovi\v{c}ck\'ach 2, 180 00 Praha 8, Czech Republic}

\begin{document}

\newcommand{\eq}[1]{eq. (\ref{#1})}
\newcommand{\MSun}{M$_\odot$}
\newcommand{\gccm}{g\,cm$^{-3}$}
\newcommand{\thlim}{\theta_\mathrm{lim}}
\newcommand{\thsrc}{\theta_\mathrm{src}}
\newcommand{\thIF}{\theta_\mathrm{IF}}
\newcommand{\Npix}{N_{_{\rm PIX}}}
\newcommand{\Fij}{F_{i,j}}
\newcommand{\Fipj}{F_{i+1,j}}
\newcommand{\Fi}{F_{i}}
\newcommand{\Fip}{F_{i+1}}
\newcommand{\Ftotip}{F_{{\rm tot},i+1}}
\newcommand{\Phiij}{\Phi_{_{{\rm EUV},i,j}}}
\newcommand{\Phiipj}{\Phi_{_{{\rm EUV},i+1,j}}}
\newcommand{\deleEUV}{\delta_{e_{_{\mathrm{EUV}}}}}
\newcommand{\deltot}{\delta_{e_{_{\mathrm{EUV}}},\mathrm{tot}}}
\newcommand{\delcel}{\delta_{e_{_{\mathrm{EUV}}},\mathrm{cell}}}

\date{Accepted . Received }

\pagerange{\pageref{firstpage}--\pageref{lastpage}} \pubyear{2018}

\maketitle

\label{firstpage}

%%%%%%%%%%%%%%%%
\begin{abstract}
The treatment of radiative transfer with multiple radiation sources is a critical challenge in simulations of star formation and the interstellar medium. 
In this paper we present the novel {\sc TreeRay} method for solving general radiative transfer problems, based on reverse ray tracing combined with tree-based accelerated integration. 
We implement {\sc TreeRay} in the adaptive mesh refinement code {\sc FLASH}, as a module of the tree solver developed by W\"{u}nsch et al. 
However, the method itself is independent of the host code and can be implemented in any grid based or particle based hydrodynamics code. 
A key advantage of {\sc TreeRay} is that its computational cost is independent of the number of sources, making it suitable for simulations with many point sources (e.g. massive star clusters) as well as simulations where diffuse emission is important. 
A very efficient communication and tree-walk strategy enables {\sc TreeRay} to achieve almost ideal parallel scalings. 
{\sc TreeRay} can easily be extended with sub-modules to treat radiative transfer at different wavelengths and to implement related physical processes. Here, we focus on ionising radiation and use the On-the-Spot approximation to test the method and its parameters. 
The ability to set the tree solver time step independently enables the speedy calculation of radiative transfer in a multi-phase interstellar medium, where the hydrodynamic time step is typically limited by the sound speed of the hot gas produced in stellar wind bubbles or supernova remnants. 
We show that complicated simulations of star clusters with feedback from multiple massive stars become feasible with {\sc TreeRay}.

\end{abstract}
%%%%%%%%%%%%%%

\begin{keywords}
galaxies: ISM -- gravitation -- hydrodynamics -- ISM: evolution -- radiative transfer
\end{keywords}

%%%%%%%%%%%%%%%%%%%%%%%
\section{Introduction}%
%%%%%%%%%%%%%%%%%%%%%%%
%Steffi: FIRST VERSION

\textsc{Need for fast RT}. The turbulent, multi-phase structure of the interstellar medium (ISM) is shaped by the complex and non-linear interplay between gravity, magnetic fields, heating and cooling, and the radiation and momentum input from stars, in particular massive stars \citep[see e.g.][]{Agertz2013, Walch2015, KO2017, Peters2017}. Therefore, an efficient treatment of radiation transport in different energy bands (from the submillimeter to X-rays), and of the associated heating and cooling processes, is essential to simulate the structure and evolution of the ISM in detail, and to compare theoretical and numerical models with observations. A fundamental consideration is that radiation is emitted from different types of sources: point sources such as stars, extended sources like cooling shock fronts, diffuse sources like dust, as well as an ambient background radiation field. Hence, a modern radiative transfer algorithm must be able to handle multiple energy bands and multiple sources in an efficient manner.

\textsc{Overview of algorithm}. There are many ways to treat the radiation from point sources in 3D: Ray tracing \citep{Mellema2006, Gritschneder2009a} with {\sc HealPix} schemes \citep{Bisbas2009, Wise2011, Baczynski2015, Kim2017, Rosen2017}, long- and short-range characteristics \citep[e.g.][]{Rijkhorst2006}; flux-limited diffusion (FLD) \citep{Krumholz2007, Skinner2013}; combined schemes which work in optically thin and thick regions \citep{Paardekooper2010, Kuiper2010, Klassen2014}; moment methods  \citep{Petkova2012, Rosdahl2013, Kannan2019}; and backward radiative transfer schemes \citep[e.g.][]{Kessel2003, Altay2013, Grond2019TREVR}, like the one developed in this work.  In comparison with ray-tracing, FLD and moment based methods tend to be computationally less expensive and their cost does not depend on the number of sources. However, typically they do not capture certain features of the radiation field, for example shadowing (although see \citealt{Rosdahl2013}). Several code comparison projects have highlighted the advantages and short-comings of different radiative transfer methods \citep[e.g.][]{Iliev2006, Iliev2009, Bisbas2015}. Ultimately, even simple radiative transfer methods are expensive, at least as expensive as all the other elements of a simulation -- (magneto-)hydrodynamics ((M)HD), self-gravity, chemistry, heating and cooling -- together. Consequently, the numerical overhead of radiative transfer severely limits the astrophysical problems that can be addressed realistically in state-of-the-art 3D simulations, even on today's largest super-computers. 

\textsc{TreeRay basics}. In response to this challenge, we have devised {\sc TreeRay}, a new tree-based, backward radiative transfer scheme, which can handle multiple sources at acceptable extra cost, when running simulations that include self-gravity anyway. The computational cost of {\sc TreeRay} is independent of the number of emitting sources (as demonstrated in Section~\ref{sec:perf:nsrc}), and indeed every cell can be a source. Therefore, \textsc{TreeRay} can readily treat, for example, the ionizing radiation from multiple massive stars in a molecular cloud \citep{Haid2019}, as well as radiating shocks or other extended sources, on-the-fly in complex 3D (M)HD simulations. {\sc TreeRay} is an extension of the octtree-based solver for gravity and diffuse radiative transfer described in detail in \citet[][hereafter Paper~I]{Wunsch2018}, which has been developed for the Eulerian, adaptive mesh refinement (AMR) code {\sc FLASH} 4\footnote{The {\sc FLASH} code is maintained by the ASC/Alliances Center for Astrophysical Thermonuclear Flashes (Flash Center for Computational Science) at the University of Chicago (http://flash.uchicago.edu/site)} \citep{Fryxell2000}. The tree-solver of Paper~I is available with the official {\sc FLASH} release. Due to its efficiency, the {\sc TreeRay} scheme allows us to treat all dynamically relevant radiative processes in full three-dimensional simulations of the multi-phase ISM with an acceptably small error. {\sc TreeRay} evolved from the original {\sc TreeCol} method \citep{Clark2012} for treating the shielding of a diffuse background radiation field in Smoothed Particle Hydrodynamics. The shielding method has been presented as an {\sc OpticalDepth} module of the tree solver in Paper~I.

\textsc{TREVR}. The recently published TREVR code \citep{Grond2019TREVR}, which uses a tree data structure and reverse ray-tracing, is, we believe, the most closely related existing code. However, the two codes differ significantly in several respects. TREVR casts rays in the directions of sources, resulting in some (albeit weak) dependence of the computational cost on source number, while {\sc TreeRay} casts {\sc HealPix} rays (strictly speaking cones) in all directions, thereby covering the whole computational domain. These two fundamentally different approaches result in different types of numerical artefacts. Whilst {\sc TreeRay} uses iteration to deal with regions irradiated by multiple sources, in situations where the absorption coefficient depends on the radiation energy density, TREVR applies limits to the time step and uses a special refinement criterion to take account of the directional dependence of absorption in tree nodes. Finally, TREVR is implemented in the Smoothed Particle Hydrodynamic (SPH) code Gasoline, while {\sc TreeRay} is implemented in the AMR code {\sc Flash}, although both codes are general and could, in principle, be adapted to work with any hydrodynamics code.

\textsc{Outline}. The plan of the paper is as follows. In Section~\ref{sec:alg} we present the general algorithm and describe its implementation and coupling to the tree solver. In Section~\ref{sec:acc}, we perform a suite of static and dynamic tests to demonstrate the viability of the scheme, and we discuss the impact of the user-specified parameters on the accuracy. We restrict ourselves to the simple On-The-Spot approximation \citep{Oesterbrock1988} ({\sc TreeRay/OnTheSpot} module) for treating the interaction of UV radiation from massive stars with the ISM. The flexibility and performance of {\sc TreeRay} are demonstrated in Section~\ref{sec:SFFeedback}, where we present a more complex simulation of multiple massive stars dispersing a molecular cloud with their ionising radiation and winds. In Section~\ref{sec:perf}, we discuss the performance and parallel scaling of the algorithm, before summarizing in Section~\ref{sec:summary}.

% LAMPRAY
% http://adsabs.harvard.edu/abs/2018arXiv180905541F
% still not published in 201007

%%%%%%%%%
\section{The algorithm}
\label{sec:alg}
%%%%%%%%%

\textsc{Code structure}. {\sc TreeRay} is implemented as a module for the {\em tree-solver} described in Paper~I. The tree-solver provides general algorithms for building an octal tree, communicating the required parts of it to other processors, and traversing the tree to calculate quantities that require integration over the computational domain. It is a generalisation of the widely used algorithm devised by \citet{Barnes1986} to solve for self-gravity in numerical codes. Individual modules (e.g. {\sc Gravity} or {\sc TreeRay}) define which quantities should be stored on the tree, and provide subroutines to be called during the tree-walk to calculate integrated quantities like the gravitational acceleration or radiation energy density. {\sc TreeRay} itself consists of a general part plus submodules that treat the different physical processes needed to solve the Radiation Transport Equation (RTE). The {\sc TreeRay/OpticalDepth} submodule (described in Paper~I) is the simplest {\sc TreeRay} submodule; instead of solving the RTE, it simply sums contributions from different directions to obtain the corresponding optical depths. Here we focus on the {\sc TreeRay/OnTheSpot} submodule, to illustrate how {\sc TreeRay} solves the RTE for ionising radiation. 

\begin{figure}
\includegraphics[width=\columnwidth]{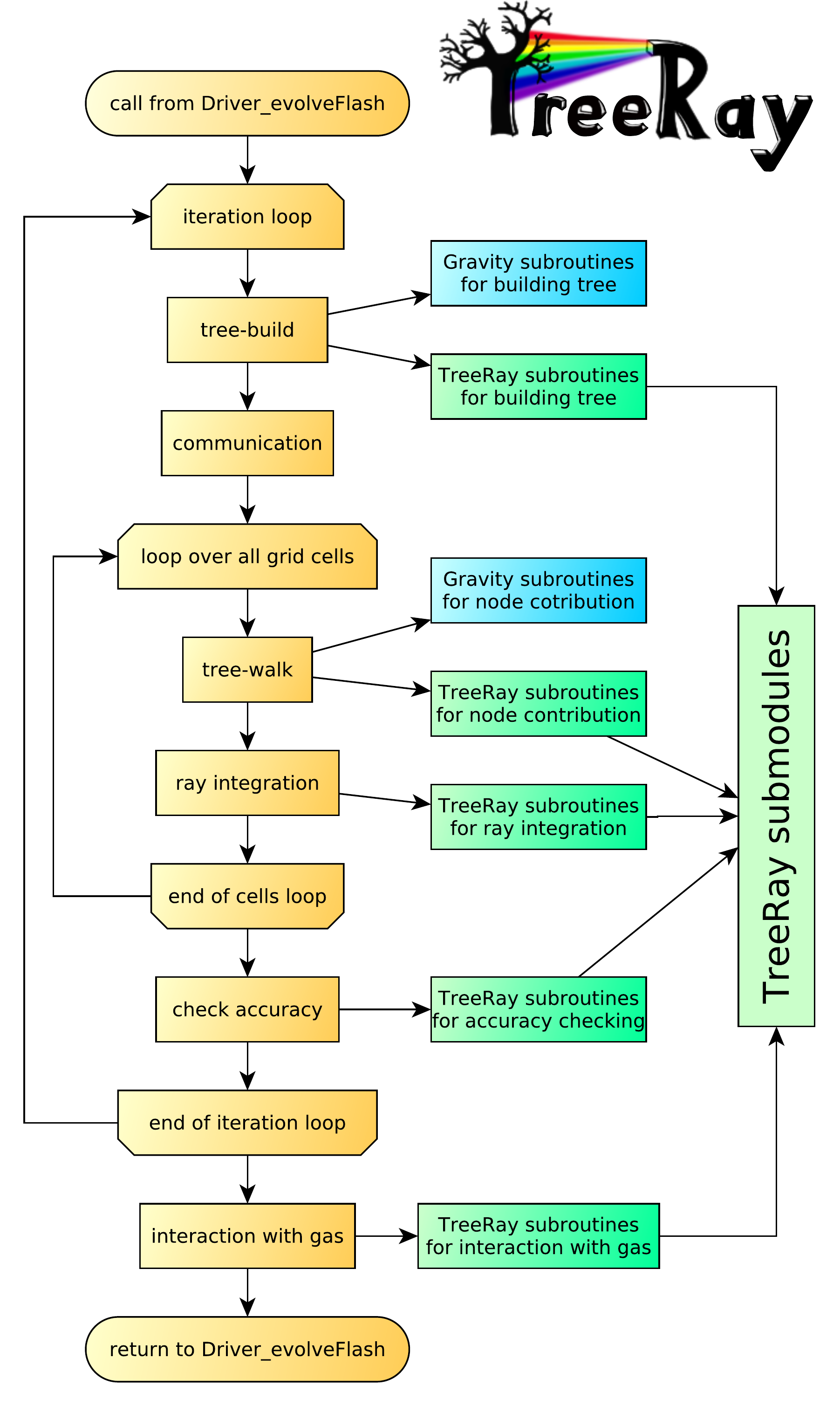}
\caption{
The tree solver flowchart and its connection to the {\sc Gravity} and {\sc TreeRay} modules. All general tree solver routines are yellow. {\sc Gravity} routines are blue. {\sc TreeRay} routines are green. }
\label{fig:flowchart}
\end{figure}

\textsc{New features in tree solver}. The simplified flowchart in Fig.~\ref{fig:flowchart} shows the connection between tree solver and {\sc TreeRay}\footnote{The full scheme of the tree solver and its modules down to the level of individual subroutines can be found at \url{http://galaxy.asu.cas.cz/pages/treeray}.}. In comparison to Paper~I, two modifications have been made. Firstly, the three main steps of the tree-solver (tree-build, communication and tree-walk) can be called several times (outer iteration loop in Fig.~\ref{fig:flowchart}) to calculate correctly absorption in regions irradiated by more than one source. Secondly, the tree walk is now called on a cell-by-cell basis (instead of block-by-block), because the temporary data structures storing quantities needed for solving the RTE are too big to be stored in memory for all the grid cells of a block.

\textsc{Algorithm flow}. The algorithm proceeds as follows. First, the tree is built to store quantities provided by the {\sc TreeRay} modules, and appropriate parts of the tree are communicated to other processors. Next, the tree is traversed for each cell (hereafter a {\it target cell}) and these quantities are mapped onto a system of rays originating at the target cell and covering the whole computational domain. Then, the RTE is solved along each ray and the radiative fluxes arriving at each target cell are calculated and converted to a radiation energy density. Finally, the whole process (starting from the tree-build) is repeated until the radiation energy density converges everywhere, i.e. until it does not change by more than a user defined fraction between successive iterations. Below, we describe individual steps in detail.

\subsection{Tree build}
\label{sec:treebuild}

\textsc{Quantities added to the tree}. The general {\sc TreeRay} module does not add any quantities to the tree; the tree-solver itself stores four numbers on each node, its mass and centre of mass. However, the {\sc TreeRay} submodules typically need to store two or three numbers for each energy band on each tree node (node index $n$). In the case of the {\sc TreeRay/OnTheSpot} submodule, we are concerned with Extreme Ultraviolet  radiation (EUV); i.e. radiation that can ionise hydrogen. The module stores the rate of emission of EUV photons within the node ($\varepsilon_{_{{\rm EUV,}n}}$), the rate of recombination of hydrogen into excited levels ($\alpha_{_{{\rm EUV,}n}}$), and the mean EUV radiation energy density from the previous iteration step ($e_{_{{\rm EUV,}n}}$):

\begin{eqnarray}
\varepsilon_{_{{\rm EUV,}n}} & = & \max \left(\sum_{c} \dot{n}_{{\rm EUV,}c} - \alpha_{\rm B} n_{{\rm H,}c}^2 dV_c , 0 \right)\,,\label{eq:os:ems}\\
\alpha_{_{{\rm EUV,}n}}   & = & \max \left(\sum_{c} \alpha_{\rm B} n_{{\rm H,}c}^2 dV_c - \dot{n}_{{\rm EUV,}c}, 0 \right)\,, \label{eq:os:abs}\\
e_{_{{\rm EUV,}n}} & = & \sum_{c} e_{_{{\rm EUV,}c}} dV_c\,.  \label{eq:os:erd}
\end{eqnarray}
Here $\dot{n}_{{\rm EUV,}c}$ is the rate at which EUV photons are emitted from hot stars in cell $c$, 
$n_{{\rm H,}c}$ is the number density of hydrogen nuclei, $\alpha_{\rm B}$ is the recombination coefficient into excited states only, $e_{_{{\rm EUV,}c}}$ is the radiation energy density calculated for cell $c$ in the previous iteration, and $dV_c$ is the cell volume. Note that the photons emitted and absorbed within cell $c$ are subtracted from each other in Eqs.~\ref{eq:os:ems} and \ref{eq:os:abs}, as this is more accurate than doing it later, during the RTE integration along rays, when both quantities have been modified by the approximations inherent in mapping the tree nodes onto the rays.

\textsc{Mapping sources onto the grid}. In order to obtain $\dot{n}_{{\rm EUV,}c}$, each source of radiation, $i$, characterized by the rate at which it emits EUV photons, $\dot{N}_{\mathrm{EUV,}i}$ and its radius, $r_i$, is mapped onto the grid before the tree-build. During the mapping, $\dot{N}_{\mathrm{EUV,}i}$ is divided between the grid cells which the source intersects, in proportion to the intersecting volumes. 

\textsc{Communication.} After the tree is built, it is communicated in the same way as described in Paper~I (Section~2.1). In summary, the code first distributes information about all the block positions and sizes to all the processors. Then, each processor runs a tree walk with target points at the closest point to the local node of all the blocks of a given remote sub-domain, and determines which local nodes will be opened when the tree walk is executed at the processor calculating that sub-domain. In this way, the local processor determines lists of the nodes needed at all the remote processors, and sends the nodes there. This ensures that all the information about the nodes to be opened during the tree walk is present at each processor, and at the same time it minimizes the amount of data communicated.

\subsection{Structure of rays}
 
\textsc{Ray-arrays}. Before the tree is walked for each target cell, three arrays (\texttt{cdMaps}, \texttt{rays}, and \texttt{raysEb}) are created in order to store the results of the tree-walk. The first one, \texttt{cdMaps} (standing for column density maps), is two dimensional and has size $\Npix\!\times l_q$, where $\Npix$ is the number of directions and $l_q$ is the number of quantities. The directions are defined with the {\sc HealPix} algorithm \citep{Gorski2005}, which tessellates a unit sphere into pixels of equal solid angle, each with a unit vector, $\hat{\boldsymbol n}_k$, which points from the sphere centre (target point) to the centre of the pixel. The value $\Npix$ is specified at the outset by setting the {\sc HealPix} level $N_{\rm side}$, with possible values $N_{\rm side} = 1, 2, 4, 8,\,.\,.\,.\,$, corresponding to $\Npix = 12 N_{\rm side}^2 = 12, 48, 192, 768,\,.\,.\,.\,$. The \texttt{cdMaps} array is used by the {\sc TreeRay/OpticalDepth} module and has already been described in Paper~I.\footnote{Note that in Paper~I this array had an additional dimension going through all the cells within a block, because the tree-walk was run on a block-by-block, rather than a cell-by-cell, basis.}

\textsc{rays}. The second array, \texttt{rays}, adds another dimension representing radial distance from the target point. We define $N_r$ {\em evaluation points} on each ray leading from the target point in the direction of vector $\hat{\boldsymbol n}_k$. Each ray is associated with a volume given by the area of the {\sc HealPix} pixel extended along the radial direction. Thus we create a system of rays pointing from each target point in $\Npix$ directions, and covering the whole computational domain. The \texttt{rays} array has size $N_r\times N_{_{\rm PIX}}\times m_q$, where $m_q$ is the number of quantities mapped onto the rays.

\textsc{raysEb}. The last array, \texttt{raysEb}, adds another dimension, which is necessary in cases involving multiple energy bands. Its size is ($N_r\times N_\mathrm{eb}\times N_{_{\rm PIX}}\times n_q$, where $N_\mathrm{eb}$ is the total number of energy bands for all sub-modules and $n_q$ is the maximum number of quantities per energy band (by default $3$). Typically, each sub-module requires one or more energy bands to treat radiation at different wavelengths (or radiation involved in different physical processes), and each energy band uses its own emission coefficient, absorption coefficient and radiation energy density, which are stored in the \texttt{raysEb} array. For simplicity, we refer to all three arrays as \texttt{ray-arrays}, considering \texttt{cdMaps} to be a special \texttt{ray-array} with only one data point in each radial direction.

\textsc{Evaluation points on rays}. Each ray intercepts $N_r$ evaluation points, where the quantities mapped onto the ray (e.g. emission and absorption coefficients) are set. Evaluation points are located so that the distances between successive points (hereafter ray {\em segment} lengths), are proportional to the distance from the target cell. This is because the size of tree nodes that need to be opened also increases -- often linearly -- with distance from the target point. Hence, starting with the first evaluation point at the target point ($r_0 \equiv 0$), the radial coordinate of the $i^{\rm \,th}$ evaluation point is
\begin{equation}
\label{eq:etaR}
r_i = \frac{\Delta x\;i^2}{2\,\eta_R^2}\,.
\end{equation}
Here $\Delta x$ is the size of the smallest cell in the simulation and $\eta_R$ is a user defined parameter that controls the resolution in the radial direction. By default, $\eta_R = 2$, which ensures that (if the Barnes-Hut criterion for node acceptance is used, see Section~\ref{sec:MACs}) the segment lengths correspond, approximately, to half the size of the nodes with which the target cell interacts during the tree walk, i.e.
\begin{equation}
\label{eq:bhmac}
|r_{i+1} - r_i| \sim \frac{\theta_\mathrm{lim}\times d}{2}\,,
\end{equation}
where $d$ is the distance between the target cell and the interacting node. The maximum ray length, $L_\mathrm{ray}$, is set automatically to the length of the three-dimensional diagonal of the computational domain, and hence the number of evaluation points along each ray is
\begin{equation}
\label{eq:Nr} 
N_R = \eta_R \times \texttt{floor}\left(\sqrt{\frac{2 L_{\rm ray}}{\Delta x}}\, \right) + 1\,,
\end{equation}
where \texttt{floor($x$)} is the largest integer less than or equal to $x$. Each module $M$ includes a user-defined parameter, $L_{\mathrm{ray,}M}$, by which the user can specify the maximum distance from the target cell to which the calculation (mapping and/or RTE solution) should be carried out.

\subsection{Tree-walk}
\label{sec:treewalk}

%%%%%
\begin{figure}
\includegraphics[width=\columnwidth]{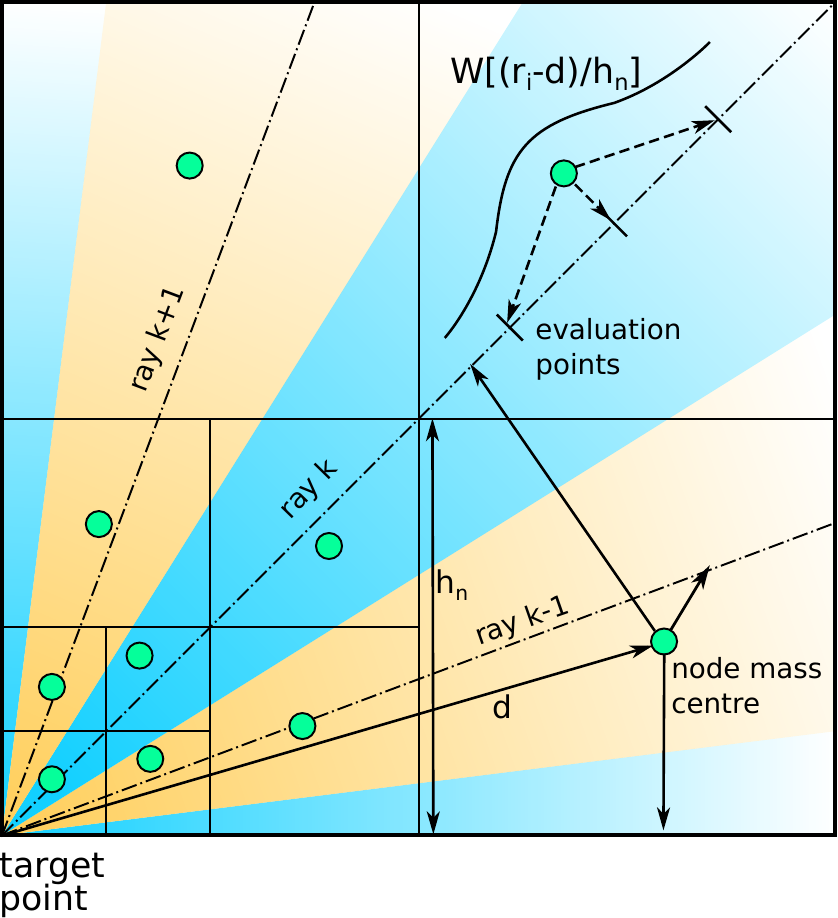}
\caption{Mapping tree nodes onto rays during the tree walk. The tree nodes that need to be opened during the tree walk for a target point in the bottom left corner are denoted by solid line squares.  The system of rays which is cast from the target point is represented by the blue and yellow areas. The green circles are node mass centres. The quantities that are stored in each of the tree nodes are distributed to all rays which a node intersects (solid arrows), weighted by the relative intersection volume. For a given ray, the quantities are distributed to the so-called evaluation points along the ray using weights given by the kernel function $W[(r_i-d)/h_n]$ (dashed arrows). }
\label{fig:mapping}
\end{figure}
%%%%%

\textsc{Mapping nodes onto rays}. When the tree is walked, the quantities stored on the tree nodes are mapped onto the \texttt{ray-arrays} in a two-step process. First, each mapped quantity, e.g. the node mass, is divided amongst different rays in proportion to the volume of the intersection between the node and the ray (see Fig.~\ref{fig:mapping}). This is implemented as described in Paper~I, using a pre-calculated table of intersection volumes between a ray and a node with given angular coordinates ($\theta$, $\phi$) and angular size $\psi$. Second, the part of the node belonging to the ray is divided between evaluation points, $r_i$, using a kernel function $W(x)$ with $x \equiv (r_i-d)/h_n)$ where $d$ is the distance of the node mass centre from the target point, and $h_n$ is the node linear size. This mapping also uses a pre-calculated table in which the weights of the evaluation points are recorded as a function of $r_i$, $d$, and $h_n$.

\textsc{Kernels}. Three kernels are presently available, as discussed in detail in Appendix \ref{app:kernels}. The first is a Gaussian kernel, $W_g(x)$ (equation~\ref{eKernGaussian}), truncated at $x=\sqrt{3}/2$. The second, $W_p(x)$ (equation~\ref{eKernPolynomial}), has the form of a piece-wise polynomial of third order and has been obtained by fitting the mean intersection between a uniform cubic node and a randomly oriented line. {\sc TreeRay} uses $W_p(x)$ by default for mapping the node masses and volumes. The third, $W_f(x)$, is tailored to the requirements of radiative transfer, and is used with the {\sc OnTheSpot} module; the considerations informing the prescription for the kernel coefficients are described in detail in Appendix \ref{app:kernels}. Due to its nature, $W_f(x)$ is only used for mapping quantities related to radiative transfer (e.g. the radiation energy); all other quantities (e.g. mass) are mapped using $W_p(x)$.

\subsection{Multipole Acceptance Criteria (MACs)}
\label{sec:MACs}

\textsc{MACs in the tree solver}. During the tree walk for a given target cell, a node is accepted if it satisfies the {\em Multipole Acceptance Criterion -- MAC}. The simplest and most commonly used MAC is the Barnes-Hut (BH) MAC \citep{Barnes1986}. With the BH~MAC, a node of size $h_n$, at distance $d$ from the target cell, is accepted if 
\begin{equation}
    h_n/d < \theta_\mathrm{lim} \ ,
\end{equation}
where $\theta_\mathrm{lim}$ is a user-specified limiting opening angle. In Paper~I we describe several data-dependent MACs that make the {\sc Gravity} module calculation more efficient. Below we define two new MACs, IF~MAC (standing for Ionisation Front) and Src~MAC (standing for Sources), designed for the {\sc TreeRay/OnTheSpot} module, and also applicable to other {\sc TreeRay} submodules. When these MACs are combined with the {\sc Gravity} module MACs to ensure accurate gravity calculation, smaller nodes are opened in regions of dense, highly structured gas, and this leads to an increase in the computational cost of the tree walk. However, this does not impact the accuracy or performance of the {\sc TreeRay} RTE solution, since the small nodes are smeared out onto the {\sc Healpix} rays with the angular resolution given by $N_{_{\rm PIX}}$.

\textsc{BH~MAC}. When the BH~MAC is used together with {\sc TreeRay}, a sensible choice is to set  $\theta_{\rm{lim}}\simeq (4\pi/\Npix)^{1/2}$. This ensures that the angular size of accepted tree nodes is approximately the same as the angle assigned to each {\sc Healpix} ray (see analysis of this criterion for the {\sc OpticalDepth} module in Paper~I, \S 3.3.1).

\textsc{New MACs}. Both the IF~MAC and the Src~MAC are controlled by limiting opening angles, respectively $\theta_\mathrm{IF}$ and $\theta_\mathrm{Src}$. If the IF~MAC is adopted, the code records for each node, $n$, both the total mass, $m_n$, and the mass of ionised gas, $m_\mathrm{ion}$ (defined as a threshold on the ionised hydrogen abundance or on the temperature). Then, during the tree walk, nodes for which $\delta_\mathrm{IF} m_n < m_\mathrm{ion} < (1-\delta_\mathrm{IF})m_n$ (where $\delta_\mathrm{IF} = 10^{-8}$) are assumed to include an ionisation front. Such nodes are accepted only if 
\begin{equation}
    h_n/d < \theta_\mathrm{IF} \ .
\end{equation}
Similarly, if the Src~MAC is adopted, nodes that include sources of radiation, i.e. $\varepsilon_n > 0$, are accepted only if 
\begin{equation}
    h_n/d < \theta_\mathrm{Src} \ .
\end{equation}

\subsection{Radiation transport equation}
\label{sec:rte}

\begin{figure*}
\includegraphics[width=\textwidth]{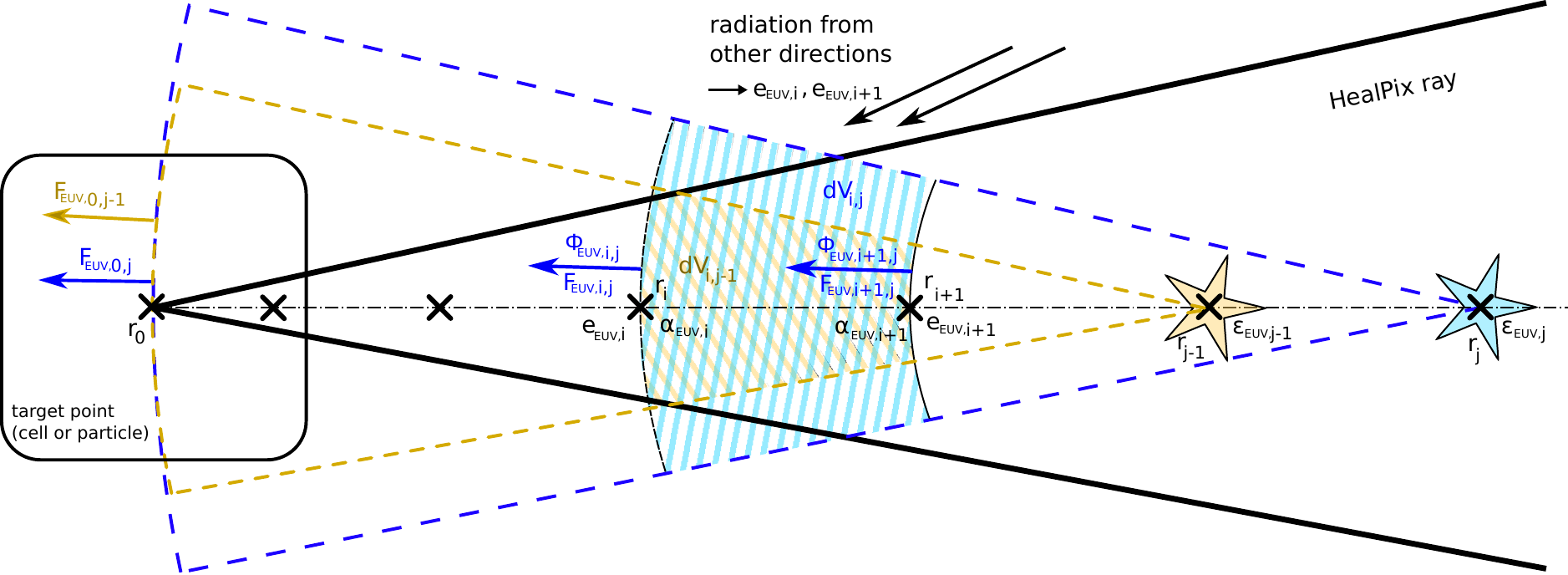}
\caption{Schematic view of {\sc TreeRay}, depicting the solution of the RTE along a single ray (bounded by thick black lines) towards a given target cell (on the left) after all the required quantities have been mapped onto the evaluation points (crosses). Each evaluation point is characterised by its distance from the target point,  $r_i$, its absorption coefficient, $\alpha_i$, the emission coefficient of all sources mapped onto it, $\varepsilon_i$, and the local radiation energy density from the previous time step, $e_i$. See Section~\ref{sec:treewalk} and Fig.~\ref{fig:mapping} for details of the mapping. For the furthest two evaluation points, $j-1$ and $j$ (marked by stars, with emission coefficients, $\varepsilon_{j-1}$ and $\varepsilon_j$), the yellow (blue) shaded area represents the volume $dV_{i,j-1}$ ($dV_{i,j}$) of the cone of radiation from source $j-1$ ($j$) in the segment between evaluation points $i$ and $i+1$ (equation~\ref{eq:dV}). The procedure described in Section~\ref{sec:rte} calculates the photon rates $\Phi_{i,j}$ and fluxes $F_{i,j}$ at each evaluation point. Fluxes $F_{0,j}$ at the target point are summed to give the radiation energy density there.}
\label{fig:rte}
\end{figure*}

\textsc{General RTE}. After the tree walk is completed, the \texttt{ray-arrays} contain all the quantities needed to solve the Radiation Transport Equation (RTE) using the reverse ray-tracing method \citep{Altay2013}. For frequency band $\nu$, the RTE takes the form\footnote{The signs on the righthand side are reversed, as compared with the standard form, because the radiation propagates in the negative direction with respect to coordinate $r$ originating at the target point.}
\begin{equation}
\label{eq:RTE}
\frac{dI_\nu}{dr} = -\varepsilon_\nu + \alpha_\nu I_\nu
\end{equation}
where $\varepsilon_\nu$ and $\alpha_\nu$ are the emission and absorption coefficients. The RTE must be solved along each of the $\Npix$ rays cast from a target point. On each ray, at each evaluation point $i$ (at distance $r_i$ from the target point along the ray) radiation transport is regulated by (i) the local emission coefficient, $\varepsilon_{\nu,i}$; (ii) the local absorption coefficient, $\alpha_{\nu,i}$; and (iii) the radiation energy density from the previous iteration step, $e_{\nu,i}$. These quantities have all been mapped onto the evaluation points during the tree walk. $e_{\nu,i}$ is only required if the emission or absorption coefficients depend on it. If the emission or absorption coefficients {\it do not} depend on $e_{\nu,i}$, it is straightforward to integrate equation~(\ref{eq:RTE}) along each of the $\Npix$ rays, and thereby obtain the (direction-dependent) radiation intensity $I_{\nu,0,k}\;(k\!=\!1\,{\rm to}\,\Npix)$ at the target point. These $I_{\nu,0,k}$ can then be summed to obtain a new estimate of the radiation energy density at the target point, and this can in turn be used to calculate radiative ionisation and/or heating rates. However, if the emission or absorption coefficients {\it do} depend on $e_{\nu,i}$, more elaborate formulations of the problem may be appropriate, and this is the case for the {\sc TreeRay/OnTheSpot} module.

\textsc{RTE in OnTheSpot}. In the {\sc TreeRay/OnTheSpot} module treating the EUV radiation, we use different quantities from those in equation~(\ref{eq:RTE}) to characterise the radiation field. $\varepsilon_{_{\rm{EUV,}i}}$ gives the rate of emission of EUV photons from sources (i.e. hot stars) associated with evaluation point $i$. $\alpha_{_{\rm{EUV,}i}}$ gives the rate of recombination of hydrogen into excited levels, per unit volume, at evaluation point $i$; in the {\sc OnTheSpot} approximation, such recombinations are exactly balanced by photo-ionisations
\footnote{Although the On-the-Spot Approximation does not strictly require ionisation/recombination equilibrium for recombinations into excited states (Case~B), it is only under extreme circumstances that the On-the-Spot Approximation is valid and ionisation/recombination equilibrium is not.}, 
so $\alpha_{_{\rm{EUV,}i}}$ is also the rate at which EUV photos are destroyed in unit volume at evaluation point $i$. $\Phi_{_{\rm{EUV,}i,j}}$ gives the rate at which EUV photons emitted into unit solid angle by the sources associated with evaluation point $j$ reach the spherical surface through evaluation point $i$ (see Fig.~\ref{fig:rte}).

\textsc{Integration procedure}. The integration procedure first calculates 
\begin{equation}
\Phi_{_{{\rm EUV},i=j,j}} = \frac{\varepsilon_{_{{\rm EUV,}j}}}{4\pi}, \qquad \Phi_{_{{\rm EUV},i\ne j,j}} = 0\,,
\end{equation}
at all the evaluation points that have sources, and sets it to zero everywhere else. Then, for all evaluation points, $j$, that have sources, starting from $N_{r}-1$, it cycles over the evaluation points from $i=j-1$ to $i=0$ and calculates
\begin{equation}
\Phiij = \Phiipj - \mathcal{R}_{i+1,j}\,,
\end{equation}
truncating it to zero if it becomes negative. $\mathcal{R}_{i+1,j}$ is the rate at which photons from evaluation point $j$ are lost due to recombinations in the ray segment between $r_{i+1}$ and $r_{i}$, and is given by
\begin{equation}\label{EQN:Rij}
\mathcal{R}_{i,j} = \frac{1}{2}\,(\alpha_{_{{\rm EUV},i}} + \alpha_{_{{\rm EUV},i+1}})\,\mathrm{d}V_{i,j} \times \zeta_{\mathrm{d},i+1} \times \zeta_{\mathrm{s},i+1,j} \times \zeta_{\mathrm{g},i,j}\,,
\end{equation}
where
\begin{equation}\label{eq:dV}
\mathrm{d}V_{i,j} = [(r_j-r_i)^3 - (r_j-r_{i+1})^3]/3
\end{equation}
is the volume of the cone of radiation from source $j$ in the segment between $r_{i}$ and $r_{i+1}$ (see Fig.~\ref{fig:rte}). The absorption rate is further corrected by three factors $\zeta_{\mathrm{s},i+1,j}$, $\zeta_{\mathrm{d},i+1}$,  and $\zeta_{\mathrm{g},i,j}$, which are defined below.

\textsc{Calculation of fluxes}. The contribution from the sources associated with evaluation point $j$ to the flux of EUV photons through evaluation point $i$ is then given by 
\begin{equation}
F_{_{{\rm EUV,}i,j}} = \Phiij / (r_j - r_i)^2 \ ,
\end{equation}
and the total flux of EUV photons at $r_i$ from all sources $j$ on the ray at $r_j > r_i$ is
\begin{equation}
F_{_{{\rm EUV},i}} = \displaystyle\sum_{j=i+1}^{N_{R}} F_{_{{\rm EUV,}i,j}} \ .
\end{equation}

\textsc{Correction factors in equation~(\ref{EQN:Rij})}. 
The first correction factor in equation~(\ref{EQN:Rij}), $\zeta_{\mathrm{s},i+1,j}$, accounts for recombinations that absorb photons coming along the same computed ray but from sources other than $j$. It is simply the ratio of the contribution to the flux from source $j$, $\Fipj$, to the total flux along the ray, $\Fip$,
\begin{equation}\label{eq:Rcorr:s}
\zeta_{\mathrm{s},i+1,j} =  \frac{F_{_{{\rm EUV,}i+1,j}}}{F_{_{{\rm EUV,}i+1}}} \ .
\end{equation}
The second correction factor in equation~(\ref{EQN:Rij}), $\zeta_{\mathrm{d},i+1}$, follows a similar logic. It accounts for recombinations that absorb photons passing through the ray segment in other directions than along the computed ray, and it is defined as the ratio between $F_{_{{\rm EUV,}i+1,j}}$ and the trace of the radiation pressure tensor $\mathbf{P}_{_{{\rm EUV},i+1}}$ (expressed in photons per unit area and time). The latter is obtained from the radiation energy density as $\mathrm{tr}(\mathbf{P}_{_{{\rm EUV},i+1}}) = e_{_{{\rm EUV,}i+1}} c / h\nu_{_{\rm EUV}}$, where $h\nu_{_{\rm EUV}}$ is the mean energy of an EUV photon and $c$ is the speed of light. The correction factor $\zeta_{\mathrm{s},i+1,j}$ is further truncated to lie between $0$ and $2$ to deliver a moderate convergence rate during the tree solver iterations, yielding
\begin{equation}\label{eq:Rcorr:d}
\zeta_{\mathrm{d},i+1} =  \min\left( 2, \frac{F_{_{{\rm EUV,}i+1}} h\nu_{_{\rm EUV}}}{e_{_{{\rm EUV,}i+1}} c} \right) \ .
\end{equation}
The third correction factor in equation~(\ref{EQN:Rij}), $\zeta_{\mathrm{g},i,j}$, corrects for the geometry of the ray segments. The segment length $r_{i+1}-r_i$, controlled by parameter $\eta_R$, can be substantially smaller than the tangential extent of the ray $\sim r_i\times N_{_{\rm side}}$. In this situation, the flux from sources close to $r_{i+1}$ would be too high, leading to an over-representation of these sources in equation~(\ref{EQN:Rij}). Therefore, we add a correction of the form
\begin{equation}\label{eq:Rcorr:g}
\zeta_{\mathrm{g},i,j} =  \min\left( 1, \frac{N_{_{\rm side}} (r_{j} - r_{i})}{2r_{j}} \right) \ .
\end{equation}
Note that the minimum in equation~(\ref{eq:Rcorr:g}) ensures that the correction is only applied to segments with $r_j - r_i < 2r_j/N_{_{\rm side}}$.

\textsc{Sum over directions}. Finally, the fluxes $F_{_{{\rm EUV,}0,k}}$ incident on the target point (i.e. i = 0) from the directions of all the {\sc HealPix} rays ($k\!=\!1\,{\rm to}\,\Npix$) are summed to give a new estimate of the radiation energy density $e_{_{{\rm EUV,}0}}^{\rm new}$. It is assumed that the target point is a small sphere with radius $r_\mathrm{tp}$.\footnote{$r_\mathrm{tp}$ is half of the grid cell size (if the target point is a grid cell) or the sink particle accretion radius (if the target point is a sink particle).} Using the approximation of single direction radiation, the radiation energy density is
\begin{equation}
    e_{_{{\rm EUV,}0}}^\mathrm{new} = \sum_{k=1}^{\Npix} \frac{F_{_{{\rm EUV,}0,k}} h\nu_{_{\rm EUV}}}{c}.
\end{equation}

\subsection{Iterations and error control}
\label{sec:err:cntrl}

\textsc{Need for iterations}. If the absorption or emission coefficients depend on the ambient radiation field --- as in the {\sc TreeRay/OnTheSpot} sub-module, where photons coming from different directions compete to be absorbed balancing recombinations --- the code must iterate to find an acceptable estimate of the EUV radiation energy density, $e_{_{\rm EUV}}$. Modules and sub-modules that do not need iterations are executed only once, during the first iteration step, to save computing time. The first iteration starts with the radiation field $e_{_{\rm EUV}}$ from the previous hydrodynamic time step. This significantly reduces the number of iterations needed, because changes in the distribution of gas and sources between time steps are typically small. In most cases, fewer than 10 iterations are needed.

\textsc{Error control}. Iteration is terminated once the fractional change in the radiation field between successive iterations, $\deleEUV$, falls below a user-defined limit $\epsilon_\mathrm{lim}$ (default value is $10^{-2}$). Currently, two ways of determining the fractional change in the radiation field are implemented. The first considers the total radiation energy,
\begin{equation}\label{eq:delta:er:tot}
\deltot = \frac{2\left( \sum e_{_{{\rm EUV,}c}}^\mathrm{new} \mathrm{d}V_{c} - \sum e_{_{{\rm EUV,}c}} \mathrm{d}V_{c} \right)}
{\sum e_{_{{\rm EUV,}c}}^\mathrm{new} \mathrm{d}V_{c} + \sum e_{_{{\rm EUV,}c}} \mathrm{d}V_{c}},
\end{equation}
where subscript $c$ denotes the quantity in a grid cell with volume $ \mathrm{d}V_{c}$, and the sums are taken over all grid cells in the computational domain. This criterion is similar to the one adopted by \citet{Dale2012}, who checked for the fractional change in the total mass of ionised gas; it works well when there are a few sources with comparable luminosity. The second criterion considers the change in the radiation field on a cell-by-cell basis,
\begin{equation}\label{eq:delta:er:cell}
\delcel = \max_{c} \left( \frac{e_{_{{\rm EUV,}c}}^\mathrm{new} - e_{_{{\rm EUV,}c}}}{e_{_{\mathrm{EUV,norm,}c}}} \right),
\end{equation}
where the maximum is taken over all grid cells $c$ in the computational domain, and the normalizing energy is
\begin{equation}
    e_{\mathrm{EUV,norm,}c} = \max(e_{_{{\rm EUV,}c}}, e_{_{\mathrm{EUV,med}}}) \ .
\end{equation}
Here $e_{\mathrm{EUV,med}}$ is the median of all the $e_{_{{\rm EUV,}c}}$ values over the whole computational domain. The median is used to avoid zero or nearly zero normalizations, because $e_{_{\mathrm EUV}}$ can have arbitrarily small values. The second criterion, using $\delcel$, is safer and is set as the default.

\textsc{Interaction with gas}. The calculated radiation energy density is used to modify the properties of the target point. If the target point is a grid cell, the code can for instance update the temperature or the chemical composition (including the ionisation degree). If $e_{_{{\rm EUV,}0}}$ is non-zero, the {\sc TreeRay/OnTheSpot} module implements two possible treatments: (i) the temperature and ionisation degree are given user defined values (e.g. $10^4$\,K and $1.0$, respectively), or (ii) $e_{_{{\rm EUV,}0}}$ is passed to the {\sc Chemistry} module \citep[see][for details, and Section~\ref{sec:SFFeedback} for a test case including chemistry]{Haid2018}.

\subsection{Sources of radiation}
\label{sec:sources}

\textsc{Source properties}. For some {\sc TreeRay} sub-modules, the emission coefficient $\varepsilon_{\nu ,c}$ in cell $c$ is derived directly from the quantities that describe the gas in the cell. Other sub-modules, including {\sc TreeRay/OnTheSpot}, use radiation sources that are independent of the grid (e.g. sink particles representing stars or star clusters). Such sources are characterised by their position, luminosity (e.g. in the case of the {\sc TreeRay/OnTheSpot} module, number of EUV photons emitted per second), and radius. These properties can either be read from a file at the start of a simulation and stay constant for the whole run, or they can be obtained from sink particles. In the latter case the sinks move and their luminosities (and in principle also their radii) can vary with time. The {\sc FeedbackSinks} module can be used to calculate the luminosities of the stars and stellar clusters that sink particles represent, as functions of their age, mass and mass accretion history \citep[see][for details]{Gatto2017, Peters2017}.

\textsc{Mapping onto the grid}. At each call of the tree solver, before the tree is built, all sources are mapped onto the grid, assuming that the sources are uniform spheres. In this way, the emission coefficient in each grid cell, which geometrically intersects with the volume of the source is set to a fraction of the source luminosity proportional to the intersection volume. In order to keep the mapping process fast, it is performed in two stages: creating a list of sources intersecting with each block, and then calculating the intersection of each cell with each source on the list.

\subsection{Boundary conditions}
\label{sec:bc}

\textsc{Boundary condition types}. Three types of boundary condition are available for the tree solver; they are independent of the boundary conditions for the hydrodynamic solver. The first two types ({\it isolated} and {\it periodic}) have already been described in Paper~I, the third one ({\it corner of symmetry}) is newly implemented and described in more detail below. {\it Isolated} boundary conditions assume that all source quantities (e.g. for gravity or emission of radiation) are zero outside the computational domain. {\it Periodic} boundary conditions invoke periodic copies of each tree node during the tree walk, and use only the closest copy to the target point. The {\sc Gravity} module uses the Ewald method to take into account all the periodic copies, but calculations with the {\sc TreeRay} sub-modules are presently limited to the nearest periodic copy.

\textsc{Corner of Symmetry}. {\em Corner of Symmetry} boundary condition allow the user to simulate one-eighth of the problem under investigation, if the problem has the appropriate symmetry. We refer to the truly simulated part of the computational domain as the \textit{original}, and we specify one corner of it as the \textit{corner}. During the mapping of the tree onto the rays we reconstruct the entire problem by spawning 7 ghosts of the \textit{original}. The first 3 ghosts are copies of the \textit{original} mirrored in the faces adjacent to the \textit{corner}. The remaining 4 ghosts are point reflections of the first 4 ghosts through the \textit{corner}. One can show that if the MACs discussed in Section~\ref{sec:MACs} hold for the \textit{original}, they also hold for all 7 ghosts. One useful application of this boundary condition is to problems with spherical symmetry such an the Spitzer test (see Section~\ref{sec:spitzer}, model (p)), giving us as an effective resolution of level $n$ while actually only computing at level $n-1$.

\subsection{Tree solver time step}
\label{sec:tsdt}

\textsc{ABU in TreeRay}. In Paper~I we introduced the {\em adaptive block update} (ABU) method, which improves the code performance by only updating blocks where the quantities calculated by the tree solver change significantly; we also illustrated the benefits when ABU is used with the {\sc Gravity} or {\sc TreeRay/OpticalDepth} modules. However, ABU does not deliver such significant benefits with the {\sc TreeRay/OnTheSpot} module, or any other module where the absorption or emission coefficients depend on the local radiation energy density. This is because calculating the radiation energy density couples together large regions of space, thereby requiring the code to update many blocks. 

\textsc{Tree solver time step}. However, in many applications the time scale for the physical processes which are calculated by the tree solver is much longer than the hydrodynamic time step. For example, in calculations with a multi-phase interstellar medium, the dense cold/warm gas is the main contributor to the gravitational potential as it contains most of the mass in the system and absorbs most of the radiation, and this gas typically moves with velocities smaller than $10$\,km/s \citep[e.g.][]{Girichidis2016}. On the other hand, the hydrodynamic time step calculated from the Courant condition can be very small, because it is defined by the hot gas (with sound speeds $\ga 300\,{\rm km/s}$) or stellar winds (with velocities $\ga 1000\,{\rm km/s}$), but the contribution of this hot or fast gas to the calculation of gravity and radiation is typically negligible. Since the tree solver is used to calculate the gas self-gravity and the EUV or longer-wavelength radiation field, it may not need to be called at every hydrodynamic time step; the decision whether this approximation is reasonable depends on the nature of the phenomena being simulated, and is left to the user. The frequency of calls to the tree solver is regulated by setting the time step, $\Delta t_\mathrm{ts}$, which can be longer than the hydrodynamic time step. A convenient way to set $\Delta t_\mathrm{ts}$ is to define a parameter
\begin{equation}
    v_\mathrm{tsdt} =  \Delta x / \Delta t_\mathrm{ts} ,
\end{equation}
where $\Delta x$ is the size of the smallest grid cell. $v_\mathrm{tsdt}$ then sets the maximum velocity with which the gas and/or sources relevant to the tree solver can move. By default, we set $v_\mathrm{tsdt} = \infty$ and hence $\Delta t_\mathrm{ts} = 0$, in which case the tree solver is called at each hydrodynamic time step. Examples of how setting $v_\mathrm{tsdt}$ to a finite value impacts the code performance and accuracy are given in Section~\ref{sec:SFFeedback}.

\subsection{Load balancing}
\label{sec:lb}

\textsc{Workload problem}. The computational cost of the tree solver is almost always dominated by the tree walk, and the most expensive operation is mapping the tree nodes onto the system of rays. If no radiation passes through a node, this operation is omitted, and as a result, the tree walk can be significantly cheaper in regions where no radiation is present. However, this may lead to a non-optimal distribution of the work load among different processors if each processor computes the same number of blocks (which is the default in {\sc Flash}). 

\textsc{Load balancing scheme}. Therefore, we implement a simple load balancing scheme, based on measuring the wall clock time needed to execute a tree walk for all grid cells in a block, $t_\mathrm{wl,blk}$. After each tree solver call, we collate the  $t_\mathrm{wl,blk}$ measurements and calculate their median value, $t_\mathrm{wl,med}$. Then we increase the workload weights, $w_\mathrm{blk}$, of blocks with $t_\mathrm{wl,blk} > t_\mathrm{wl,med}$ by the factor $t_\mathrm{wl,blk} / t_\mathrm{wl,med}$. These workload weights are then used by {\sc Flash} in the next redistribution of blocks amongst the processors; each processor is assigned a number of blocks such that the sum of their weights is approximately the same for all processors. Note that this scheme improves the code performance only if the computational time is dominated by the tree solver. Therefore we use this scheme, in combination with the tree solver time step, only for time steps when the tree solver is called. All other time steps are calculated with a default flat block distribution, i.e. the same number of blocks on each processor.

%%%%%%%%%%%%%%%%%%%%%%%%%%%%%%%%%%%%%%
\section{Accuracy and performance tests}
\label{sec:acc}

In this section we describe four tests of the {\sc TreeRay} algorithm illustrating its strengths and weaknesses, using idealised configurations of gas and sources. A more complex test, involving physical processes commonly included in astrophysical simulations of star formation and its feedback, is presented in Section~\ref{sec:SFFeedback}. All tests in Sections~\ref{sec:acc} and \ref{sec:SFFeedback} have been run on the IT4I/Salomon supercomputer cluster\footnote{\url{https://docs.it4i.cz/salomon/}} consisting of 1008 compute nodes, each equipped with two 12-core Intel Xeon E5-2680v3 @ 2.5 GHz processors and 2GB of RAM per core. The nodes are interconnected with the InfiniBand FDR56 network using the 7D Enhanced hypercube topology.

\subsection{Spitzer test}
\label{sec:spitzer}

\begin{table*}
\caption{Accuracy and performance of the Spitzer test.}
\label{tab:acc:spitzer}
\begin{center}
\begin{tabular}{l|l|c|c|r|c|c|c|l|c|c|c|c}
\hline
model & $l_r$ & $\theta_\mathrm{lim}$ & $\theta_\mathrm{IF}$, $\theta_\mathrm{Src}$ & $N_{_{\rm PIX}}$ & $\deleEUV$
& $\epsilon_\mathrm{lim}$ & $\eta_R$ & $R_\mathrm{src}$ & $e_\mathrm{IF}$ & $t_\mathrm{iter}$ & $t_\mathrm{tr}$ 
& $t_\mathrm{hydro}$ \\
\hline
(a) fiducial   & $5$ & $0.5$  & $\infty$ & $48$  & cell & $10^{-2}$ & $2$ & $1$         & $0.025$ & $0.24$ & $158$  & $3.4$ \\
(b) $\deleEUV = \mathrm{tot}$ & $5$ & $0.5$  & $\infty$ & $48$  & tot  & $10^{-2}$ & $2$ & $1$         & $0.024$ & $0.25$ & $78$   & $3.4$ \\
(c) $l_{r} = 4$ & $4$ & $0.5$  & $\infty$ & $48$  & cell & $10^{-2}$ & $2$ & $1$         & $0.011$ & $0.03$ & $9.2$  & $0.3$ \\
(d) $l_{r} = 6$ & $6$ & $0.5$  & $\infty$ & $48$  & cell & $10^{-2}$ & $2$ & $1$         & $0.040$ & $2.5$  & $3100$ & $44$  \\
(e) $\Npix = 12$ & $5$ & $1.0$  & $\infty$ & $12$  & cell & $10^{-2}$ & $2$ & $1$         & $0.173$ & $0.05$ & $65$   & $5.6$ \\
(f) $\Npix = 192$ & $5$ & $0.25$ & $\infty$ & $192$ & cell & $10^{-2}$ & $2$ & $1$         & $0.046$ & $1.5$  & $940$  & $3.4$ \\
(g) $\epsilon_\mathrm{lim} = 10^{-1}$ & $5$ & $0.5$  & $\infty$ & $48$  & cell & $10^{-1}$ & $2$ & $1$         & $0.024$ & $0.24$ & $125$  & $3.4$ \\
(h) $\epsilon_\mathrm{lim} = 10^{-3}$ & $5$ & $0.5$  & $\infty$ & $48$  & cell & $10^{-3}$ & $2$ & $1$         & $0.025$ & $0.24$ & $250$  & $3.4$ \\
(i) $\eta_R = 1$ & $5$ & $0.5$  & $\infty$ & $48$  & cell & $10^{-2}$ & $1$ & $1$         & $0.203$ & $0.21$ & $180$  & $3.5$ \\
(j) $\eta_R = 8$ & $5$ & $0.5$  & $\infty$ & $48$  & cell & $10^{-2}$ & $8$ & $1$         & $0.033$ & $0.39$ & $260$  & $3.4$ \\
(k) $R_\mathrm{src} = 4$\,gc & $5$ & $0.5$  & $\infty$ & $48$  & cell & $10^{-2}$ & $2$ & $4$         & $0.026$ & $0.24$ & $150$  & $3.4$ \\
(l) $100$\,sources & $5$ & $0.5$  & $\infty$ & $48$  & cell & $10^{-2}$ & $2$ & $8^{\star}$ & $0.023$ & $0.24$ & $240$  & $4.5$ \\
(m) $\theta_\mathrm{IF} = 0.5, \Npix = 12$ & $5$ & $1.0$  & $0.5$    & $12$  & cell & $10^{-2}$ & $2$ & $1$         & $0.034$ & $0.11$ & $100$  & $4.1$ \\
(n) $\theta_\mathrm{IF} = 0.5, \Npix = 48$ & $5$ & $1.0$  & $0.5$    & $48$  & cell & $10^{-2}$ & $2$ & $1$         & $0.029$ & $0.12$ & $90$   & $3.7$ \\
(o) $\theta_\mathrm{IF} = 0.25, \Npix = 192$ & $5$ & $1.0$  & $0.25$   & $192$ & cell & $10^{-2}$ & $2$ & $1$         & $0.048$ & $0.34$ & $240$  & $3.7$ \\
(p) COS & $4^{\star\star}$ & $0.5$ & $\infty$ & $48$ & cell & $10^{-2}$ & $2$ & $1$ & $0.065$ & $0.12$ & $80$ & $0.8$ \\
\hline
\end{tabular}
\end{center}
\begin{flushleft}
Column 1 gives the model name. The following columns list:
\begin{itemize}
\item $l_{r}$: the refinement level defining the grid resolution($4\rightarrow 64^3$, $5\rightarrow 128^3$, $6\rightarrow 256^3$)
\item $\theta_\mathrm{lim}$: limiting opening angle for BH~MAC
\item $\theta_\mathrm{IF}, \theta_\mathrm{Src}$: limiting opening angles for IF~MAC and Src~MAC, respectively (see Section~\ref{sec:MACs})
\item $N_{_{\rm PIX}}$: number of rays (defining the angular resolution)
\item $\deleEUV$: error control method (either $\delcel$ given by equation~\ref{eq:delta:er:cell}, or $\deltot$ given by equation~\ref{eq:delta:er:tot}; see Section~\ref{sec:err:cntrl})
\item $\epsilon_\mathrm{lim}$: maximum allowed relative error
\item $\eta_R$: resolution in the radial direction; $\eta_R$ is inversely proportional to the distance between evaluation points on rays (see equation~\ref{eq:etaR})
\item $R_\mathrm{src}$: size of the source (in grid cells)
\item $e_\mathrm{IF}$: relative error in the ionisation front position at $t=1.5$\,Myr
\item $t_\mathrm{iter}$: processor time for a single iteration step (in core hours)
\item $t_\mathrm{tr}$; processor time spent by the tree solver in the whole run (in core hours)
\item $t_\mathrm{hydro}$; processor time spent by the hydrodynamic solver in the whole run (in core hours)
\end{itemize}
$^\star$ in model (k), 100 sources of radius $1$ grid cell were distributed randomly in a sphere with radius $8$ grid cells.\\
$^{\star\star}$ model (p) uses {\em corner of symmetry}, i.e. only one octet is calculated, the grid cell size is the same as in model (a)
\end{flushleft}
\end{table*}

\begin{figure}
\includegraphics[width=\columnwidth]{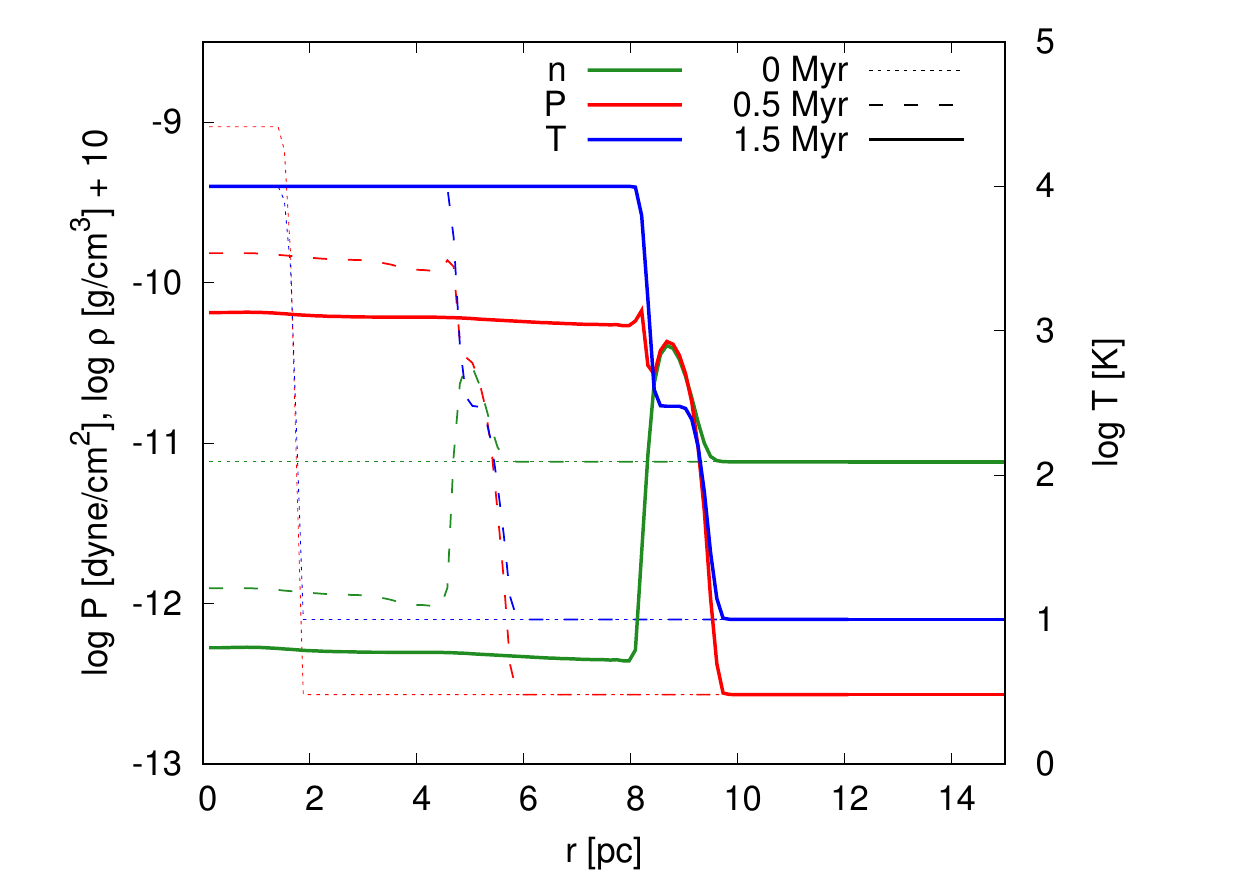}
\caption{Spitzer test: the evolution of the fiducial run (model a). The green, red and blue lines show the mean radial variation of, respectively, the gas particle density, the pressure and the temperature. Different line types distinguish different stages in the evolution: initial conditions (dotted), 0.5 Myr (dashed) and $1.5$\,Myr (solid).}
\label{fig:spitz:fid}
\end{figure}

\begin{figure*}
\includegraphics[width=0.9\textwidth]{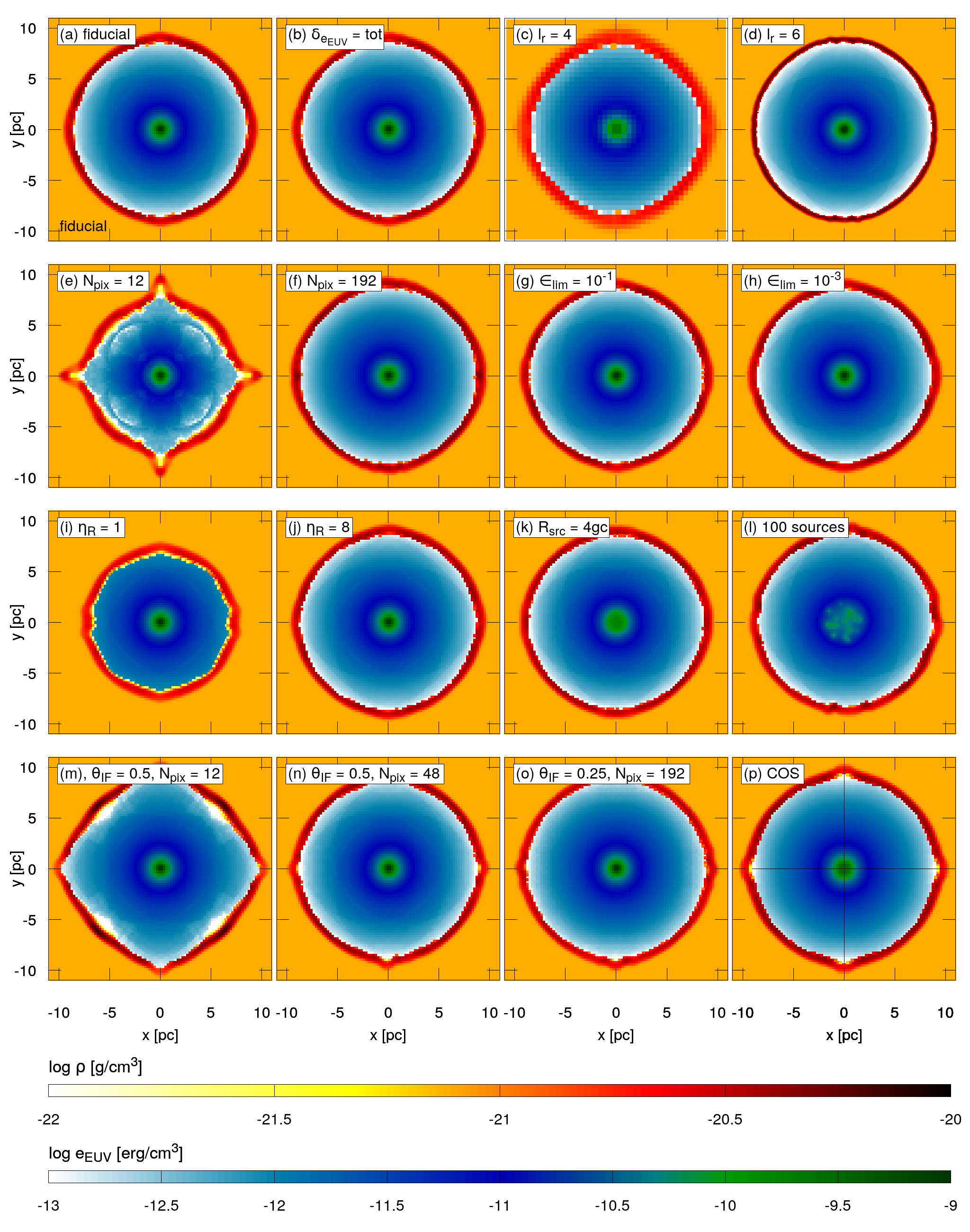}
\caption{Spitzer test: the distribution of the radiation energy (blue-green) and the gas density (yellow-red), in the plane $z=0$, at time $1.5$\,Myr, for models (a) through (p), as noted in the top left corner of each panel; see Table~\ref{tab:acc:spitzer} for the model parameters. The logarithm of the radiation energy is shown in the region with non-zero ionisation degree, the logarithm of the gas density is shown in the remaining parts (i.e. for the neutral gas only).}
\label{fig:spitz:matrix}
\end{figure*}

\begin{figure}
\includegraphics[width=\columnwidth]{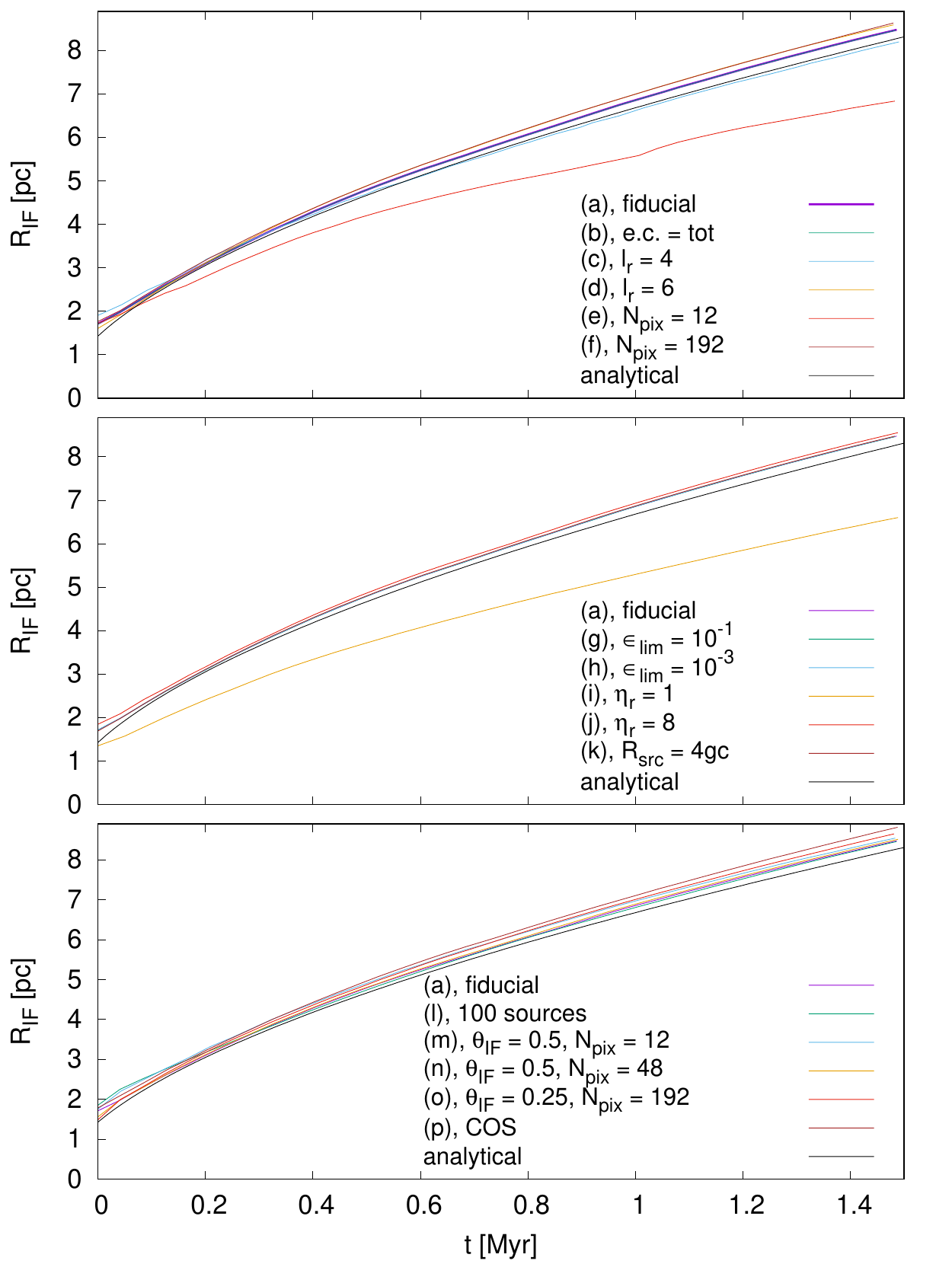}
\caption{Spitzer test: the evolution of the ionisation front radius. The top panel shows models (a) -- (f); the middle panel shows models (a) and (g) through (k); and the bottom panel shows models (a) and (l) through (p). The black line shown on all three panels is the analytic solution given by equation~(\ref{eq:spitzer}).}
\label{fig:spitz:evol}
\end{figure}

\begin{figure}
\includegraphics[width=\columnwidth]{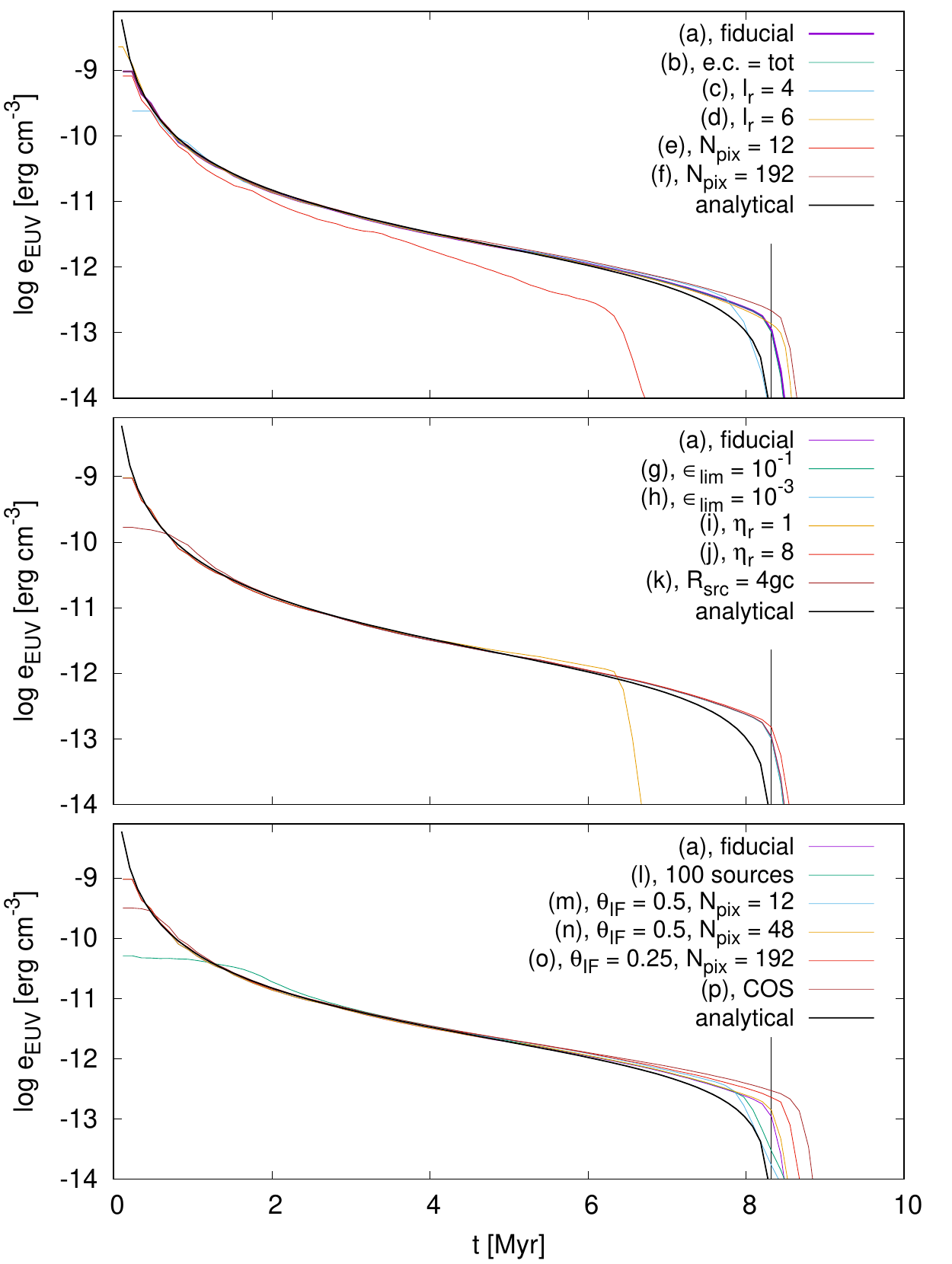}
\caption{Spitzer test: the mean radial profile of the radiation energy density at $1.5$\,Myr. The top panel shows models (a) through (f); the middle panel shows models (a) and (g) through (k); and the bottom panel shows models (a) and (l) through (p). The black line on all three panels is the radiation energy density in the analytical solution (equation \ref{eq:erSpitzer}).}
\label{fig:spitz:rp}
\end{figure}

\textsc{Spitzer bubble}. The Spitzer bubble is one of the simplest models of the interaction of ionising radiation with gas. In this model, the UV radiation from a young, massive star ionises and heats the surrounding gas, creating an H{\sc ii} region, i.e. an over-pressured, expanding bubble of photo-ionised gas bounded by a sharp ionisation front \citep[e.g.][]{Spitzer1978, Whitworth1979, Deharveng2010}. For a typical O-star, the radiative energy input can be very large, $\ga 10^4\,{\rm L_\odot}$. However, only a small fraction of this radiative energy is converted into kinetic energy \citep[$\lesssim$ 0.1\%;][]{Walch2012a}. The expanding ionisation front drives a shock into the surrounding neutral gas, sweeping it up into a dense cold shell, and this may trigger the formation of a second generation of stars \citep[e.g.][]{Elmegreen1977,Walch2013}. H{\sc ii} regions are able to disrupt low-mass molecular couds long before the massive stars explode as supernovae \citep{Whitworth1979, Walch2012a, Dale2012, Haid2019}. 

\textsc{Spitzer bubble expansion}. The Spitzer test describes the spherically symmetric expansion of an H{\sc ii} region into a uniform ambient medium with density $\rho_{\rm o}$, which supposedly resembles a cold molecular cloud. It is used as a standard test for radiative transfer schemes \citep[e.g.][]{Mellema2006, Krumholz2007, Iliev2009, Mackey2010, Mackey2011, Raga2012, Rosdahl2013, Bisbas2015, Raskutti2017}. The analytic solution for the time evolution of the D-type ionization front is given by \citep{Spitzer1978}
\begin{equation} \label{eq:spitzer}
R_\mathrm{IF,anl}(t) = R_{\rm S} \left(1+ \frac{7}{4}\frac{c_i t}{R_{\rm S}} \right)^{4/7},
\end{equation}
where 
\begin{equation}
\label{eq:RStr}
R_{\rm S} = \left(\frac{3\dot{N}_{_{\rm LyC}} m_p^2}{4 \pi \alpha_{\rm B} X^2 \rho_{\rm o}^2} \right)^{1/3}
\end{equation}
is the Str{\o}mgren radius \citep{Stromgren1939}; $c_i$ is the sound speed in the ionized gas inside the bubble; $\dot{N}_{_{\rm LyC}}$ is the rate at which the central star emits hydrogen ionizing photons (i.e. photons with energy $E_{\nu} \ge 13.6\,{\rm eV}$); $m_{\rm p}$ is the proton mass, $X$ is the mass fraction of hydrogen; and $\alpha_{\rm B}\!=\!2.7 \times 10^{-13}\,{\rm cm}^{3}\,{\rm s}^{-1}$ is the Case-B recombination coefficient for an isothermal H{\sc ii} region with temperature $T_{\rm i}\!=\!10^4\,{\rm K}$. We assume that the gas is composed of hydrogen, with mass fraction $X=0.70$, and helium, with mass fraction $Y=0.30$; ionisation of helium is ignored.

\textsc{Spitzer test setup}. In \citet{Bisbas2015} and \citet{Haid2018}, we have already demonstrated that {\sc TreeRay/OnTheSpot}, as implemented in the {\sc FLASH} code, delivers high accuracy in the Spitzer test. Here we explore how the code behaviour depends on the various control parameters. We set up a cubic computational domain of size $30\times 30\times 30\,\mathrm{pc}^3$, which is filled with cold, dense molecular gas, having uniform temperature, $T_n=10\,{\rm K}$, and uniform density, $\rho_0=7.63\times 10^{-22}\,{\rm g}\,{\rm cm}^{-3}$; the hydrogen is assumed to be molecular, and the helium atomic, so the mean molecular weight is 2.35, and the total number-density of gas particles $n_0\approx 195\,{\rm cm}^{-3}$. At $t=0$, a radiation source at the centre starts emitting ionising photons at rate $\dot{N}_{_{\rm LyC}}=10^{49}\,{\rm s}^{-1}$. In the ionized gas the temperature is set to $T_i=10^4\,{\rm K}$ and the mean molecular weight to $\mu_i=0.678$, corresponding to ionised hydrogen and atomic helium.  All the gas has adiabatic index $\gamma = 5/3$. If, during the evolution, the temperature in the neutral gas exceeds $300\,{\rm K}$, it is instantaneously cooled to $300\,{\rm K}$; consequently the shell of shock-compressed neutral gas immediately ahead of the expanding IF is effectively isothermal at $T_s=300\,{\rm K}$ (strictly speaking, the adiabatic index should be slightly below $\gamma=5/3$ at this temperature, because the rotational degrees of freedom of H$_{2}$ are starting to be excited, but we ignore this detail). This temperature limit is chosen in order to resolve the thickness of the shell and prevent numerical instabilities in the hydrodynamic (PPM -- Piecewise Parabolic Method) solver that would otherwise occur. The sound speeds in the molecular cloud, the shell and the H{\sc ii} region, are $c_n = 0.187\,{\rm km\,s^{-1}}$, $c_s = 1.30\,{\rm km\,s^{-1}}$ and $c_i = 14.3\,{\rm km\,s^{-1}}$, respectively. These parameters result in a Str{\o}mgren radius of $1.434\,{\rm pc}$.

\textsc{Varied parameters}. We present results for $16$ models, denoted (a) through (p) (see Table~\ref{tab:acc:spitzer}) with different combinations of the following six parameters: the radiation-field error control limit ($\delcel$ or $\deltot$, see Section~\ref{sec:err:cntrl}), the grid resolution (given by the refinement level $l_r$), the angular resolution (given by the limiting opening angle $\theta_\mathrm{lim}$, the number of {\sc HealPix} rays $N_{_{\rm PIX}}$, and the limiting opening angles $\theta_\mathrm{IF}$ and $\theta_\mathrm{Src}$ for the IF~MAC and Src~MAC, respectively), the radiation-energy error limit ($\epsilon_\mathrm{lim}$), the radial resolution of rays ($\eta_R$), and the source radius ($R_\mathrm{src}$).  In model (l), the single radiation source is replaced by $100$ sources with luminosities $\dot{N}_{_{\rm LyC}}=10^{47}\,{\rm s}^{-1}$ distributed in a sphere of radius $8$ grid cells, instead of a single source of that radius.  This is done to demonstrate that {\sc TreeRay} can faithfully handle a large number of radiation sources.  The last model (p), is calculated with the {\em corner of symmetry} boundary condition, i.e. only one octant of the total domain is computed (see Section~\ref{sec:bc}). For each model we evaluate the error in the ionisation front position, $e_\mathrm{IF} \equiv (R_\mathrm{IF,num} - R_\mathrm{IF,anl})/R_\mathrm{IF,anl}$, at the end of the run ($1.5$\,Myr), the processor time for a single iteration step, $t_\mathrm{iter}$, and the processor times in the tree solver, $t_\mathrm{tr}$, and in the hydrodynamic solver, $t_{\rm hydro}$, for the whole run.

\textsc{Fiducial run}. Fig.~\ref{fig:spitz:fid} shows the evolution of model (a), the fiducial run, displaying the radial profiles of the gas density, pressure and temperature, averaged over all directions, at three times throughout the evolution. By the end of the evolution ($t\!=\!1.5\,{\rm Myr}$), the pressure in the shell approaches the value within the H{\sc ii} region indicating that the shell thickness is approximately resolved. Fig.~\ref{fig:spitz:evol} shows the position of the ionisation front, $R_\mathrm{IF,num}$, as a function of time, and compares it with the prediction of equation~\ref{eq:spitzer}. The fractional error in the ionisation front position, $e_\mathrm{IF}\equiv |R_\mathrm{IF,num} - R_\mathrm{IF,anl}|/R_\mathrm{IF,anl}$, at $t\!=\!1.5\,{\rm Myr}$, is shown in Table~\ref{tab:acc:spitzer}. For the fiducial run, this error is $\sim\!2.5\,\%$. Fig.~\ref{fig:spitz:rp} shows the radial profile of the radiation energy density, $e_{_{\rm EUV}}$, and compares it with the analytically obtained radiation energy density, $e_{_{\mathrm{EUV,anl}}}$, denoted by the black lines. The latter is determined by combining equations (\ref{eq:spitzer}) and (\ref{eq:RStr}), and replacing $\rho_{\rm o}$ with the density in the H{\sc ii} region, $\rho_i$, which is assumed to be uniform between $r = 0$ and $R_\mathrm{IF}$:
\begin{equation}
\label{eq:erSpitzer}
e_{_{\mathrm{EUV,anl}}}(r,t) = \frac{E_{\nu} \dot{N}_{_{\rm LyC}}}{4\pi r^2 c} \times \left[
1 - \frac{r^3}{R_{\rm S}^3} \left(
1 + \frac{7}{4}\frac{c_i t}{R_{\rm S}} \right)^{-12/7}
\right] \ .
\end{equation}

\textsc{Error control}. Figs~\ref{fig:spitz:matrix} and \ref{fig:spitz:evol} show that most of the other models (all but models (e), (i) and (m)) also agree very well with the analytic solution: the error in the ionisation front position is better than $5\,\%$ (corresponding to $\sim\!1.7$ grid cells at the standard resolution), and the discrepancy in the radiation energy density is negligible. Models (b), (g) and (h) evolve almost identically to model (a), indicating that the choice of error control method has almost no impact, and that the calculation is accurate even with the fractional error limit $\epsilon_\mathrm{lim} = 0.1$. All three models use the same processor time per iteration, $t_\mathrm{iter}$, as the fiducial run, which is to be expected since the error control condition does not affect the calculation within an iteration. The total time spent in the tree solver, $t_\mathrm{tr}$, is approximately two times lower for model (b), as $\deltot$ requires fewer iterations. Conversely, model (h) has $t_\mathrm{tr}$ approximately $\sim\!60\,\%$ higher than model (a), because the lower error limit leads to a larger number of iterations. In the case of model (g), with increased $\epsilon_{\rm lim}=0.1$, $t_\mathrm{tr}$ drops by $\sim\!20\,\%$ relative to model (a), reaching approximately $2$ iterations per hydrodynamic time step.

\textsc{Grid resolution}. A comparison of model (a) with models (c) and (d) shows that a lower (higher) resolution leads to a smoother and more blurred (denser and better resolved) shell. However, even the low resolution model (c), where $R_\mathrm{IF, num}$ is only resolved with $17$ grid cells at $t=1.5$\,Myr, and with $\sim 3$ grid cells at $t=0$, results in an H{\sc ii} region with the correct radius and shape. The processor time per iteration, $t_\mathrm{iter}$, scales with the number of grid cells to the power $\sim\!1.1$, between models (c), (a) and (d) (i.e. slightly super-linearly). Similarly, the total time in the tree solver, $t_\mathrm{tr}$, scales with the number of grid cells to the power $\sim\!1.4$ between the same models (again, slightly above the theoretical $(4/3)$-power, where the extra $(1/3)$ derives from the shorter time steps required by the Courant condition at higher resolution).

\textsc{Angular resolution}. Models (e) and (f) test the dependence on the angular resolution by varying $N_{_{\rm PIX}}$ and $\theta_\mathrm{lim}$, setting them so that the typical tree node angular size is similar to the angular size of the rays, as suggested in Paper~I. Model (f) evolves in a similar way to model (a), but $t_\mathrm{iter}$ and $t_\mathrm{tr}$ are $\sim\!6$ times larger. This shows that the fiducial run is well converged with regard to angular resolution, and that the scaling is in agreement with Paper~I where we found that $t_\mathrm{tr}$ scales with $\theta_\mathrm{lim}$ somewhere between $\sim\!\theta_\mathrm{lim}^{-2}$ and $\sim\!\theta_\mathrm{lim}^{-3}$. On the other hand, model (e), with very low angular resolution, $N_{_{\rm PIX}}=12$, exhibits significant departures from the analytical solution. The volume of the H{\sc ii} region reaches only $\sim 53$\% of the correct value, and the radiation energy is distributed non-spherically, with lower values along the Cartesian diagonals. Consequently the shell expands more slowly along the diagonals than along the axes. Along the axes the faster expansion is also accelerated by numerical instabilities in the inadequately resolved shell. The main reason for the depressed radiation-energy density along the diagonals is that the very large $\theta_\mathrm{lim}$ allows the acceptance of very large tree nodes by the BH~MAC. Such large nodes include the source in one corner and a part of the shell in the opposite one, leading to a poor estimate of the rate of absorption of photons within the node, and hence poor estimates of the radiation fluxes and energies. 

\textsc{New MACs}. The new criteria for accepting nodes in the tree walk, IF~MAC and Src~MAC, also control the angular resolution, setting it higher in regions where increased resolution is needed. The new MACs are tested in models (m), (n) and (o). Model (m) with $\theta_\mathrm{IF}\!=\!\theta_\mathrm{Src}\!=\!0.5$ and $\Npix\!=\!12$ demonstrates that it is not worth increasing the angular resolution in the tree walk without also increasing the number of rays. Even though model (m) does not suffer from the problem of accepting too large tree nodes (unlike model (e)), the radiation field shows significant departures from spherical symmetry, and its computational cost is similar to the much more accurate model (n). Model (n) with $\theta_\mathrm{IF}\!=\!\theta_\mathrm{Src}\!=\! 0.5$ and $\Npix\!=\!48$ evolves in a similar way to the fiducial run (a), and its computational cost is approximately $40\,\%$ lower. Model (o), with even higher angular resolution, $\theta_\mathrm{IF}\!=\!\theta_\mathrm{Src}\!=\! 0.25$ and $\Npix\!=\!192$, shows improvement in the spherical symmetry of the radiation field, like model (f), but it is calculated in approximately one quarter of the time. This demonstrates the benefits of the new MACs, particularly at higher angular resolution.

\textsc{Radial resolution}. Models (i) and (j) explore the code behaviour at different radial resolutions, controlled by the parameter $\eta_R$. Model (j), with four times more evaluation points on the rays, gives almost identical results to model (a), indicating that the fiducial run with $\eta_R = 2$ is well converged. The much higher density of evaluation points results in a $\sim\!40\,\%$ increase in the computational time. On the other hand, model (i), with $\eta_R = 1$, takes almost the same amount of time as model (a), and the error in the position of the ionisation front increases to $7$\%.

\textsc{Source size}. Models (k) and (l) show that the H{\sc ii} region still evolves correctly if the source of ionising radiation is distributed throughout a larger volume. Model (l) exhibits moderate departures from spherical symmetry; these are attributable to significant departures from spherical symmetry in the ionising flux from $100$ discrete sources distributed {\it randomly} in a sphere of radius $1.9\,{\rm pc}$. The time per iteration, $t_\mathrm{iter}$, is essentially the same as in model (a), demonstrating a unique property of our algorithm, namely that the computational cost does not depend on the number of radiation sources (see Section~\ref{sec:perf:nsrc} for a more detailed discussion).

\textsc{Corner-of-symmetry}. Finally, model (p) tests the special 'Corner of Symmetry' boundary conditions, which allow the user to simulate one-eighth of a spherically symmetric problem (see Section~\ref{sec:bc}). Model (p) runs at approximately twice the speed of model (a), and produces broadly similar results. However, the shell exhibits numerical artefacts along the axes resulting from the directionally split hydrodynamic solver. As a result, the error on the radius of the ionisation front is greater, $e_\mathrm{IF}\sim 6.5\,\%$.

%%%%%%%%%%%%%%%%%%%%%%%
\subsection{Blister-type HII region}
\label{sec:blister}

\begin{figure*}
\begin{center}
\includegraphics[width=1\linewidth]{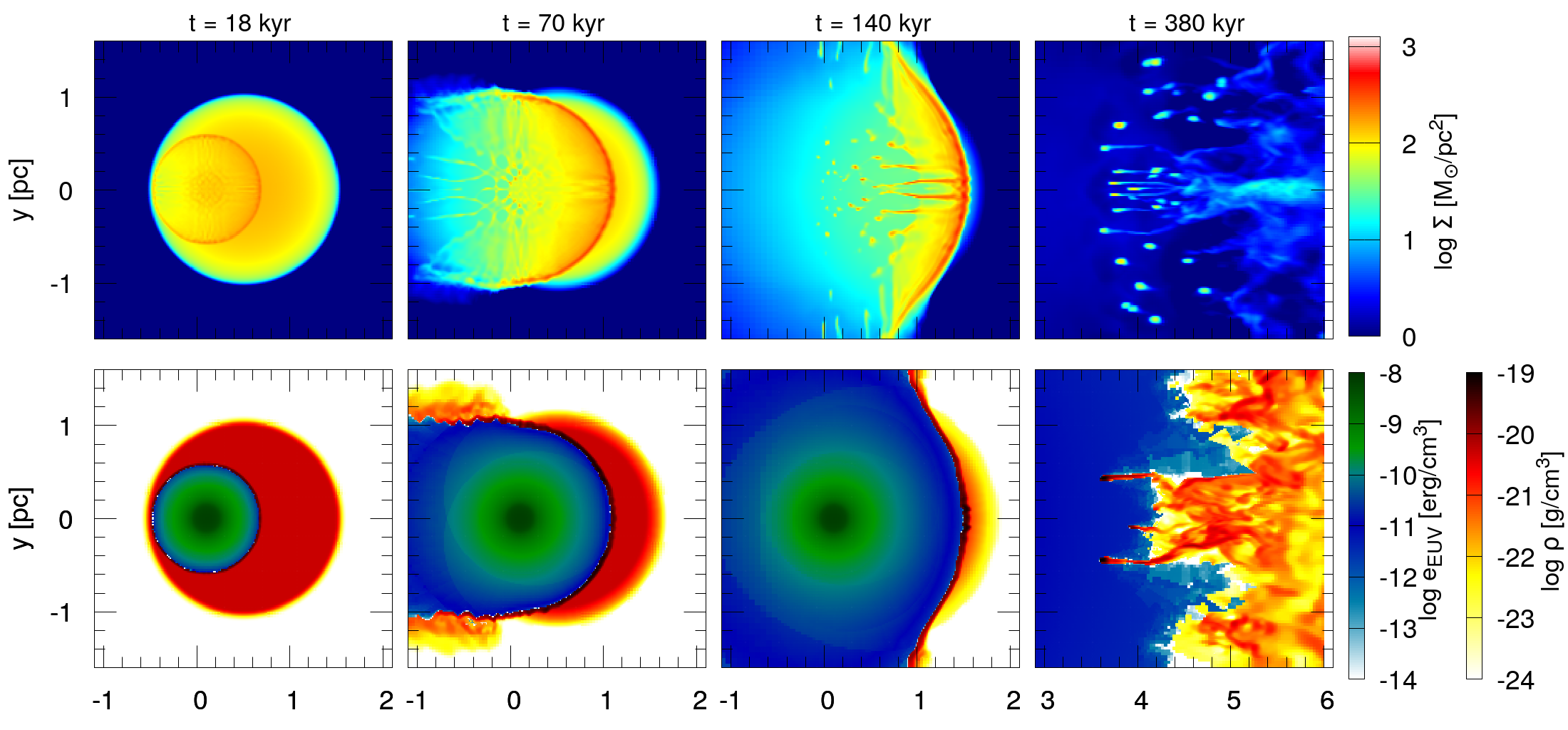}
\caption{Blister-type HII region test, at times $t\!=\!18$, $70$, $140$ and $380\,{\rm kyr}$. The top panels show the logarithm of the gas column density. The bottom panels show the logarithm of the radiation energy in the region with non-zero ionisation degree, and the logarithm of the gas density in the remaining parts (i.e. for the neutral gas only). Note that the computational domain has dimensions $8\times4\times4\,{\rm pc}^3$, and is much larger than the region shown in these maps.}
\label{fig:blister}
\end{center}
\end{figure*}

\textsc{Blister HII region}. In order to test the algorithm in a situation that is not spherically symmetric, we model a spherical cloud with an ionising star located inside it, but not at its centre. This model was first discussed by \citet{Whitworth1979} and \citet{TenorioTagle1979} who suggested that, as soon as the ionisation front reaches the edge of the cloud on one side, the H{\sc ii} region bursts out of the cloud and the remainder of the cloud on the other side is accelerated by the rocket effect \citep{Kahn1954}. This scenario was later studied numerically by \citet{Bisbas2009} (hereafter B09), \citet{GendelevKrumholz2012} and others. Here, we set up the cloud and the radiation source with the same physical parameters as in B09, and compare our results with theirs.

\textsc{Initial conditions}. The radiation source is at the centre of the coordinate system and emits ionising photons at rate $\dot{N}_{_{\rm LyC}} = 10^{49}\,{\rm s^{-1}}$. The spherical cloud has mass $M_0 = 300$\,\MSun, radius $R_0 = 1$\,pc and uniform density $\rho_0 = 4.85\times 10^{-21}$\,g\,cm$^{-3}$; its centre is at $(0.4,0,0)\,{\rm pc}$. The cloud is embedded in a rarefied ambient gas with density $\rho_\mathrm{amb} = 10^{-24}$\,g\,cm$^{-3}$, and the computational domain has dimensions $8\times4\times4\,{\rm pc}^3$. The neutral gas has temperature $T_n = 100$\,K, but otherwise both neutral and ionised (i.e. irradiated) phases have the same properties (molecular weights, adiabatic index, hydrogen mass fraction) as in Section~\ref{sec:spitzer}. All the gas that is not irradiated is immediately returned to the temperature $T_n$, at each time step. Self-gravity is switched off. The simulation is run for $0.5\,{\rm Myr}$, corresponding to $1574$ time steps, and the computational cost is $\sim\!10000$ core hours.

\textsc{Numerical parameters}. The model is calculated on an AMR grid with minimum and maximum refinement levels of $4$ and $6$, respectively (i.e. the highest resolution corresponds to $512\times 256 \times 256$). A simple density-based criterion is used to refine/derefine blocks wherever the maximum density exceeds $10^{-21}\,{\rm g\,cm^{-3}}$, or drops below $5\times 10^{-22}\,{\rm g\,cm^{-3}}$. The hydrodynamic boundary conditions are set to outflow. The parameters controlling the {\sc TreeRay} accuracy are set as follows: the tree solver uses both the IF~MAC and the Src~MAC with $\thIF = 0.25$, $\thsrc = 0.25$ and $\thlim = 1.0$; the number of rays is $\Npix = 192$; the radial resolution is $\eta_R = 2$; and the maximum allowed relative error is $\epsilon_\mathrm{lim} = 0.01$. 

\textsc{Blister evolution}. The evolution of the blister-type H{\sc ii} region is shown in Fig.~\ref{fig:blister}. The ionised region expands spherically until $t = 18\,{\rm kyr}$, when it reaches the edge of the cloud on the lefthand side. After that time, the ionised gas flows out of the cloud on this side, while opening a growing cavity within it. On the righthand side, the H{\sc ii} region continues to expand into the remainder of the cloud, opening a growing cavity. The originally spherical shell of swept-up gas is at first converted into an hemispherical shell (at $t\sim 70$\,kyr), and later into an almost flat layer (after $t\sim 140$\,kyr). When all the cloud material has been swept up in a given direction from the source, accretion onto the shell stops and the shell starts to accelerate and become Rayleigh-Taylor unstable. As a result, the layer breaks into a large number of cloudlets. These cloudlets were called ``cometary knots'' in B09, due to their almost spherical core and a tail created by  ablation by the ionised gas streaming around them and away from the source. 

\textsc{Comparison with B09}. A qualitative comparison with the SPH simulation of the same model in B09 (see their Figures~13 and 14) shows an almost perfect agreement. The location and the shape of the ionisation front and other large scale features are indistinguishable. The formation of cometary knots is also reproduced remarkably well, given that the angular resolution of the {\sc TreeRay} simulation ($\thIF = 0.25$) is more than an order of magnitude coarser than in B09, where the angular separation between neighbouring rays is set by the local SPH-particle smoothing lengths, typically $\theta\lesssim (0.1\,\mathrm{pc})/(4\,\mathrm{pc}) = 0.025$. This demonstrates the ability of the reverse ray-tracing method to deliver high spatial resolution of the radiation field at the ionisation front, even with relatively large angles between neighbouring rays. However, there are some differences in the small scale structures. Firstly, the {\sc TreeRay} simulation generates larger perturbations of the shell along the grid axes, and the higher noise there seeds the Rayleigh-Taylor instability through the odd-even decoupling mechanism identified by \citet{Quirk1994}. Secondly, by $t=380\,{\rm kyr}$, the number the cometary knots is substantially lower than in B09, because many of them have evaporated, due to their lower density, which in turn is caused by the lower spatial resolution of the grid code.

%%%%%%%%%%%%%%%%%%%%%%%%%%%%
\subsubsection{Rabbit hole test}
\label{sec:rabbithole}

\textsc{Penetration depth problem}. If the ionisation front has a complex structure, as in the previous test, it is questionable whether the radiation field computed by a code with limited angular resolution can properly follow the ionisation front geometry. A particularly difficult configuration for {\sc TreeRay} is a radiation source shining into a deep, narrow hole. Such holes form frequently in astrophysical applications, for instance when a swept-up shell becomes unstable and breaks apart \citep[e.g. as in][]{Walch2013}. In order to quantify the accuracy of the code in this situation, we implement a test called the {\it rabbit hole}, in which we measure the penetration of the radiation field as a function of the width of the hole.

\textsc{Rabbit hole setup}. To mimic a hole within a dense shell, we set up an elongated box containing two media: the walls of the hole are formed of cold dense gas with sound speed $c_{{\rm c}} = 0.25\,{\rm km\,s}^{-1}$ and $\rho_{{\rm c}}$\,=\,10$^{-18}$\,g\,cm$^{-3}$; the inside of the hole is filled with warm rarefied gas with $c_{{\rm w}} = 10\,{\rm km\, s}^{-1}$ and $\rho_{{\rm w}}$\,=\,10$^{-24}$\,g\,cm$^{-3}$. The two media are not in pressure equilibrium, but, since we only calculate the first time step, this is irrelevant. However, the extreme density contrast renders this test particularly difficult, because nodes that are far away from a given target point are large. Hence, the denser the cold medium is, the more mass will be accumulated in these large and distant nodes, and this can lead to an artificial blocking of the radiation. 

\begin{figure}
\begin{center}
\includegraphics[width=\linewidth]{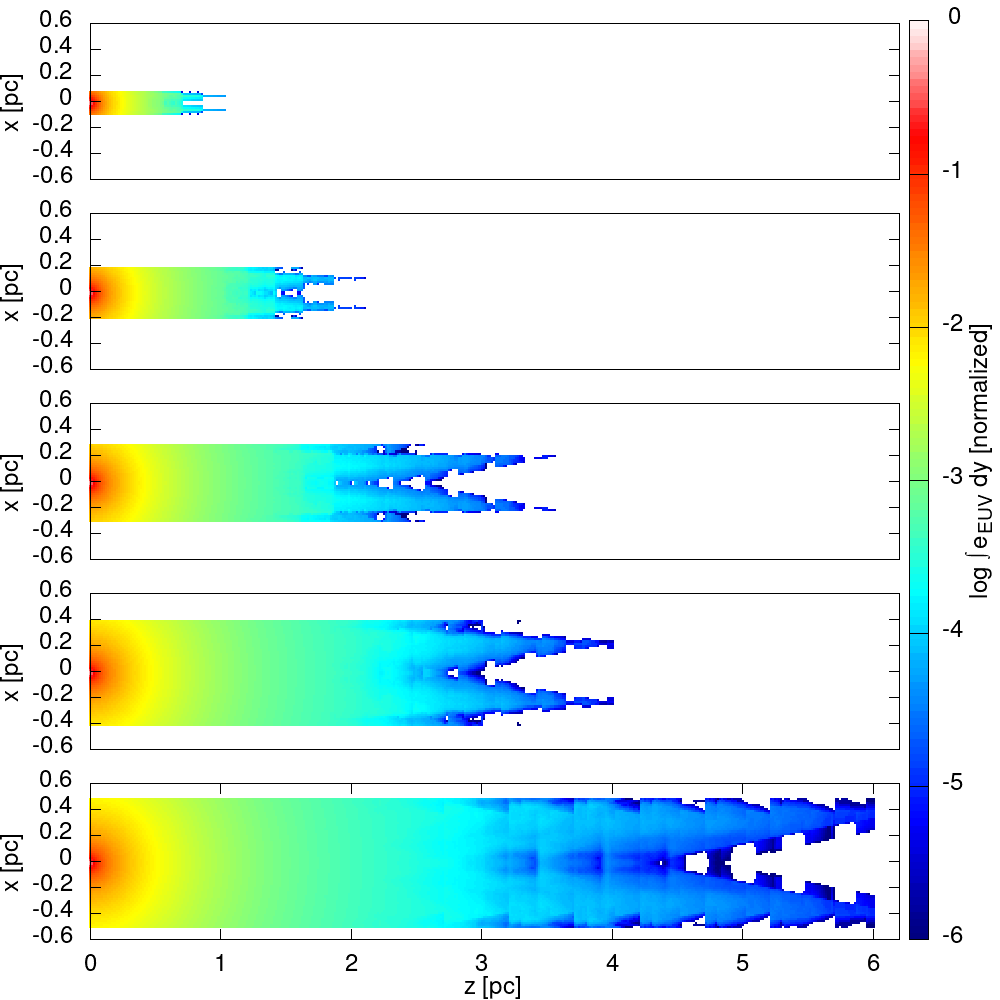}
\includegraphics[width=\linewidth]{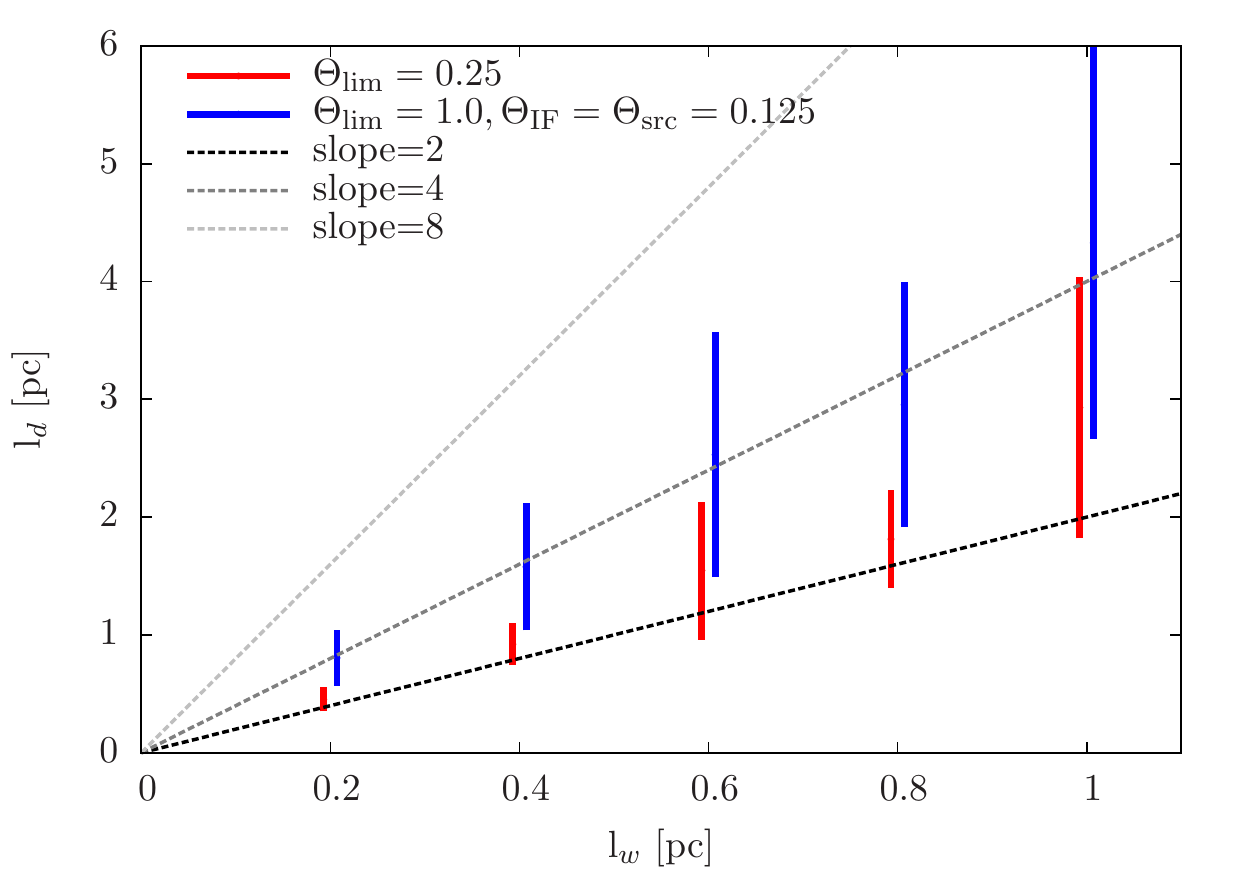}
\caption{{\it Top 5 panels:} The radiation energy density integrated along the line of sight ($y$-direction) with $\theta_{\rm lim}=1.0$, $\theta_{\rm IF}=\theta_{\rm src} = 0.125$ and -- from top to bottom -- $l_{\rm w} = 0.2, 0.4, 0.6, 0.8$ and $1.0\,{\rm pc}$. Radiation only reaches the end of the computational domain ($z=6\,{\rm pc}$) for $l_{\rm w}\!=\!1\,{\rm pc}$. {\it Bottom panel:} The range of depths up to which the radiation propagates as a function of the width of the rabbit hole. The `error-bars' indicate the minimum and maximum depth, as found by different simulations with the hole pointing in the $x-$ or $z-$directions. The dotted lines show $l_{\rm d} = ml_{\rm w}$ with slopes $m\!=\!2,\,4\;{\rm and}\;8$; $m\!=\!4$ corresponds to the theoretical expectation, $l_{\rm d} = l_{\rm w}/\theta_{\rm lim}$ with $\theta_{\rm lim} = 0.25$, and $m\!=\!8$ corresponds to the theoretical expectation, $l_{\rm d} = l_{\rm w}/\theta_{\rm IF}$ with $\theta_{\rm IF}=\theta_{\rm src} = 0.125$. The results with physical MACs are significantly better than those obtained with $\theta_{\rm lim} = 0.25$, but a slope of $m\!=\!8$ cannot be achieved for the high density contrast simulated here. }
\label{fig:rabbit2}
\end{center}
\end{figure}

\textsc{Test parameters}. The computational domain is $6\times 2\times 2\,{\rm pc}^3$. The hole has a square cross-section with side $l_{\rm w}$ and stretches from 0 to $6\,{\rm pc}$. The ionising source is placed at the entrance to the hole, $(x,y,z)=(0,0,0)$, and emits ionising photons at rate $N_{\rm LyC} = 10^{49}\,{\rm s}^{-1}$. The test is performed with the hole pointing in the $x-$direction and then in the $z-$direction (as illustrated in Fig.~\ref{fig:rabbit2}, top panels), and with uniform resolution ($384\times 128\times 128\simeq 6.3\times 10^6$ cubic cells with side length $\sim\!0.16\,{\rm pc}$).

\textsc{Penetration depth measurement}. The Str{\o}mgren radius for a star with $N_{\rm LyC} = 10^{49}\,{\rm s}^{-1}$, in a uniform medium with $\rho_{{\rm w}}$\,=\,10$^{-24}$\,g\,cm$^{-3}$, is $72\,{\rm pc}$ (see equation~\ref{eq:RStr}), i.e. the radiation should shine right through the rabbit hole. However, for narrow holes (small $l_{\rm w}$), {\sc TreeRay} is unable to obtain the correct solution without reducing the angular resolution to intolerably low values. This is shown in Fig.~\ref{fig:rabbit2} (bottom panel), where we plot the maximum depth, $l_{\rm d}$, up to which we measure a non-zero radiation energy density as a function of $l_{\rm w}$, for two different settings of the control parameters.

\textsc{Expected performance}. Theoretically, we expect $l_d$ to fall in the range $l_w/(2\theta_\mathbf{lim}) \lesssim l_d \lesssim l_w/\theta_\mathbf{lim}$. The lower limit corresponds to the situation where the {\sc HealPix} cone widens symmetrically from the target cell towards the source, and touches the walls on both sides of the hole at distance $l_w/(2\theta_\mathbf{lim})$. The upper limit corresponds to the situation where one border of the cone is parallel to the wall of the hole and the opposite border touches the wall at $l_w/\theta_\mathbf{lim}$.

\textsc{Actual performance}. To test this behaviour, we use two settings. In the first setting we adopt a constant opening angle of $\theta_{\rm lim}\!=\!0.25$ corresponding to $\Npix\!=\!192$. For this Oct-Tree resolution, we expect $l_{\rm d} = l_{\rm w}/\theta_{\rm  lim}  = 4 l_{\rm w}$, which is slightly better than the performance actually achieved. Moreover, simply using more rays without implementing the source and ionization-front MACs does not improve this result. In the second setting, we adopt a larger $\theta_{\rm lim}\!=\!1.0$ but also implement the physical MACs, with $\theta_{\rm IF} = \theta_{\rm src} = 0.125$, corresponding to {\sc Healpix} level 8, i.e. $\Npix=768$ rays. Despite the large value of $\theta_{\rm lim}$, the results with the additional physical MACs outperform the more expensive simulations with $\theta_{\rm lim} =0.25$ (see bottom panel of Fig.~\ref{fig:rabbit2}), and in the run with $l_w=1\,{\rm pc}$ the radiation field reaches the end of the computational domain. However, the theoretical scaling of $l_{\rm d} = 8 l_{\rm w}$ is not achieved, due to the high density contrast (a factor of $10^6$) between the warm gas within the hole and the surrounding cold gas. 

%%%%%%%%%%%%%%%%%%%%%%%%%%%%%%%%%%%%%
\subsection{Radiation driven implosion (RDI)}
\label{sec:rdi}

\begin{figure*}
\includegraphics[width=\textwidth]{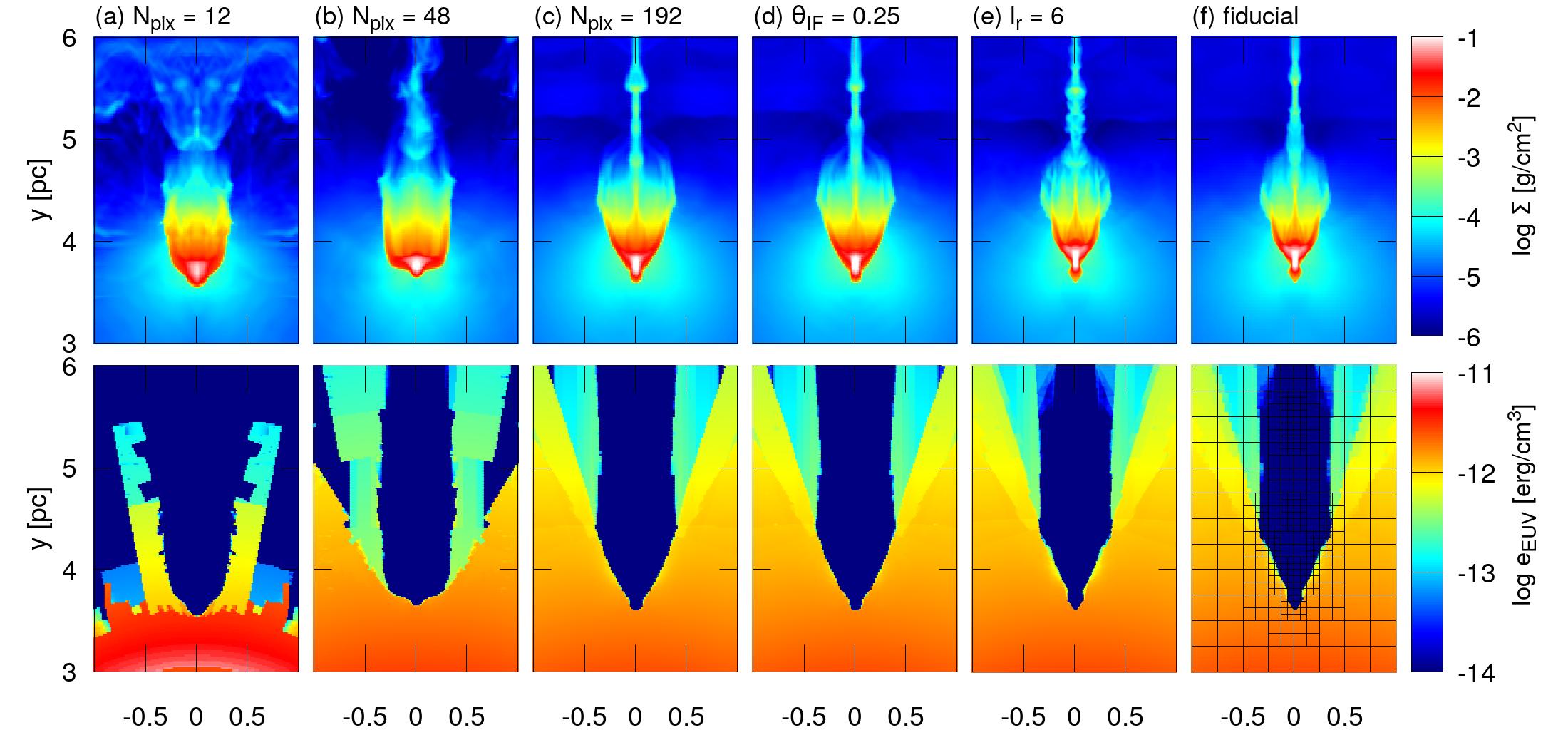}
\caption{Radiation driven implosion test, comparison of models (a) through (f). The top panels show the logarithm of the gas column density. The bottom panels show the logarithm of the radiation energy density on the mid-plane. All models are plotted at time $180$\,kyr.}
\label{fig:rdi:cmp}
\end{figure*}

\begin{figure*}
\includegraphics[width=\textwidth]{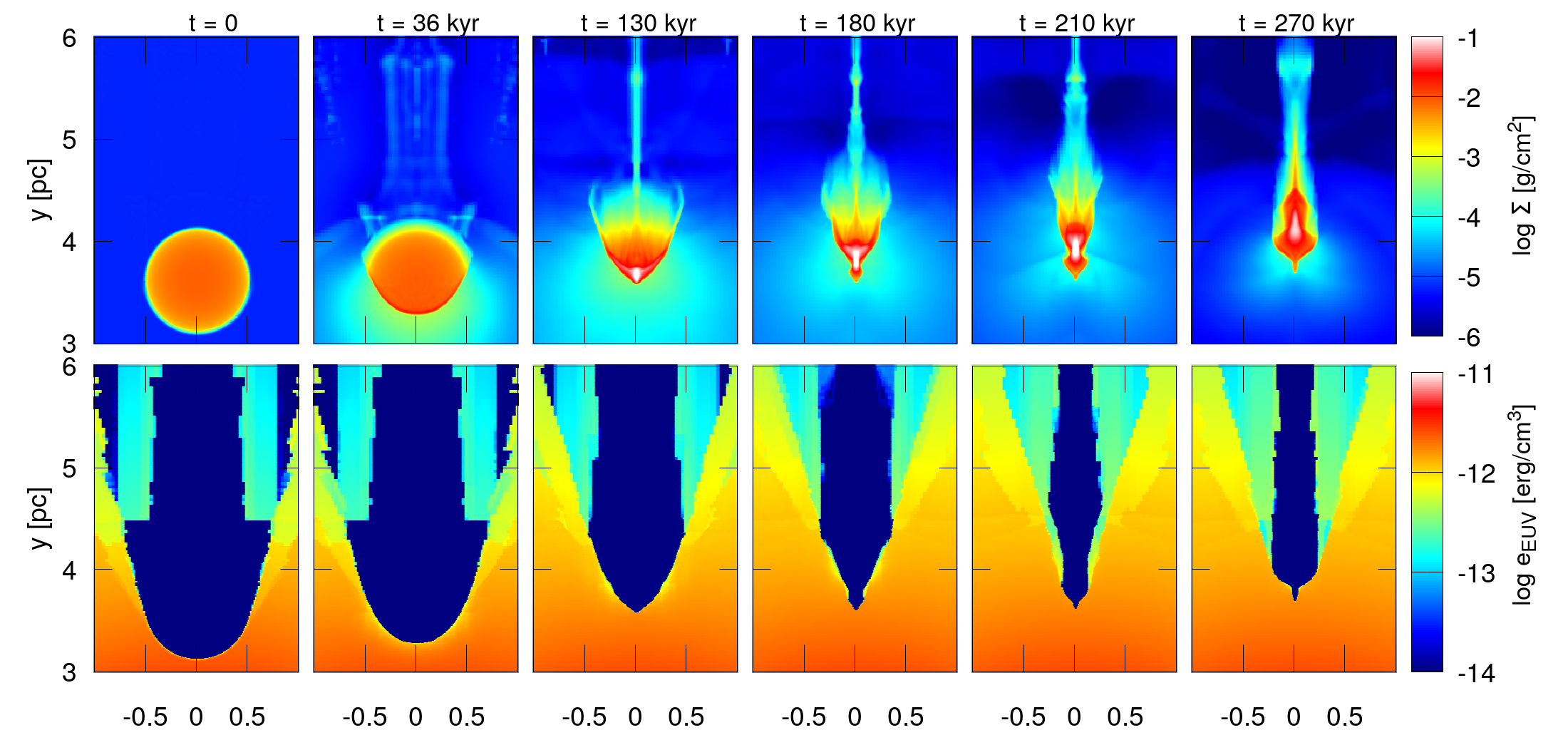}
\caption{Radiation driven implosion, evolution of the fiducial model (f). The top panels show the logarithm of the gas column density at times $0$, $36$, $130$, $180$, $210$ and $270$\,kyr. The bottom panels show the logarithm of the radiation energy density on the mid-plane at the same times.}
\label{fig:rdi:evol}
\end{figure*}

\begin{figure}
\includegraphics[width=\columnwidth]{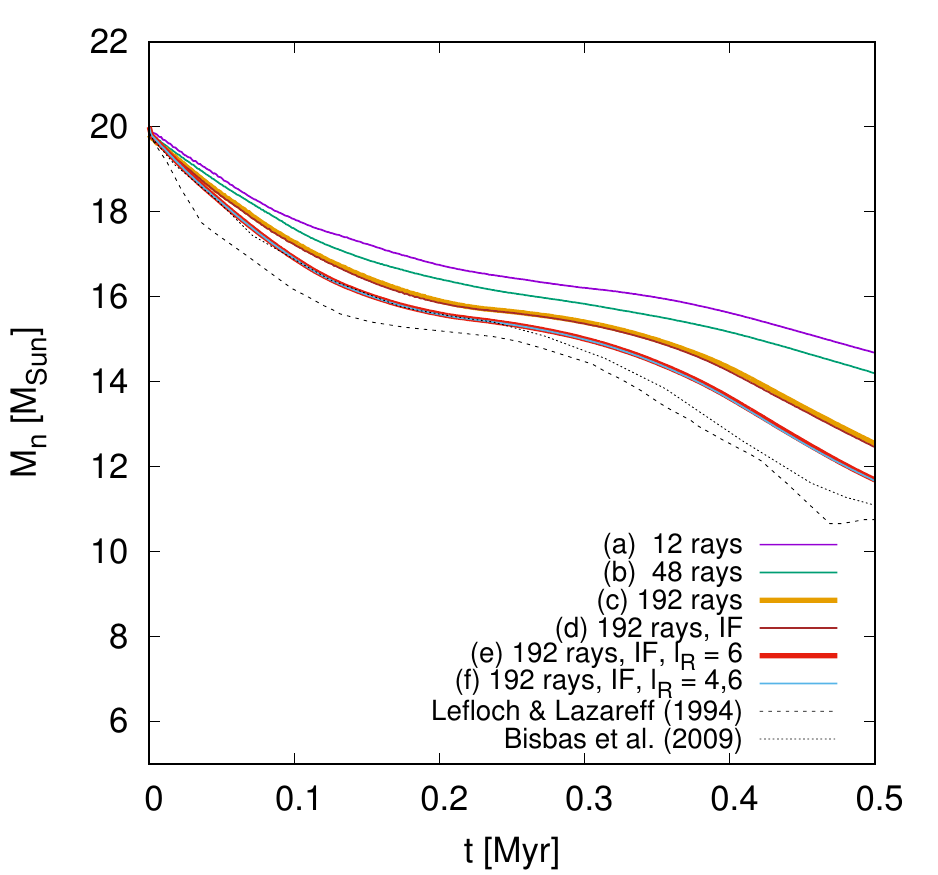}
\caption{Radiation driven implosion: evolution of the neutral gas mass for models (a) through (f), compared with B09 and \citet{LeflochLazareff1994}.}
\label{fig:rdi:Mcold}
\end{figure}

\begin{table}
\caption{Accuracy and performance of the RDI test.}
\label{tab:acc:rdi}
\begin{center}
\begin{tabular}{l|c|c|c|c|c|c|r}
\hline
model & $l_r$ & $\theta_\mathrm{lim}$ & $N_{_{\rm PIX}}$ & $\theta_\mathrm{IF}$ & $e_{M}$ & $t_\mathrm{iter}$ & $t_\mathrm{tr}$ \\
\hline
(a) $\Npix = 12$ & $5$   & $1$    & $12$  & $\infty$  & $0.24$ & $0.49$ & $840$   \\
(b) $\Npix = 48$ & $5$   & $0.5$  & $48$  & $\infty$  & $0.22$ & $1.1$  & $1900$  \\
(c) $\Npix = 192$ & $5$   & $0.25$ & $192$ & $\infty$  & $0.12$ & $6.1$  & $10500$ \\
(d) $\theta_\mathrm{IF} = 0.25$ & $5$   & $1$    & $192$ & $0.25$    & $0.11$ & $2.6$  & $4800$  \\
(e) $l_r = 6$ & $6$   & $1$    & $192$ & $0.25$    & $0.05$ & $26$   & $45000$ \\
(f) fiducial & $4,6$ & $1$    & $192$ & $0.25$    & $0.05$ & $1.7$  & $6300$ \\
\end{tabular}
\end{center}
\begin{flushleft}
Column 1 gives the model name. The following columns list:
\begin{itemize}
\item $l_{r}$: the refinement level defining grid resolution `5' $\rightarrow 128^2\times 384$; `6' $\rightarrow$  $256^2\times 792$; `4,6' $\rightarrow$ AMR with minimum and maximum refinement levels $4$ and $6$, respectively.)
\item $\theta_\mathrm{lim}$: the limiting opening angle for the BH MAC
\item $N_{_{\rm PIX}}$: the number of rays (defining the angular resolution)
\item $\theta_\mathrm{IF}$: the limiting opening angle for the IF MAC
\item $e_\mathrm{M}$: the relative error in the neutral gas mass at $t=0.5$\,Myr
\item $t_\mathrm{iter}$: the processor time for a single iteration step (in core-hours)
\item $t_\mathrm{tr}$; the processor time in the tree solver for the whole simulation (in core-hours)
\end{itemize}
\end{flushleft}
\end{table}

\textsc{RDI}. In this test we study a compact, dense, neutral cloud illuminated by ionising radiation from a single direction. The astronomical motivation is the cometary globules, commonly observed in Galactic {\sc Hii} regions, with bright rims on the side irradiated by a nearby hot star star (or stars) and tails pointing in the opposite direction \citep[see e.g.][]{LeflochLazareff1995,  Deharveng2010, Schneider2012, Getman2012}.  It has been suggested that, in such a configuration, star formation can be triggered by the {\em Radiation Driven Implosion} (RDI) mechanism \citep{Bertoldi1989} and this mechanism has been extensively studied analytically \citep[e.g.][]{Mellema1998, Miao2009, MackeyLim2010} and numerically using SPH codes \citep[e.g.][and references therein]{KesselDeynetBurkert2003, Gritschneder2010, Bisbas2011, Dale2012}, grid-based hydrodynamic codes \citep[e.g.][]{LeflochLazareff1994, Mellema1998} and magneto-hydrodynamic codes \citep{MackeyLim2010}.

\textsc{RDI and TreeRay}. It is usually assumed that the size of the cloud is small, compared with the distance to the radiation source. Technically, this can be arranged, either by using a plane-parallel radiation field (as in the Lefloch \& Lazareff setup), or by setting a large distance between the source and the cloud \citep[as in][hereafter B09]{Bisbas2009}. We choose the latter option here, even though {\sc TreeRay} can easily be modified to treat a plane-parallel radiation field, and we postpone description of this feature to a future paper. This test is relatively hard for algorithms with limited angular resolution, due to the small angular size of the cloud, as seen from the source. This is why it was chosen by B09, to demonstrate the ability of their algorithm to split rays adaptively, so that the ray separation is everywhere similar to the resolution of the hydrodynamic solver. {\sc TreeRay} achieves a similar resolution at the irradiated border of the cloud, by using reverse ray-tracing, which ensures a small separation between neighbouring rays at the point of the flux calculation.

\textsc{RDI setup}. We use a similar setup to B09, which was chosen to resemble as closely as possible the setup defined by \citet{LeflochLazareff1994}. A spherical cloud with mass $M = 20$\,\MSun, radius $R = 0.5$\,pc, and uniform density $\rho_0 = 2.6\times 10^{-21}$\,g\,cm$^{-3}$, is illuminated by a source at distance $D = 3.5$\,pc, emitting ionising photons at rate $N_{_{\rm LyC}} = 3.2\times 10^{48}$\,s$^{-1}$. The neutral gas (i.e. the cloud and the gas in its shadow) has temperature $T_n = 100$\,K and is composed of pure atomic hydrogen, i.e. $\mu_n = 1$. The ionised gas has  temperature $T_i = 10^4$\,K and $\mu_i = 0.5$. The computational domain is $2\times 2\times 6\,{\rm pc}^3$. Initially, the source is located at $(x,y,z)=(1,1,0)$\,pc, and the cloud centre at $(1,1,3.6)$\,pc. The whole computational domain (apart from the cloud) is filled with a rarefied gas having density $\rho_\mathrm{amb} = 10^{-24}$\,g\,cm$^{-3}$. We calculate 6 models denoted (a) through (f), for which we vary the angular resolution (parameters $\theta_\mathrm{lim}$ and $N_{_{\rm PIX}}$), the grid resolution (refinement level $l_r$) and the MAC criterion (IF MAC on or off; see Table~\ref{tab:acc:rdi}). Models (a) through (e) use a uniform grid, i.e. the minimum and maximum refinement levels are the same, while model (f) uses adaptive mesh refinement (AMR) with refinement levels $4$ to $6$, so the coarsest grid is $64^2\times 192$ and the finest one is $256^2\times 768$. The AMR criterion refines a block if the maximum density within it exceeds $10^{-21}$\,g\,cm$^{-3}$, and de-refines it if the maximum density drops below $5\times 10^{-22}$\,g\,cm$^{-3}$, thereby ensuring that only the cloud and its immediate surroundings are calculated on the highest resolution. A typical grid structure can be seen in the bottom right panel of Fig.~\ref{fig:rdi:cmp}. We evaluate the morphology of the cloud and its shadow, and compute the mass of the neutral gas as a function of time, comparing these quantities with B09 and \citet{LeflochLazareff1994}. The error in the neutral gas mass is $e_M = (M_\mathrm{n} - M_\mathrm{n,B09})/M_\mathrm{n,B09}$, where $M_\mathrm{n}$ is the neutral gas mass at $t=0.5$\,Myr and $M_\mathrm{n,B09}$ is the reference value from B09.

\textsc{RDI evolution}. We select model (f) as the fiducial model, since -- along with model (e) -- it gives the best accuracy at reasonable computational cost. Fig.~\ref{fig:rdi:evol} shows the logarithms of column density (top) and radiation energy density (bottom) for model (f) at a sequence of times. The column density can be compared directly with Figure~15 in B09, and we see good agreement between the two codes. Qualitatively, the cloud evolves in the same way as in previous studies. Initially, the radiation ionises the outer layers of the cloud in the direction of the source and a shock starts to propagate into the remaining neutral gas, compressing it from the sides. At $\sim 130$\,kyr a dense core is formed on the cloud axis near the ionisation front and it re-expands due to its internal thermal pressure, while at the same time being ablated by radiation on the side facing the source. Eventually, a cometary tail is formed at $\sim 200$\,kyr. The bottom panels show the radiation energy density and the shadow behind the cloud. Even though a certain amount of radiation diffuses artificially into the shadow region (due to smoothing the edges of the cloud into larger tree nodes), the overall shape of the shadow looks reasonably good. The mass of neutral gas (see Fig.~\ref{fig:rdi:Mcold}) follows almost exactly the curve from B09 up to $t\!\simeq\!0.25$\,Myr, and then becomes slightly higher, leading to a discrepancy $\sim\!5$\% at $t\!\simeq\!0.5$\,Myr. This discrepancy is due to insufficient spatial resolution of the dense core, which starts to be ablated by the ionising radiation after $200$\,kyr. The rate at which gas is ionised is very sensitive to the density of the neutral gas close to the ionisation front, and the SPH code used in B09 uses many more resolution elements to describe the core density profile; in our simulation it is only a few grid cells in diameter. 

\textsc{RDI comparison}. Fig.~\ref{fig:rdi:cmp} shows the logarithms of the column density (top) and the radiation energy density (bottom), for models (a) through (f), at $t = 180$\,kyr. Models (a) through (c) explore the effect of changing the angular resolution. As expected, a relatively high angular resolution is needed to compute this configuration faithfully. In Model (a), with $\theta_\mathrm{lim}\!=\!1,\;N_{_{\rm PIX}}\!=\!12$, the radiation energy is incorrect by tens of percent, which is mainly due to the very large sizes of tree nodes. In model (b) with $\theta_\mathrm{lim}\!=\!0.5,\;N_{_{\rm PIX}} = 48$, the radiation field is approximately correct in the ionised regions, but the shadow is too wide, and the irradiated side of the cloud is too flat. In both models (a) and (b), the mass of neutral gas is higher than in B09 by $\sim 20$\%. In model (c), with $\theta_\mathrm{lim}\!=\!0.25,\;N_{_{\rm PIX}}\!=\!192$, the shape of the cloud and its shadow closely match the results of B09, and the error in the neutral gas mass is about $10\%$. The computational cost increases by a factor between 5 and 6 for each reduction of $\theta_\mathrm{lim}$ by a factor of 2. Model (d) behaves almost exactly like model (c), but computationally it is almost 5 times cheaper, demonstrating how effective the IF MAC can be. Model (e) refines the spatial resolution by a factor of $2$ in each direction (as compared with models (a) through (d)), and this reduces the error to $e_M\sim 5$\% but increases the computational cost by a factor of $\sim 48$ relative to model (d). Part of this (a factor $\sim 16$) is due to the higher number of grid cells and the shorter time step. The remainder (a factor of $\sim 3$) is partly due to the higher number of tree nodes that need to be opened, and partly due to the larger number of evaluation points on each ray.

\subsection{Cloud irradiated by two sources}
\label{sec:2src}

\begin{figure*}
\begin{center}
\includegraphics[width=1.0\textwidth]{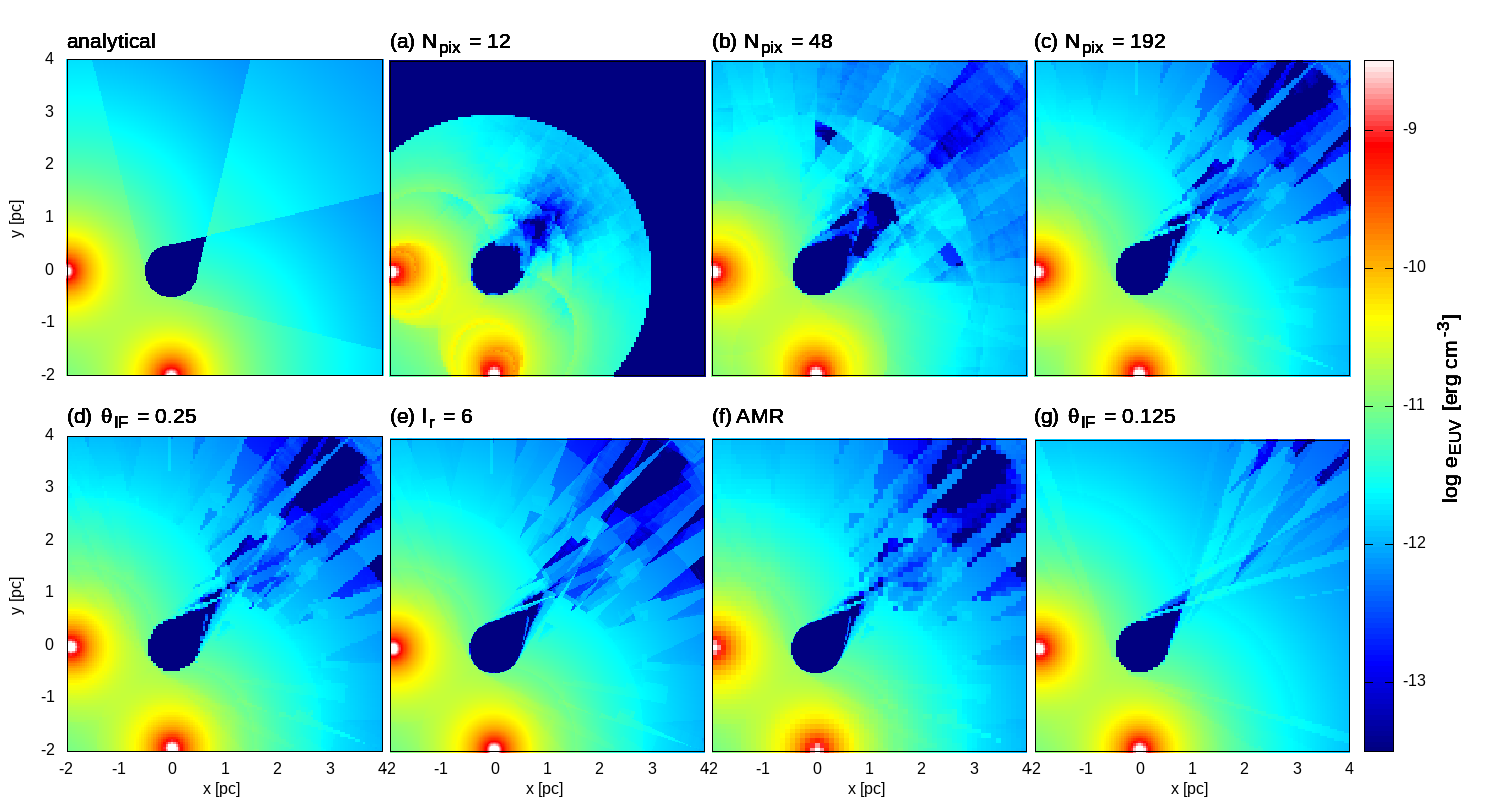}
\caption{A spherical cloud irradiated by two sources. The top left panel shows the radiation energy density on the $z\!=\!0\,$ plane, calculated analytically. The remaining panels show the same quantity for models (a) through (g).}
\label{fig:2src}
\end{center}
\end{figure*}

\begin{table}
\caption{Accuracy and performance of the two-source test.}
\label{tab:acc:2src}
\begin{center}
\begin{tabular}{l|c|l|r|l|l}
\hline
model & $l_r$ & $\theta_\mathrm{lim}$ & $N_{_{\rm PIX}}$ & $\theta_\mathrm{IF}, \theta_\mathrm{Src}$ & $t_\mathrm{iter}$ \\
\hline
(a) $\Npix = 12$ & $5$ &   $1$  & $12$  & $\infty$     &  $0.095$  \\
(b) $\Npix = 48$ & $5$ &   $0.5$  & $48$  & $\infty$   &  $0.3$    \\
(c) $\Npix = 192$ & $5$ &   $0.25$  & $192$ & $\infty$  &  $1.7$    \\
(d) $\theta_\mathrm{IF} = 0.25$ & $5$ &   $1$  & $192$ & $0.25$       &  $0.65$   \\
(e) $l_r = 6$ & $6$ &   $1$  & $192$ & $0.25$       &  $6.8$    \\
(f) AMR & $4,6$ & $1$  & $192$ & $0.25$       &  $0.2$    \\
(g) $\theta_\mathrm{IF} = 0.125$ & $5$   & $1$  & $768$ & $0.125$      &  $2.1$   \\
\end{tabular}
\end{center}
\begin{flushleft}
Column 1 gives the model name. The following columns list:
\begin{itemize}
\item $l_{r}$: the refinement level defining the grid resolution (`5' $\rightarrow 128^3$; `6' $\rightarrow$ $256^3$; `4,6' $\rightarrow$ AMR with minimum and maximum refinement levels $4$ and $6$, respectively.)
\item $\theta_\mathrm{lim}$: the limiting opening angle for the BH MAC
\item $N_{_{\rm PIX}}$: the number of rays (defining the angular resolution)
\item $\theta_\mathrm{IF}$, $\theta_\mathrm{Src}$: the limiting opening angles for the IF MAC and the Src MAC, respectively
\item $t_\mathrm{iter}$: the processor time for a single iteration step (in core-hours)
\end{itemize}
\end{flushleft}
\end{table}

\textsc{Two-source motivation}. This test assesses the fidelity of the code when treating a cloud that is irradiated by two identical sources, from different directions. Similarly to the previous tests (Sections~\ref{sec:rabbithole} and \ref{sec:rdi}), it is sensitive to the angular resolution, represented by $\theta_{\rm lim}$ and $\Npix$, and to the choice of MAC. Additionally, it evaluates the iteration and error control procedures (see Section~\ref{sec:err:cntrl}), since their failure would corrupt the symmetry of the radiation field with respect to the plane perpendicular to the line connecting the two sources. 

\textsc{Two-source setup}. The computational domain is $-2\!\leq\!x\!\leq\!4\,{\rm pc}$, $-2\!\leq\!y\!\leq\!4\,{\rm pc}$ and $-3\!\leq\!z\!\leq\!3\,{\rm pc}$. The two sources are located at $(x,y,z)\!=\!(-2,0,0)$\,pc and $(0,-2,0)$\,pc, and each emits ionising photons at rate $N_{_{\rm LyC}} = 3.2\times 10^{48}$\,s$^{-1}$. The cloud has mass $M = 20$\,\MSun\,, radius $R = 0.5$\,pc, density $2.6\times 10^{-21}\rm{g\,cm^{-3}}$, and is located at the centre of coordinates. Outside the cloud, the computational domain is filled with rarefied gas with density $10^{-24}$\,g\,cm$^{-3}$. The other parameters are given in Table~\ref{tab:acc:2src} for all 7 calculated models, (a) through (g). Models (a) through (f) have the same parameters as the corresponding models for the RDI test; model (g) has a very high angular resolution given by $\theta_\mathrm{IF} = \theta_\mathrm{Src} = 0.125$ and $\Npix = 768$ (as used in the Rabbit hole test, Section~\ref{sec:rabbithole}). All the models were run for a single time step to let the {\sc TreeRay} iteration process converge; the time evolution was not explored.

\textsc{Two-source results}. We only evaluate this test qualitatively, by comparing the computed radiation field with the analytic solution. Fig.~\ref{fig:2src} shows the radiation energy density on the $z\!=\!0$ plane for all models, and for the analytic solution, assuming a completely opaque cloud. We see that none of the models exhibit significant deviation from symmetry about the line $x\!=\!y$. Model (a), with the lowest angular resolution, exhibits the largest deviations from the analytic solution. The shadow behind the cloud has the wrong shape, some radiation leaks into the shadowed region, and the radiation energy drops to zero at distances $D\ga\!3$\,pc from the cloud. The last effect is caused by the large tree opening angle $\theta_\mathrm{lim} =1$: at large distance from a target cell, the two sources and parts of the cloud are merged into a single node with high absorption coefficient. The incorrect shadow geometry is the result of both a low number of rays and a high $\theta_\mathrm{lim}$. In model (b) the drop to zero disappears, but distortion of the shadow geometry is still significant. The shadow geometry is approximately correct in model (c) with $\theta_\mathrm{lim} = 0.25$ and $192$ rays. Nevertheless, even this relatively high angular resolution is not sufficient to reproduce the radiation energy in the top right corner, which should be relatively high, because this region is irradiated by both sources. Model (d) shows that almost the same radiation field as in model (c) can be obtained at lower computational costs by invoking the physical MACs with $\theta_\mathrm{IF} = \theta_\mathrm{Src} = 0.25$. Models (e) and (f) show that the grid resolution and the AMR have negligible effect on the radiation field. Finally, model (g) with very high angular resolution (shown in the top right corner) better reproduces the radiation field (though not perfectly) showing that the method converges to the correct solution with increasing angular resolution.

\section{Star formation and feedback with TreeRay}
\label{sec:SFFeedback}

\begin{figure*}
\begin{center}
\includegraphics[width=1\linewidth]{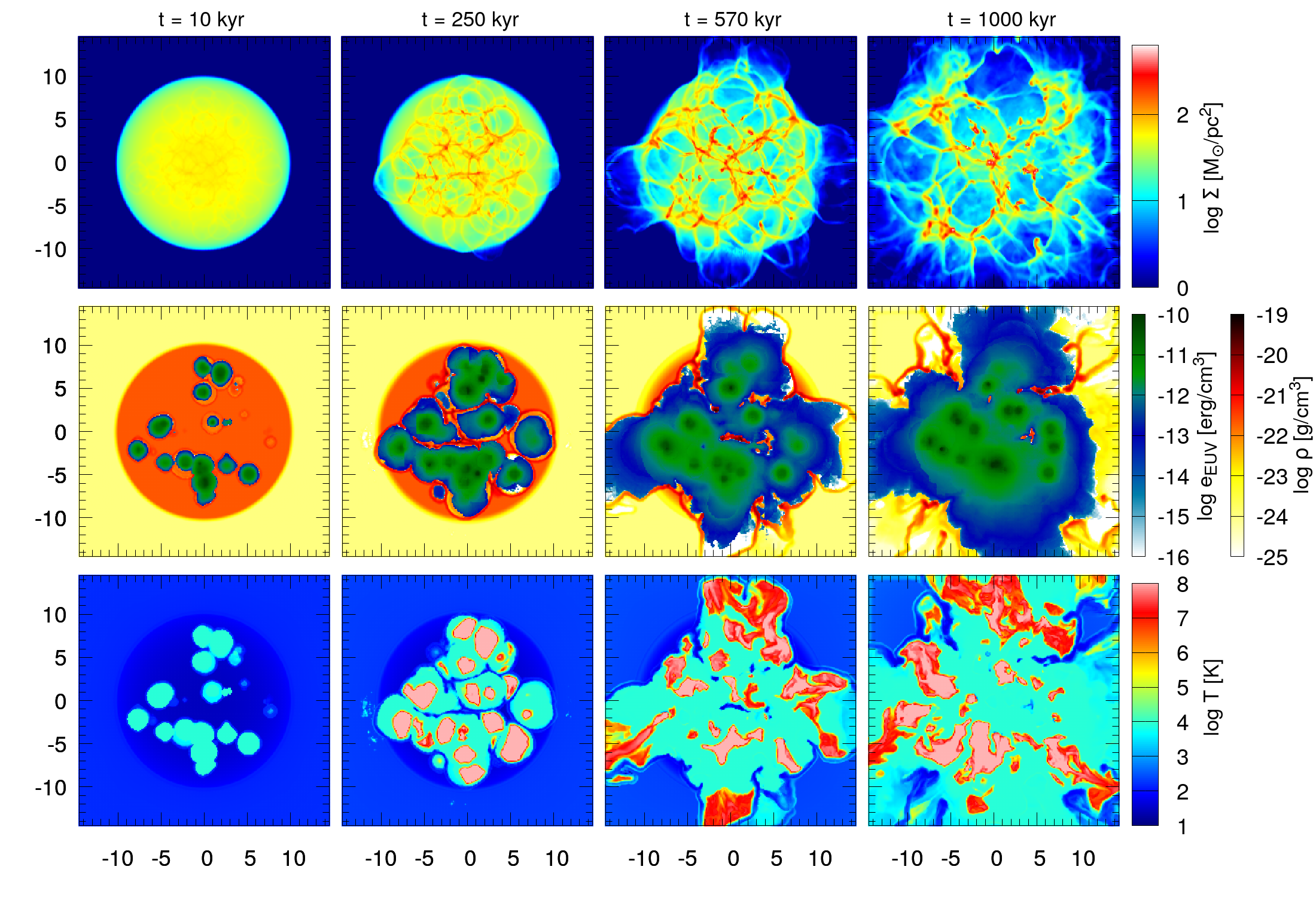}
\caption{Star formation and feedback test: the fiducial model (a) at times $t\!=\!10$, $250$, $570$ and $1000$\,kyr (from left to right). The top row shows the logarithm of the column density. The middle row shows the logarithm of the radiation energy, in the region with non-zero ionisation degree, and the logarithm of the gas density, in the remaining parts (i.e. for the neutral gas only). The bottom row shows the logarithm of the gas temperature. Panels in the middle and bottom rows show the quantities on the $z=0$ plane.}
\label{fig:sff:evol}
\end{center}
\end{figure*}

\begin{figure}
\begin{center}
\includegraphics[width=1\linewidth]{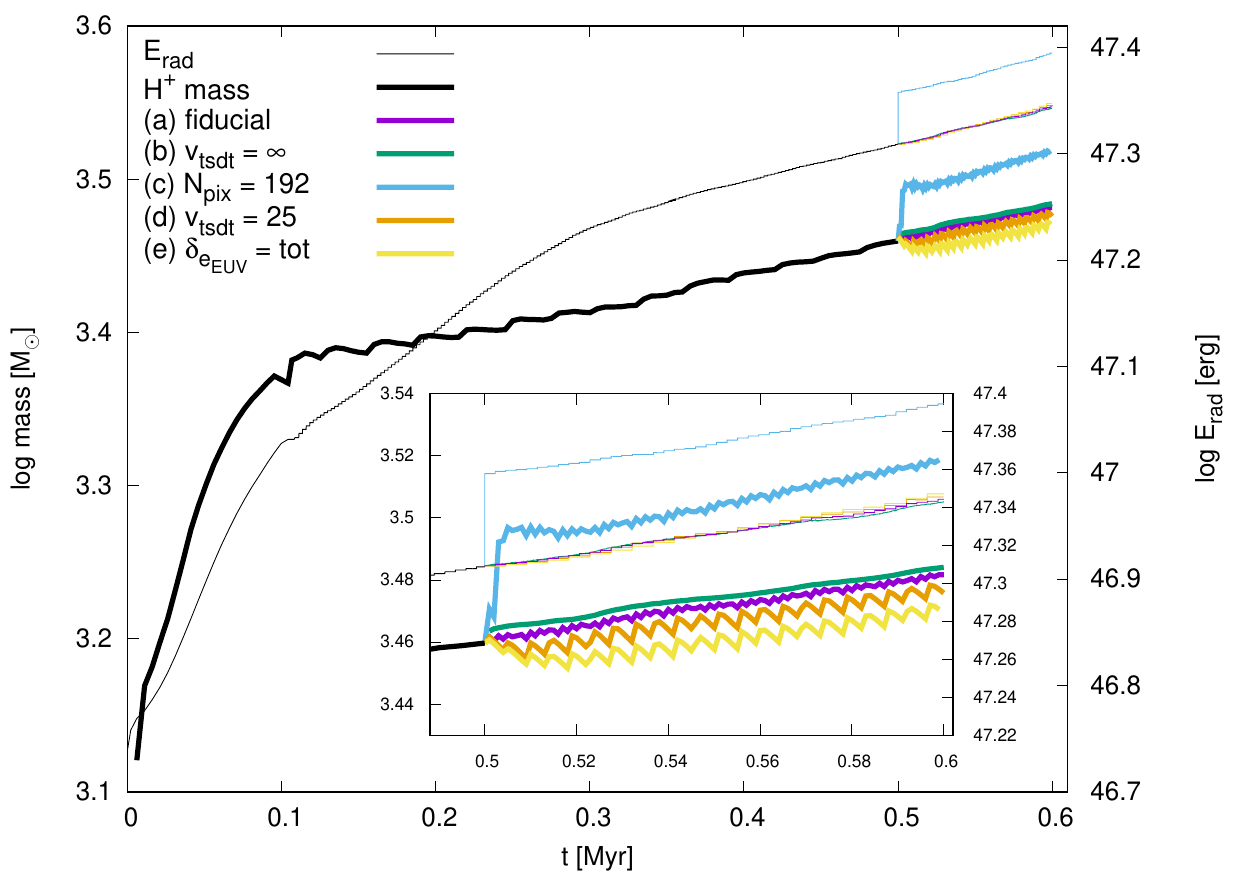}
\caption{Star formation and feedback: the evolution of the total radiation energy, $E_\mathrm{rad}$, (thin lines, righthand ordinate), and the mass of ionised gas, $M_\mathrm{H^{+}}$,  (thick lines, lefthand ordinate), in the whole computational domain, for models (a) through (e). The main figure shows the first $500$\,kyr of evolution for model (a) only (black lines). The inset shows the evolution between $500$ and $600$\,kyr for all models. Note that, even at the resolution of the inset, the differences in $E_\mathrm{rad}$ between models (a), (b), (d) and (e) cannot be resolved.}
\label{fig:sff:cmp}
\end{center}
\end{figure}

\begin{figure}
\begin{center}
\includegraphics[width=1\linewidth]{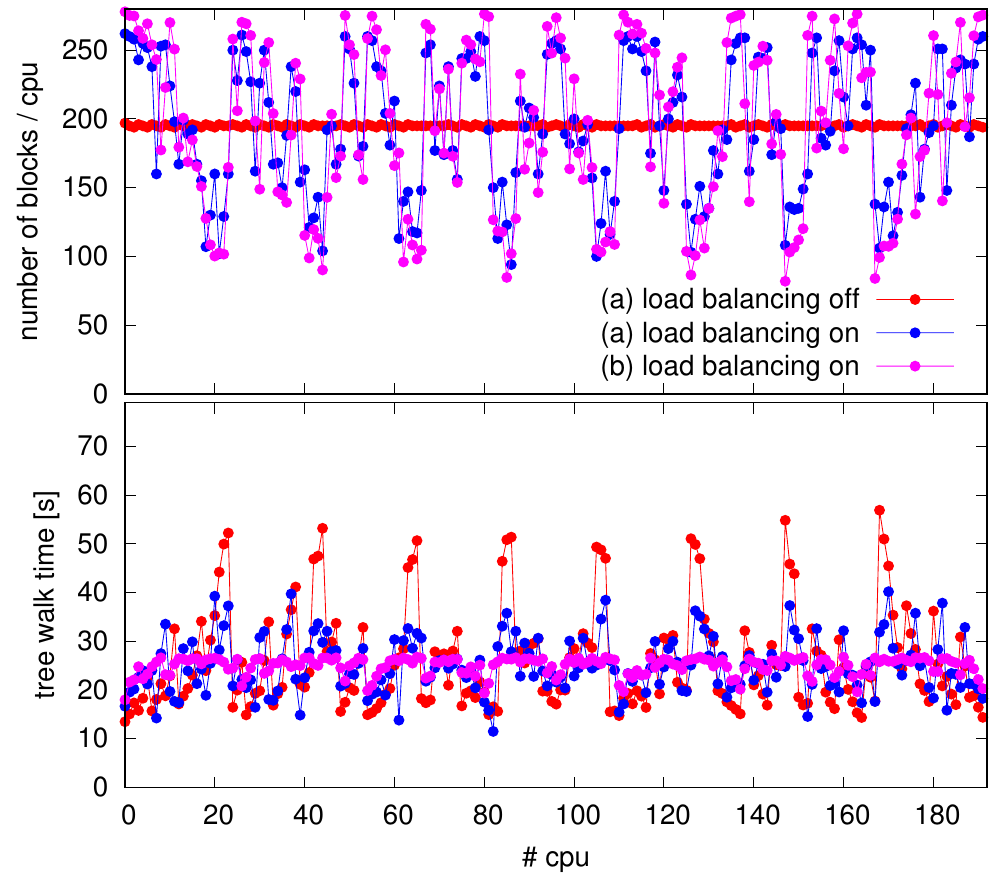}
\caption{The effect of load balancing in the Star Formation and Feedback test. {\it Top panel:} the number of blocks on each processor for model (a) with load balancing off (red, flat distribution), model (a) with load balancing on (blue), and model (b) with load balancing on (magenta). {\it Bottom panel:} the duration of the tree walk on each processor for the same three cases. The measurements shown have been made at time $t\!=\!500$\,kyr, but are representative of the majority of the time.}
\label{fig:sff:lb}
\end{center}
\end{figure}

\begin{table*}
\caption{Star formation \& feedback test.}
\label{tab:sff}
\begin{center}
\begin{tabular}{l|c|c|l|r|l|c|l|c|c|c|c|r|r|r|r|r}
\hline
model & $N_\mathrm{src}$ & $l_r$ & $\theta_\mathrm{lim}$ & $N_{_{\rm PIX}}$ & $\theta_\mathrm{IF}$ & $v_\mathrm{tsdt}$ & $\deleEUV$  & $n_\mathrm{steps}$ & $n_\mathrm{tr}$ & $e_\mathrm{Erad}$ & $e_\mathrm{M_{H^{+}}}$ & $t_\mathrm{hydro}$ & $t_\mathrm{chem}$ & $t_\mathrm{part}$ & $t_\mathrm{tr}$ & $t_\mathrm{tot}$ \\
\hline
(a) fiducial & 100 & $6$ &   $1.0$  & $48$   & $0.5$.     & $50$     & cell & 4253 & 309   & --       & --      & 1430 & 1000 & 980  & 950 & 4600  \\
(b) $v_\mathrm{tsdt} = \infty$ & 100 & $6$ &   $1.0$  & $48$   & $0.5.  $   & $\infty$ & cell & 4227 & 36278 & $0.005$ & $0.011$ & 2030 & 1520 & 1180 & $89000$ & $ 94500$  \\
(c) $\Npix = 192$ & 100 & $6$ &   $1.0$  & $192$  & $0.25$     & $50$     & cell & 4997 & 358   & $0.12$   & $0.085$ & 1700 & 1160 & 1140  & 4500 & 8800 \\
(d) $v_\mathrm{tsdt} = 25$ & 100 & $6$ &   $1.0$  & $48$   & $0.5$.     & $25$     & cell & 4253  & 180   & $0.008$ & $0.016$ & 1460 & 990 & 960  & 630 & 4300 \\
(e) $\deleEUV = \mathrm{tot}$ & 100 & $6$ &   $1.0$  & $48$   & $0.5$.     & $25$     & tot  & 4288  & 22    & $0.01$   & $0.032$ & 1440 & 1010  & 970  &  240 & 3900 \\
(f) 1 source & 1   & $6$ &   $1.0$  & $48$   & $0.5$.     & $50$    & cell & 2690  & 156   & --       & --      & 890  & 670  & 370  & 2000 & 4060 \\
\end{tabular}
\end{center}
\begin{flushleft}
Column 1 gives the model name. The following columns list:
\begin{itemize}
\item $N_\mathrm{src}$: the number of sources
\item $l_{r}$: the refinement level defining grid resolution (`5' $\rightarrow 128^3$; `6' $\rightarrow 256^3$; `4,6' $\rightarrow$ AMR with minimum and maximum refinement levels $4$ and $6$, respectively.)
\item $\theta_\mathrm{lim}$: the limiting opening angle for the BH MAC
\item $N_{_{\rm PIX}}$: the number of rays (defining the angular resolution)
\item $\theta_\mathrm{IF}$, : the limiting opening angle for the IF MAC, parameter $\theta_\mathrm{Src}$ of the Src MAC is set to the same value
\item $v_\mathrm{tsdt}$: the velocity limit in km/s for the adaptive tree solver time step 
\item $\deleEUV$: error control method (either $\delcel$ given by equation~\ref{eq:delta:er:cell}, or $\deltot$ by equation~\ref{eq:delta:er:tot}; see Section~\ref{sec:err:cntrl})
\item $n_\mathrm{steps}$: the number of hydrodynamic time steps for the whole run
\item $n_\mathrm{tr}$: number of tree solver iterations for the whole run
\item $e_\mathrm{Erad}$: the maximum fractional difference in the total radiation energy, relative to model (a)
\item $e_\mathrm{M_{H^{+}}}$: maximum relative difference of the total mass of the ionised gas w.r.t. model (a)
\item $t_\mathrm{hydro}$: the processor time spent in the hydro module, for the whole run (in core-hours)
\item $t_\mathrm{chem}$: the processor time spent in the chemistry module, for the whole run (in core-hours)
\item $t_\mathrm{part}$: the processor time in the particles module, for the whole run (in core-hours)
\item $t_\mathrm{tr}$: the processor time spent in the tree solver, for the whole run (in core-hours)
\item $t_\mathrm{tot}$: the processor time in all modules, for the whole run (in core-hours)
\end{itemize}
\end{flushleft}
\end{table*}

\textsc{Star Formation and Feedback motivation}. The purpose of the final test is to demonstrate the combination of {\sc TreeRay} with other physical modules that are often used in simulations of star formation with feedback, and to evaluate the code performance under realistic conditions. We derive this test setup from model CNM~60 (i.e. $60$\,M$_{\odot}$ star located in the cold neutral medium) of \citet{Haid2018} who explore the relative impact of radiation and stellar winds in different environments. In the CNM~60 model, a source of radiation and stellar wind representing a $60$\,M$_\odot$ star is placed in a dense cold neutral medium, resulting in the formation of an expanding H{\sc ii} region with a stellar wind bubble at its centre. A challenging aspect of this test is the combination of relatively complex physics (related to radiation and cold gas chemistry) with the high velocity of stellar winds, $\ga 1000$ km/s (obliging the hydrodynamic solver to take very short time steps, due to the Courant-Friedrichs-Lewy criterion). In order to demonstrate the ability of {\sc TreeRay} to deal efficiently with a large number of sources, we split the single source used by \citet{Haid2018} into $100$ smaller sources, which together emit ionising photons at the same net rate, and together deliver the same total wind power (with the same wind velocity, so the mass loss rate from each individual star is simply divided by 100). The 100 sources are distributed randomly in a sphere of radius $8$\,pc, within a cloud of radius $10$\,pc, to represent a toy-model star cluster.

\textsc{Physical processes}. The physical processes included in this test and the corresponding {\sc Flash} modules are the following. In addition to the standard {\sc Flash} PPM hydrodynamic solver, we use the tree solver to calculate the gas self-gravity, and the {\sc TreeRay/OpticalDepth} module to include the ambient interstellar radiation field (see Paper~I for both modules). The ionising radiation is treated using the {\sc TreeRay/OnTheSpot} module described here, and its coupling to the chemistry module is given in \citet{Haid2018}. The {\sc Chemistry} module implements a network with 7 active species (H$_2$, H, H$^{+}$, CO, C$^{+}$, O, and $e^{-}$; see \citealt{Walch2015, Glover2010, Nelson1997} for details). The sources are modelled as {\sc Flash} sink particles \citep{Federrath2010}, and move under the influence of the gravitational field, but accretion onto sinks is switched off. Stellar winds are treated with the procedure described by \citet{Gatto2017} and implemented in the {\sc FeedbackSinks} module.

\textsc{Physical parameters}. Apart from the number of sources, the parameters are similar to \citet{Haid2018}. The computational domain is a $30\times 30\times 30\,{\rm pc}^3$ cube and the grid is uniform with refinement level $l_r = 6$ (corresponding to $256^3$ grid cells). The hydrodynamic boundary conditions are set to ``diode'', and the gravitational boundary conditions are ``isolated''. The interstellar radiation field, from which the heating is calculated by the {\sc TreeRay/OpticalDepth} module, has strength $G_0 = 1.7$ \citep{Habing1968, Draine1978}. At the centre of the computational domain is a cold neutral cloud with mass $1.3\times 10^4\,{\rm M}_\odot$, radius $10$\,pc, density $2.1\times 10^{-22}$\,g\,cm$^{-3}$, and temperature $20$\,K. The remainder of the computational domain is filled with a rarefied ambient medium having density $10^{-24}$\,g\,cm$^{-3}$.  The sources are positioned randomly in a sphere of radius $8\,{\rm pc}$, centred on the centre of the cloud. Each of the $100$ sources has ionising output $N_{\rm LyC} = 2.4\times 10^{48}$\,s$^{-1}$, surface temperature $T = 45000$\,K, wind mass loss rate $3\times 10^{-8}$\,M$_\odot$\,yr$^{-1}$, and wind velocity $2700$\,km/s. 

\textsc{Technical parameters}. We evaluate this test by comparing runs computed with different {\sc TreeRay} parameters. The model is most interesting, and computationally most demanding, when (i) a significant fraction of the cloud is ionised, and (ii) there is hot shocked stellar-wind gas. This happens after several hundred kyr, and therefore we only compare the runs during the time period from 500 to 600\,kyr. The first 500\,kyr are only calculated once, with model (a). The parameters of the fiducial model (a) are selected as a compromise between accuracy and performance. Model (a) uses moderate angular resolution with $\Npix = 48$ and {\sc TreeRay}-specific MACs with $\theta_\mathrm{IF} = 0.5$ and $\theta_\mathrm{Src} = 0.5$. This allows us to adopt a large general opening angle $\theta_\mathrm{lim} = 1.0$. Since the presence of very hot gas results in a very short hydrodynamic time step, we apply the tree solver time step (see Section~\ref{sec:tsdt}) and set the velocity limit  to $v_\mathrm{tsdt} = 50$\,km/s. Model (b) does not use the tree solver time step ($v_\mathrm{tsdt} = \infty$) and the tree solver is called at each hydrodynamic time step; hence, (b) is by far the most expensive model. Model (c) differs from the fiducial model by invoking higher angular resolution, with $\Npix = 192$ and $\theta_\mathrm{IF} = \theta_\mathrm{Src} = 0.25$. Models (d) and (e) explore the behaviour when the low velocity limit for the tree solver time step is reduced to $v_\mathrm{tsdt} = 25$\,km/s; model (e) also adopts the less demanding error control criterion involving the total radiation energy. Finally, model (f) involves only a single source of radiation and stellar wind, located at the centre of the cloud, with a total photon emissivity and mass loss rate equal to the sum of the 100 sources of the fiducial model; this model is intended to reproduce model CNM~60 from \citet{Haid2018}. The parameters of all models are summarized in Table~\ref{tab:sff}.

\textsc{Model evolution}. Fig.~\ref{fig:sff:evol} illustrates the first Myr of evolution for the fiducial model (a). At early times, H{\sc ii} regions appear around the sources and start to expand at $\sim 6$\,km/s, in agreement with the Spitzer solution (Eqs. \ref{eq:spitzer} and \ref{eq:RStr}). The H{\sc ii} regions around neighbouring sources eventually merge. At $t \simeq 70$\,kyr, some hot gas, resulting from shocked stellar winds, appears and quickly expands due to its high pressure; consequently the H{\sc ii} regions are squeezed between the hot wind bubbles on the inside and the surrounding shells of swept-up cold neutral gas on the outside. The bubbles and H{\sc ii} regions continue to expand and merge, and at the same time the cloud slowly collapses due to its self-gravity. Eventually, at $t \simeq 250$\,kyr, some bubbles reach the cloud edge and break out, in the process known as a {\it champagne flow} \citep{TenorioTagle1979}. Thereafter, the warm photo-ionised H{\sc ii}, and the shocked hot wind-gas, start to flow out of the cloud, with the shocked hot wind-gas also being accelerated by the buoyancy force. The remainder of the cloud decays into a network of filaments, which expands slowly outwards, accelerated by the pressure of the warm and hot gas, and by the rocket effect \citep{OortSpitzer1955}. Some structures formed in the later stages of evolution resemble the {\it elephant trunks} that are frequently observed in star forming regions \citep[see e.g.][]{Hillenbrand1993, McLeod2015}.\\

\textsc{Comparison of models}. Fig.~\ref{fig:sff:cmp} compares the evolution of models (a) through (e) between 500 and 600\,kyr. The thin lines show the total radiation energy in the computational domain ($E_\mathrm{rad}$); its maximum fractional difference relative to model (a), $e_\mathrm{E_{rad}}$, is given in Table~\ref{tab:sff}. Models (a), (b), (d) and (e) have almost the same $E_\mathrm{rad}$ with relative differences of order 1\% or smaller. For model (c), $E_\mathrm{rad}$ is higher by $\sim$10\%, because its higher angular resolution allows the radiation to follow better the curved surfaces of irregular shells, resulting in slightly larger H{\sc ii} regions. The thick lines in Fig.~\ref{fig:sff:cmp} show the total mass of ionised gas, $M_\mathrm{H^{+}}$, and its fractional difference relative to model (a), $e_\mathrm{M_{H^{+}}}$, is again given in Table~\ref{tab:sff}. For model (c) $M_\mathrm{H^{+}}$ is higher than for model (a) by $\sim$8.5\%, again because of higher resolution. Models (a), (b), (d) and (e) show $M_\mathrm{H^{+}}$ differing by of order 1\%, with higher $M_\mathrm{H^{+}}$ for models with a shorter time between calls to the tree solver (i.e. higher $v_\mathrm{tsdt}$). This is because during time steps when the tree solver is not called, the shells continue to expand and a small fraction of the ionised gas gets into regions that are not irradiated, where it recombines and so $M_\mathrm{H^{+}}$ drops unphysically (see periods of decrease in the saw-tooth pattern of models (d) and (e)). We conclude that the angular resolution (i.e. parameters $\Npix$, $\theta_\mathrm{IF}$ and $\theta_\mathrm{Src}$) has an impact on the accuracy of calculations of this type. In contrast, the tree solver time step parameter, $v_\mathrm{tsdt}$, seems to have little impact, provided $v_\mathrm{tsdt} \gtrsim 50$\,km/s. Indeed, even models (d) and (e) with $v_\mathrm{tsdt} = 25$\,km/s give satisfactory results, and would be suitable for quick tests scanning the parameter space.

\textsc{Performance}. Table~\ref{tab:sff} shows the total CPU times spent in the four computationally most demanding modules: the hydrodynamic solver ($t_\mathrm{hydro}$), the {\sc Chemistry} module ($t_\mathrm{chem}$), the particle module ($t_\mathrm{part}$), and the tree solver including {\sc TreeRay} ($t_\mathrm{tr}$). In model (a), the times taken by these four modules are comparable, with $t_\mathrm{tr}\sim 2t_\mathrm{hydro}/3$. The small $t_\mathrm{tr}$ is largely due to setting a finite $v_\mathrm{tsdt}$; the tree solver is called only $309$ times, while the hydrodynamic solver is called $4253$ times (columns $n_\mathrm{tr}$ and $n_\mathrm{step}$ in Table~\ref{tab:sff}, respectively). The benefits of setting a finite $v_\mathrm{tsdt}$ are further illustrated by model (b), where $v_\mathrm{tsdt}$ is not set to a finite value, and therefore defaults to $\infty$; the results are indistinguishable from model (a) but the tree solver takes $\sim$90 times more time than in model (a). Model (c), with finer angular resolution, is approximately two times slower than model (a), and the tree solver takes approximately three times longer than the hydrodynamic solver. In models (d) and (e), the tree solver is called less often than in model (a), and $t_\mathrm{tr}$ is proportionally smaller; model (e) uses the total radiation energy as the error control, and {\sc TreeRay} does not need to iterate at all. Model (f), with a single source of radiation and wind, needs almost two times fewer hydrodynamic time steps, due to the lower maximum temperature of the hot shocked gas. This also results in a smaller number of tree solver calls than in model (a). However, each tree solver iteration takes more time, because the ionisation front has larger surface area, and consequently a higher number of tree nodes must be opened. 

\textsc{Load balancing}. Models (a) through (f) have been calculated with load balancing switched on (see Section~\ref{sec:lb}). In order to evaluate the impact of load balancing on the tree solver performance, we calculate a few time steps of model (a), starting at 500\,kyr, with load balancing off. The top panel of Fig.~\ref{fig:sff:lb} compares the number of blocks per processor for model (a) with load balancing on and off, and for model (b) with load balancing on. The bottom panel shows, for the same three cases, the time spent in a single tree walk on each processor. It can be seen that models with load balancing on have smaller variations in the tree walk time, and that the variation is much smaller for model (b) where the tree solver is called at every time step. Since processors that finish the tree walk earlier have to wait for the slowest processor, we estimate that in model (a) the load balancing decreases the tree walk time from $\sim$\,60\,seconds to $\sim$\,40\,seconds, saving approximately 30\% of the tree solver time.

\section{Scaling tests}
\label{sec:perf}

\subsection{Weak and strong scaling}
\label{sec:strong_sc}

\textsc{Hardware for scaling tests}. We carry out weak and strong scaling tests for the {\sc TreeRay} algorithm based on the Spitzer test (see Section~\ref{sec:spitzer}). The tests are run on the HPC system COBRA, hosted by the Rechenzentrum Garching at the Max-Planck Computing and Data Facility. COBRA has Intel Xeon 'Skylake' processors. In total there are 3188 compute nodes with 40 cores @ 2.4 GHz each and a memory of more than 2.4 GByte per core. The available memory per core is thus at least 2.2 GByte. 

\begin{figure}
\begin{center}
\includegraphics[width=\linewidth]{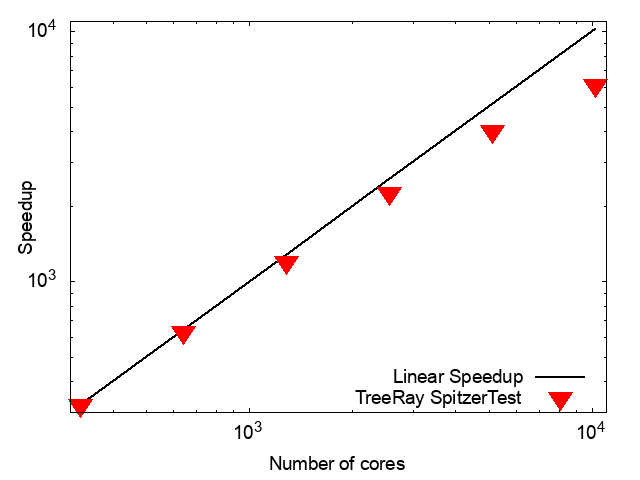}
\caption{Strong scaling on up to 10240 cores, showing almost linear speedup for the Spitzer test.}
\label{fig:scale_strong}
\end{center}
\end{figure}

\textsc{Strong scaling}. Fig.~\ref{fig:scale_strong} shows the results of the strong scaling test. We plot the time in seconds measured for the tree build, communication, walk, and radiative transfer calculation as measured for 10 time steps during the Spitzer test. The spatial resolution is set to 512$^3$ cells, and therefore the average number of blocks per core changes from 819.2 on 320 cores, to 25.6 on 10240 cores. The scaling is very good, showing an almost ideal, linear speedup with the number of cores.

\begin{figure}
\begin{center}
\includegraphics[width=\linewidth]{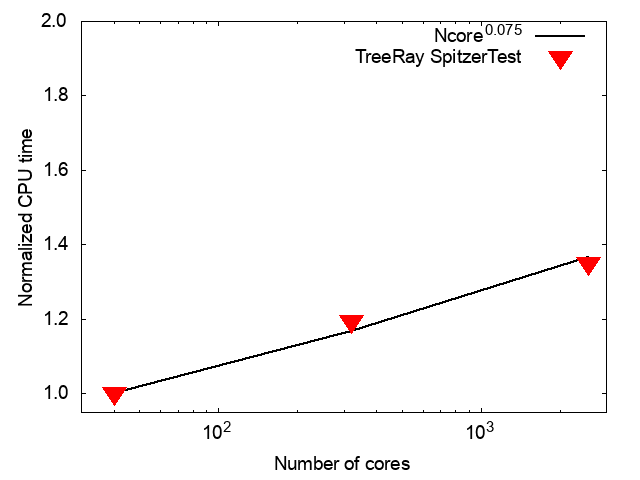}
\caption{Log-linear plot showing the results of a weak scaling test on 40 to 20480 cores. The scaling within one node is not ideal, but for more than 40 cores, the scaling is close to ideal. A power-law fit gives a very weak dependence on the number of cores $\propto N_{_{\rm core}}^{\,0.075}$.}
\label{fig:scale_weak}
\end{center}
\end{figure}

\textsc{Weak scaling}. Fig.~\ref{fig:scale_weak} shows the result of the weak scaling test. The simulations are chosen such that there is the same average number of blocks per core of 102.4. In order to achieve this, we change the resolution of the Spitzer test from $128^3$ cells on $40$ cores to $512^3$ cells on $2560$ cores. The processor time in the tree solver only depends very weakly on the number of cores, $N_{_{\rm core}}$, and can be approximated by a power-law, $\propto N_{_{\rm core}}^{\,0.075}$. This is close to an ideal weak scaling for which the same amount of wall-clock time should be used for all simulations. 

\begin{figure}
\begin{center}
\includegraphics[width=\linewidth]{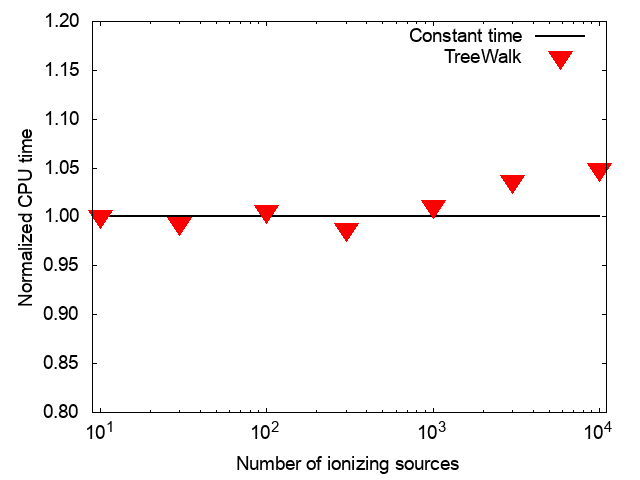}
\caption{Scaling with the number of sources. The setup consists of a uniform density medium that contains from $N_{\rm src}=10$ up to 10$^4$ randomly distributed ionizing sources. The log-linear plot shows the integration time normalized to the time it takes to calculate 10 time steps for only 10 sources, i.e. 10 Spitzer bubbles (left-most point). As $N_{\rm src}$ is increased by a factor of 1000, the integration time only increases by $\sim$ 5\%. This means that the integration time is almost independent of the number of sources.}
\label{fig:scale_nsource}
\end{center}
\end{figure}

\subsection{Scaling with the number of sources}
\label{sec:perf:nsrc}

\textsc{Number-of-sources scaling test setup}. To test the extent to which the performance of the code degrades as the number of ionising sources is increased, we set up a $30\times 30\times 30\,{\rm pc}^3$ computational domain, containing atomic hydrogen with uniform density $\rho=7.63\times 10^{-22}\,{\rm g}\,{\rm cm}^{-3}$ and uniform temperature $T$=10\,K. $N_{\rm src}$ sources are placed randomly in the computational domain, and each source emits ionising photons at rate  $N_{\rm LyC} = 5\times 10^{48} \,{\rm s}^{-1}$ into an injection region with radius $r_{\rm inj} = 0.32$ pc (which corresponds to about 1.4 cells). $N_{\rm src}$ is set to 10, 32, 100, 316, 1000, 3162 and 10\,000. All setups are evolved for 10 time steps, on 32 cores, in order to measure the exact integration time.

\textsc{Number-of-sources scaling results}. Fig.~\ref{fig:scale_nsource} shows the resulting simulation times as a function of $N_{\rm src}$, normalised to the simulation time for $N_{\rm src}=10$. For $N_{\rm src}\la 10^3$ there is essentially no difference in the simulation times. For $N_{\rm src}> 10^3$, the simulation time increases slightly and becomes $\sim$ 5\% longer for $N_{\rm src}\!=\!10^4$ than for $N_{\rm src}\!=\!10$. We conclude that the simulation time is very nearly independent of the number of sources, as expected from the algorithm design. Therefore, it is an excellent basis for implementing radiation transport schemes where every grid cell represents a source of radiation, e.g. emission from hot gas or dust.

\section{Summary}
\label{sec:summary}

In this paper we describe {\sc TreeRay}, a new, fast algorithm for treating radiation transport in gaseous media. It is based on the combination of reverse ray tracing \citep[e.g.][]{Altay2013} and a tree-based \citep{Barnes1986} accelerated integration scheme. In general, the incident flux of radiation is computed for every grid cell, but it can also be computed for any other target point in the computational domain, for example the position of a sink particle. From every target point, reverse ray tracing is executed in $\Npix$ directions (hence the angular resolution is user-defined), and the directions are interpreted as cones with equal solid angle, based on the {\sc HealPix} scheme \citep{Gorski2005}. Due to the equal solid-angle pixelation, every direction's contribution carries equal weight. In the limit of infinite angular resolution, {\sc TreeRay} converges to the long characteristics method, which is very accurate but usually prohibitively expensive for time-dependent astrophysical simulations. 

{\sc TreeRay} treats all the gas in the computational domain, and can capture the shadows of even quite small and dense objects, with the limitation that structures at large distances from a target point are smoothed out over a solid angle $\,\sim\!4\pi /\Npix\!$. The smoothing is controlled by the {\sc HealPix} resolution (user-specified $\Npix\!$) and the limiting opening angles set for the Multipole Acceptance Criteria (user-specified $\theta_{\rm lim},\,\theta_{\rm IF},\,\theta_{\rm Src}$). The number of evaluation points at which the radiative transfer equation is integrated, along a given ray, is of secondary importance.  

A key strength of {\sc TreeRay} is that its computational cost is essentially independent of the number of radiation sources. This enables {\sc TreeRay} to treat big star clusters with many radiation sources, and extended sources like radiatively cooling shock fronts or cool dust clouds, without incurring an unacceptable computational overhead.  

Furthermore {\sc TreeRay} scales extremely well with the number of processors, which is due to the communication and local tree-walk strategy of the scheme \citep[see also][]{Wunsch2018}. We demonstrate {\it both} an almost ideal weak scaling up to $\sim 2.5\times 10^3$ cores, {\it and} an almost ideal strong scaling on up to $\sim 10^4$ cores (which is usually even harder to achieve). 

{\sc TreeRay} can easily be extended to include additional radiative transfer sub-modules. Additional sub-modules that are already in preparation include the transfer of non-ionising radiation including the emission from dust and the associated radiation pressure (Klepitko et al., in prep.); the multi-wavelength transfer of X-rays originating from point sources such as high-mass X-ray binaries \citep[Gaches et al., in prep., based on the diffuse X-ray radiative transfer scheme with the {\sc TreeRay/OpticalDepth} module developed by][]{Mackey2019}; and the multi-wavelength transfer of Far Ulraviolet (FUV) and EUV radiation, including the dissociation of molecules (Walch et al., in prep.). We plan in future to include the option to have directionally dependent absorption coefficients, similar to those introduced in \citep{Grond2019TREVR}.

%%%%%%%%%%%%%%%%%%%%%%%%%%%%
\section*{Acknowledgments}%
%%%%%%%%%%%%%%%%%%%%%%%%%%%%

% {{{
We thank the anonymous referee for constructive and useful comments that helped to improve the paper.
This study has been supported by project 19-15008S of the Czech Science Foundation and by the institutional project RVO:67985815. 
SW, SH, and AK gratefully acknowledge the European Research Council under the European Community's Framework Programme FP8 via the ERC Starting Grant RADFEEDBACK (project number 679852). 
SW and FD further thank the Deutsche Forschungsgemeinschaft (DFG) for funding through SFB~956 ``The conditions and impact of star formation'' (sub-project C5), and SW thanks the Bonn-Cologne-Graduate School. 
DS acknowledges the DFG for funding through SFB~956 ``The conditions and impact of star formation'' (sub-project C6).
APW acknowledges the support of a consolidated grant (ST/K00926/1) from the UK Science and Technology Facilities Council. 
This work was supported by The Ministry of Education, Youth and Sports from the Large Infrastructures for Research, Experimental Development and Innovations project ``IT4Innovations National Supercomputing Center -- LM2015070''. 
The software used in this work was in part developed by the DOE NNSA-ASC OASCR Flash Center at the University of Chicago.
We particularly thank the Max-Planck Data \& Computing Facility for being able to carry out the scaling tests presented in this paper on the HPC cluster COBRA.
% }}}

%%%%%%%%%%%%%%%%%%%%%%%%%%%%
\section*{Data availability}
%%%%%%%%%%%%%%%%%%%%%%%%%%%%

The {\sc TreeRay} module source code and the simulation data underlying this article will be shared on reasonable request to the corresponding author. We plan to include {\sc TreeRay} into a future version of the {\sc Flash} code.

%%%%%%%%%%%%%%%%%%%%%%%%%%
\bibliographystyle{mnras}%%
\bibliography{references}%

\begin{thebibliography}{}
\makeatletter
\relax
\def\mn@urlcharsother{\let\do\@makeother \do\$\do\&\do\#\do\^\do\_\do\%\do\~}
\def\mn@doi{\begingroup\mn@urlcharsother \@ifnextchar [ {\mn@doi@}
  {\mn@doi@[]}}
\def\mn@doi@[#1]#2{\def\@tempa{#1}\ifx\@tempa\@empty \href
  {http://dx.doi.org/#2} {doi:#2}\else \href {http://dx.doi.org/#2} {#1}\fi
  \endgroup}
\def\mn@eprint#1#2{\mn@eprint@#1:#2::\@nil}
\def\mn@eprint@arXiv#1{\href {http://arxiv.org/abs/#1} {{\tt arXiv:#1}}}
\def\mn@eprint@dblp#1{\href {http://dblp.uni-trier.de/rec/bibtex/#1.xml}
  {dblp:#1}}
\def\mn@eprint@#1:#2:#3:#4\@nil{\def\@tempa {#1}\def\@tempb {#2}\def\@tempc
  {#3}\ifx \@tempc \@empty \let \@tempc \@tempb \let \@tempb \@tempa \fi \ifx
  \@tempb \@empty \def\@tempb {arXiv}\fi \@ifundefined
  {mn@eprint@\@tempb}{\@tempb:\@tempc}{\expandafter \expandafter \csname
  mn@eprint@\@tempb\endcsname \expandafter{\@tempc}}}

\bibitem[\protect\citeauthoryear{{Agertz}, {Kravtsov}, {Leitner}  \&
  {Gnedin}}{{Agertz} et~al.}{2013}]{Agertz2013}
{Agertz} O.,  {Kravtsov} A.~V.,  {Leitner} S.~N.,   {Gnedin} N.~Y.,  2013,
  \mn@doi [\apj] {10.1088/0004-637X/770/1/25}, \href
  {http://adsabs.harvard.edu/abs/2013ApJ...770...25A} {770, 25}

\bibitem[\protect\citeauthoryear{{Altay} \& {Theuns}}{{Altay} \&
  {Theuns}}{2013}]{Altay2013}
{Altay} G.,  {Theuns} T.,  2013, \mn@doi [\mnras] {10.1093/mnras/stt1067},
  \href {http://adsabs.harvard.edu/abs/2013MNRAS.434..748A} {434, 748}

\bibitem[\protect\citeauthoryear{{Baczynski}, {Glover}  \&
  {Klessen}}{{Baczynski} et~al.}{2015}]{Baczynski2015}
{Baczynski} C.,  {Glover} S.~C.~O.,   {Klessen} R.~S.,  2015, \mn@doi [\mnras]
  {10.1093/mnras/stv1906}, \href
  {https://ui.adsabs.harvard.edu/abs/2015MNRAS.454..380B} {454, 380}

\bibitem[\protect\citeauthoryear{{Barnes} \& {Hut}}{{Barnes} \&
  {Hut}}{1986}]{Barnes1986}
{Barnes} J.,  {Hut} P.,  1986, \mn@doi [\nat] {10.1038/324446a0}, \href
  {http://adsabs.harvard.edu/abs/1986Natur.324..446B} {324, 446}

\bibitem[\protect\citeauthoryear{{Bertoldi}}{{Bertoldi}}{1989}]{Bertoldi1989}
{Bertoldi} F.,  1989, \mn@doi [\apj] {10.1086/168055}, \href
  {http://adsabs.harvard.edu/abs/1989ApJ...346..735B} {346, 735}

\bibitem[\protect\citeauthoryear{{Bisbas}, {W{\"u}nsch}, {Whitworth}  \&
  {Hubber}}{{Bisbas} et~al.}{2009}]{Bisbas2009}
{Bisbas} T.~G.,  {W{\"u}nsch} R.,  {Whitworth} A.~P.,   {Hubber} D.~A.,  2009,
  \mn@doi [\aap] {10.1051/0004-6361/200811522}, \href
  {http://adsabs.harvard.edu/abs/2009A%26A...497..649B} {497, 649}

\bibitem[\protect\citeauthoryear{{Bisbas}, {W{\"u}nsch}, {Whitworth}, {Hubber}
  \& {Walch}}{{Bisbas} et~al.}{2011}]{Bisbas2011}
{Bisbas} T.~G.,  {W{\"u}nsch} R.,  {Whitworth} A.~P.,  {Hubber} D.~A.,
  {Walch} S.,  2011, \mn@doi [\apj] {10.1088/0004-637X/736/2/142}, \href
  {http://adsabs.harvard.edu/abs/2011ApJ...736..142B} {736, 142}

\bibitem[\protect\citeauthoryear{{Bisbas}, {Haworth}, {Williams}, {Mackey},
  {Tremblin}, {Raga}, {Arthur}  \& {and 12 co-authors}}{{Bisbas}
  et~al.}{2015}]{Bisbas2015}
{Bisbas} T.~G.,  {Haworth} T.~J.,  {Williams} R.~J.~R.,  {Mackey} J.,
  {Tremblin} P.,  {Raga} A.~C.,  {Arthur} S.~J.,   {and 12 co-authors} 2015,
  \mn@doi [\mnras] {10.1093/mnras/stv1659}, \href
  {http://adsabs.harvard.edu/abs/2015MNRAS.453.1324B} {453, 1324}

\bibitem[\protect\citeauthoryear{{Clark}, {Glover}  \& {Klessen}}{{Clark}
  et~al.}{2012}]{Clark2012}
{Clark} P.~C.,  {Glover} S.~C.~O.,   {Klessen} R.~S.,  2012, \mn@doi [\mnras]
  {10.1111/j.1365-2966.2011.20087.x}, \href
  {http://adsabs.harvard.edu/abs/2012MNRAS.420..745C} {420, 745}

\bibitem[\protect\citeauthoryear{{Dale}, {Ercolano}  \& {Bonnell}}{{Dale}
  et~al.}{2012}]{Dale2012}
{Dale} J.~E.,  {Ercolano} B.,   {Bonnell} I.~A.,  2012, \mn@doi [\mnras]
  {10.1111/j.1365-2966.2012.21205.x}, \href
  {http://adsabs.harvard.edu/abs/2012MNRAS.424..377D} {424, 377}

\bibitem[\protect\citeauthoryear{{Deharveng} et~al.,}{{Deharveng}
  et~al.}{2010}]{Deharveng2010}
{Deharveng} L.,  et~al., 2010, \mn@doi [\aap] {10.1051/0004-6361/201014422},
  \href {http://adsabs.harvard.edu/abs/2010A%26A...523A...6D} {523, A6+}

\bibitem[\protect\citeauthoryear{{Draine}}{{Draine}}{1978}]{Draine1978}
{Draine} B.~T.,  1978, \mn@doi [\apjs] {10.1086/190513}, \href
  {http://adsabs.harvard.edu/abs/1978ApJS...36..595D} {36, 595}

\bibitem[\protect\citeauthoryear{{Elmegreen} \& {Lada}}{{Elmegreen} \&
  {Lada}}{1977}]{Elmegreen1977}
{Elmegreen} B.~G.,  {Lada} C.~J.,  1977, \mn@doi [\apj] {10.1086/155302}, \href
  {http://adsabs.harvard.edu/abs/1977ApJ...214..725E} {214, 725}

\bibitem[\protect\citeauthoryear{{Federrath}, {Banerjee}, {Clark}  \&
  {Klessen}}{{Federrath} et~al.}{2010}]{Federrath2010}
{Federrath} C.,  {Banerjee} R.,  {Clark} P.~C.,   {Klessen} R.~S.,  2010,
  \mn@doi [\apj] {10.1088/0004-637X/713/1/269}, \href
  {http://adsabs.harvard.edu/abs/2010ApJ...713..269F} {713, 269}

\bibitem[\protect\citeauthoryear{{Fryxell} et~al.,}{{Fryxell}
  et~al.}{2000}]{Fryxell2000}
{Fryxell} B.,  et~al., 2000, \mn@doi [\apjs] {10.1086/317361}, \href
  {http://adsabs.harvard.edu/abs/2000ApJS..131..273F} {131, 273}

\bibitem[\protect\citeauthoryear{{Gatto} et~al.,}{{Gatto}
  et~al.}{2017}]{Gatto2017}
{Gatto} A.,  et~al., 2017, \mn@doi [\mnras] {10.1093/mnras/stw3209}, \href
  {http://adsabs.harvard.edu/abs/2017MNRAS.466.1903G} {466, 1903}

\bibitem[\protect\citeauthoryear{{Gendelev} \& {Krumholz}}{{Gendelev} \&
  {Krumholz}}{2012}]{GendelevKrumholz2012}
{Gendelev} L.,  {Krumholz} M.~R.,  2012, \mn@doi [\apj]
  {10.1088/0004-637X/745/2/158}, \href
  {http://adsabs.harvard.edu/abs/2012ApJ...745..158G} {745, 158}

\bibitem[\protect\citeauthoryear{{Getman}, {Feigelson}, {Sicilia-Aguilar},
  {Broos}, {Kuhn}  \& {Garmire}}{{Getman} et~al.}{2012}]{Getman2012}
{Getman} K.~V.,  {Feigelson} E.~D.,  {Sicilia-Aguilar} A.,  {Broos} P.~S.,
  {Kuhn} M.~A.,   {Garmire} G.~P.,  2012, \mn@doi [\mnras]
  {10.1111/j.1365-2966.2012.21879.x}, \href
  {http://adsabs.harvard.edu/abs/2012MNRAS.426.2917G} {426, 2917}

\bibitem[\protect\citeauthoryear{{Girichidis} et~al.,}{{Girichidis}
  et~al.}{2016}]{Girichidis2016}
{Girichidis} P.,  et~al., 2016, \mn@doi [\mnras] {10.1093/mnras/stv2742}, \href
  {http://adsabs.harvard.edu/abs/2016MNRAS.456.3432G} {456, 3432}

\bibitem[\protect\citeauthoryear{{Glover}, {Federrath}, {Mac Low}  \&
  {Klessen}}{{Glover} et~al.}{2010}]{Glover2010}
{Glover} S.~C.~O.,  {Federrath} C.,  {Mac Low} M.-M.,   {Klessen} R.~S.,  2010,
  \mn@doi [\mnras] {10.1111/j.1365-2966.2009.15718.x}, \href
  {http://adsabs.harvard.edu/abs/2010MNRAS.404....2G} {404, 2}

\bibitem[\protect\citeauthoryear{{G{\'o}rski}, {Hivon}, {Banday}, {Wandelt},
  {Hansen}, {Reinecke}  \& {Bartelmann}}{{G{\'o}rski}
  et~al.}{2005}]{Gorski2005}
{G{\'o}rski} K.~M.,  {Hivon} E.,  {Banday} A.~J.,  {Wandelt} B.~D.,  {Hansen}
  F.~K.,  {Reinecke} M.,   {Bartelmann} M.,  2005, \mn@doi [\apj]
  {10.1086/427976}, \href {http://adsabs.harvard.edu/abs/2005ApJ...622..759G}
  {622, 759}

\bibitem[\protect\citeauthoryear{{Gritschneder}, {Naab}, {Burkert}, {Walch},
  {Heitsch}  \& {Wetzstein}}{{Gritschneder} et~al.}{2009}]{Gritschneder2009a}
{Gritschneder} M.,  {Naab} T.,  {Burkert} A.,  {Walch} S.,  {Heitsch} F.,
  {Wetzstein} M.,  2009, \mn@doi [\mnras] {10.1111/j.1365-2966.2008.14185.x},
  \href {http://adsabs.harvard.edu/abs/2009MNRAS.393...21G} {393, 21}

\bibitem[\protect\citeauthoryear{{Gritschneder}, {Burkert}, {Naab}  \&
  {Walch}}{{Gritschneder} et~al.}{2010}]{Gritschneder2010}
{Gritschneder} M.,  {Burkert} A.,  {Naab} T.,   {Walch} S.,  2010, \mn@doi
  [\apj] {10.1088/0004-637X/723/2/971}, \href
  {http://adsabs.harvard.edu/abs/2010ApJ...723..971G} {723, 971}

\bibitem[\protect\citeauthoryear{{Grond}, {Woods}, {Wadsley}  \&
  {Couchman}}{{Grond} et~al.}{2019}]{Grond2019TREVR}
{Grond} J.~J.,  {Woods} R.~M.,  {Wadsley} J.~W.,   {Couchman} H.~M.~P.,  2019,
  \mn@doi [\mnras] {10.1093/mnras/stz525}, \href
  {https://ui.adsabs.harvard.edu/abs/2019MNRAS.485.3681G} {485, 3681}

\bibitem[\protect\citeauthoryear{{Habing}}{{Habing}}{1968}]{Habing1968}
{Habing} H.~J.,  1968, \bain, \href
  {https://ui.adsabs.harvard.edu/abs/1968BAN....19..421H} {19, 421}

\bibitem[\protect\citeauthoryear{{Haid}, {Walch}, {Seifried}, {W{\"u}nsch},
  {Dinnbier}  \& {Naab}}{{Haid} et~al.}{2018}]{Haid2018}
{Haid} S.,  {Walch} S.,  {Seifried} D.,  {W{\"u}nsch} R.,  {Dinnbier} F.,
  {Naab} T.,  2018, \mn@doi [\mnras] {10.1093/mnras/sty1315}, \href
  {http://adsabs.harvard.edu/abs/2018MNRAS.478.4799H} {478, 4799}

\bibitem[\protect\citeauthoryear{{Haid}, {Walch}, {Seifried}, {W{\"u}nsch},
  {Dinnbier}  \& {Naab}}{{Haid} et~al.}{2019}]{Haid2019}
{Haid} S.,  {Walch} S.,  {Seifried} D.,  {W{\"u}nsch} R.,  {Dinnbier} F.,
  {Naab} T.,  2019, \mn@doi [\mnras] {10.1093/mnras/sty2938}, \href
  {https://ui.adsabs.harvard.edu/abs/2019MNRAS.482.4062H} {482, 4062}

\bibitem[\protect\citeauthoryear{{Hillenbrand}, {Massey}, {Strom}  \&
  {Merrill}}{{Hillenbrand} et~al.}{1993}]{Hillenbrand1993}
{Hillenbrand} L.~A.,  {Massey} P.,  {Strom} S.~E.,   {Merrill} K.~M.,  1993,
  \mn@doi [\aj] {10.1086/116774}, \href
  {https://ui.adsabs.harvard.edu/abs/1993AJ....106.1906H} {106, 1906}

\bibitem[\protect\citeauthoryear{{Iliev} \& {et al.}}{{Iliev} \& {et
  al.}}{2006}]{Iliev2006}
{Iliev} I.~T.,  {et al.} 2006, \mn@doi [\mnras]
  {10.1111/j.1365-2966.2006.10775.x}, \href
  {http://adsabs.harvard.edu/abs/2006MNRAS.371.1057I} {371, 1057}

\bibitem[\protect\citeauthoryear{{Iliev} et~al.,}{{Iliev}
  et~al.}{2009}]{Iliev2009}
{Iliev} I.~T.,  et~al., 2009, \mn@doi [\mnras]
  {10.1111/j.1365-2966.2009.15558.x}, \href
  {http://adsabs.harvard.edu/abs/2009MNRAS.400.1283I} {400, 1283}

\bibitem[\protect\citeauthoryear{{Kahn}}{{Kahn}}{1954}]{Kahn1954}
{Kahn} F.~D.,  1954, \bain, \href
  {http://adsabs.harvard.edu/abs/1954BAN....12..187K} {12, 187}

\bibitem[\protect\citeauthoryear{{Kannan}, {Vogelsberger}, {Marinacci},
  {McKinnon}, {Pakmor}  \& {Springel}}{{Kannan} et~al.}{2019}]{Kannan2019}
{Kannan} R.,  {Vogelsberger} M.,  {Marinacci} F.,  {McKinnon} R.,  {Pakmor} R.,
    {Springel} V.,  2019, \mn@doi [\mnras] {10.1093/mnras/stz287}, \href
  {https://ui.adsabs.harvard.edu/abs/2019MNRAS.485..117K} {485, 117}

\bibitem[\protect\citeauthoryear{{Kessel-Deynet} \& {Burkert}}{{Kessel-Deynet}
  \& {Burkert}}{2003a}]{Kessel2003}
{Kessel-Deynet} O.,  {Burkert} A.,  2003a, \mn@doi [\mnras]
  {10.1046/j.1365-8711.2003.05737.x}, \href
  {http://adsabs.harvard.edu/abs/2003MNRAS.338..545K} {338, 545}

\bibitem[\protect\citeauthoryear{{Kessel-Deynet} \& {Burkert}}{{Kessel-Deynet}
  \& {Burkert}}{2003b}]{KesselDeynetBurkert2003}
{Kessel-Deynet} O.,  {Burkert} A.,  2003b, \mn@doi [\mnras]
  {10.1046/j.1365-8711.2003.05737.x}, \href
  {http://adsabs.harvard.edu/abs/2003MNRAS.338..545K} {338, 545}

\bibitem[\protect\citeauthoryear{{Kim} \& {Ostriker}}{{Kim} \&
  {Ostriker}}{2017}]{KO2017}
{Kim} C.-G.,  {Ostriker} E.~C.,  2017, \mn@doi [\apj]
  {10.3847/1538-4357/aa8599}, \href
  {https://ui.adsabs.harvard.edu/abs/2017ApJ...846..133K} {846, 133}

\bibitem[\protect\citeauthoryear{{Kim}, {Kim}, {Ostriker}  \& {Skinner}}{{Kim}
  et~al.}{2017}]{Kim2017}
{Kim} J.-G.,  {Kim} W.-T.,  {Ostriker} E.~C.,   {Skinner} M.~A.,  2017, \mn@doi
  [\apj] {10.3847/1538-4357/aa9b80}, \href
  {https://ui.adsabs.harvard.edu/abs/2017ApJ...851...93K} {851, 93}

\bibitem[\protect\citeauthoryear{{Klassen}, {Kuiper}, {Pudritz}, {Peters},
  {Banerjee}  \& {Buntemeyer}}{{Klassen} et~al.}{2014}]{Klassen2014}
{Klassen} M.,  {Kuiper} R.,  {Pudritz} R.~E.,  {Peters} T.,  {Banerjee} R.,
  {Buntemeyer} L.,  2014, \mn@doi [\apj] {10.1088/0004-637X/797/1/4}, \href
  {http://adsabs.harvard.edu/abs/2014ApJ...797....4K} {797, 4}

\bibitem[\protect\citeauthoryear{{Krumholz}, {Klein}  \& {McKee}}{{Krumholz}
  et~al.}{2007}]{Krumholz2007}
{Krumholz} M.~R.,  {Klein} R.~I.,   {McKee} C.~F.,  2007, \mn@doi [\apj]
  {10.1086/510664}, \href {http://adsabs.harvard.edu/abs/2007ApJ...656..959K}
  {656, 959}

\bibitem[\protect\citeauthoryear{{Kuiper}, {Klahr}, {Dullemond}, {Kley}  \&
  {Henning}}{{Kuiper} et~al.}{2010}]{Kuiper2010}
{Kuiper} R.,  {Klahr} H.,  {Dullemond} C.,  {Kley} W.,   {Henning} T.,  2010,
  \mn@doi [\aap] {10.1051/0004-6361/200912355}, \href
  {http://adsabs.harvard.edu/abs/2010A%26A...511A..81K} {511, A81}

\bibitem[\protect\citeauthoryear{{Lefloch} \& {Lazareff}}{{Lefloch} \&
  {Lazareff}}{1994}]{LeflochLazareff1994}
{Lefloch} B.,  {Lazareff} B.,  1994, \aap, \href
  {http://adsabs.harvard.edu/abs/1994A%26A...289..559L} {289, 559}

\bibitem[\protect\citeauthoryear{{Lefloch} \& {Lazareff}}{{Lefloch} \&
  {Lazareff}}{1995}]{LeflochLazareff1995}
{Lefloch} B.,  {Lazareff} B.,  1995, \aap, \href
  {http://adsabs.harvard.edu/abs/1995A%26A...301..522L} {301, 522}

\bibitem[\protect\citeauthoryear{{Mackey} \& {Lim}}{{Mackey} \&
  {Lim}}{2010a}]{Mackey2010}
{Mackey} J.,  {Lim} A.~J.,  2010a, \mn@doi [\mnras]
  {10.1111/j.1365-2966.2009.16181.x}, \href
  {http://adsabs.harvard.edu/abs/2010MNRAS.403..714M} {403, 714}

\bibitem[\protect\citeauthoryear{{Mackey} \& {Lim}}{{Mackey} \&
  {Lim}}{2010b}]{MackeyLim2010}
{Mackey} J.,  {Lim} A.~J.,  2010b, \mn@doi [\mnras]
  {10.1111/j.1365-2966.2009.16181.x}, \href
  {http://adsabs.harvard.edu/abs/2010MNRAS.403..714M} {403, 714}

\bibitem[\protect\citeauthoryear{{Mackey} \& {Lim}}{{Mackey} \&
  {Lim}}{2011}]{Mackey2011}
{Mackey} J.,  {Lim} A.~J.,  2011, \mn@doi [\mnras]
  {10.1111/j.1365-2966.2010.18043.x}, \href
  {http://adsabs.harvard.edu/abs/2011MNRAS.412.2079M} {412, 2079}

\bibitem[\protect\citeauthoryear{{Mackey}, {Walch}, {Seifried}, {Glover},
  {W{\"u}nsch}  \& {Aharonian}}{{Mackey} et~al.}{2019}]{Mackey2019}
{Mackey} J.,  {Walch} S.,  {Seifried} D.,  {Glover} S. C.~O.,  {W{\"u}nsch} R.,
    {Aharonian} F.,  2019, \mn@doi [\mnras] {10.1093/mnras/stz902}, \href
  {https://ui.adsabs.harvard.edu/abs/2019MNRAS.486.1094M} {486, 1094}

\bibitem[\protect\citeauthoryear{{McLeod}, {Dale}, {Ginsburg}, {Ercolano},
  {Gritschneder}, {Ramsay}  \& {Testi}}{{McLeod} et~al.}{2015}]{McLeod2015}
{McLeod} A.~F.,  {Dale} J.~E.,  {Ginsburg} A.,  {Ercolano} B.,  {Gritschneder}
  M.,  {Ramsay} S.,   {Testi} L.,  2015, \mn@doi [\mnras]
  {10.1093/mnras/stv680}, \href
  {https://ui.adsabs.harvard.edu/abs/2015MNRAS.450.1057M} {450, 1057}

\bibitem[\protect\citeauthoryear{{Mellema}, {Raga}, {Canto}, {Lundqvist},
  {Balick}, {Steffen}  \& {Noriega-Crespo}}{{Mellema}
  et~al.}{1998}]{Mellema1998}
{Mellema} G.,  {Raga} A.~C.,  {Canto} J.,  {Lundqvist} P.,  {Balick} B.,
  {Steffen} W.,   {Noriega-Crespo} A.,  1998, \aap, \href
  {http://adsabs.harvard.edu/abs/1998A%26A...331..335M} {331, 335}

\bibitem[\protect\citeauthoryear{{Mellema}, {Arthur}, {Henney}, {Iliev}  \&
  {Shapiro}}{{Mellema} et~al.}{2006}]{Mellema2006}
{Mellema} G.,  {Arthur} S.~J.,  {Henney} W.~J.,  {Iliev} I.~T.,   {Shapiro}
  P.~R.,  2006, \mn@doi [\apj] {10.1086/505294}, \href
  {http://adsabs.harvard.edu/abs/2006ApJ...647..397M} {647, 397}

\bibitem[\protect\citeauthoryear{{Miao}, {White}, {Thompson}  \&
  {Nelson}}{{Miao} et~al.}{2009}]{Miao2009}
{Miao} J.,  {White} G.~J.,  {Thompson} M.~A.,   {Nelson} R.~P.,  2009, \mn@doi
  [\apj] {10.1088/0004-637X/692/1/382}, \href
  {http://adsabs.harvard.edu/abs/2009ApJ...692..382M} {692, 382}

\bibitem[\protect\citeauthoryear{{Nelson} \& {Langer}}{{Nelson} \&
  {Langer}}{1997}]{Nelson1997}
{Nelson} R.~P.,  {Langer} W.~D.,  1997, \mn@doi [\apj] {10.1086/304167}, \href
  {http://adsabs.harvard.edu/abs/1997ApJ...482..796N} {482, 796}

\bibitem[\protect\citeauthoryear{{Oort} \& {Spitzer}}{{Oort} \&
  {Spitzer}}{1955}]{OortSpitzer1955}
{Oort} J.~H.,  {Spitzer} Jr. L.,  1955, \mn@doi [\apj] {10.1086/145958}, \href
  {http://adsabs.harvard.edu/abs/1955ApJ...121....6O} {121, 6}

\bibitem[\protect\citeauthoryear{{Osterbrock}}{{Osterbrock}}{1988}]{Oesterbrock1988}
{Osterbrock} D.~E.,  1988, \mn@doi [\pasp] {10.1086/132188}, \href
  {http://adsabs.harvard.edu/abs/1988PASP..100..412O} {100, 412}

\bibitem[\protect\citeauthoryear{{Paardekooper}, {Kruip}  \&
  {Icke}}{{Paardekooper} et~al.}{2010}]{Paardekooper2010}
{Paardekooper} J.~P.,  {Kruip} C.~J.~H.,   {Icke} V.,  2010, \mn@doi [\aap]
  {10.1051/0004-6361/200913821}, \href
  {https://ui.adsabs.harvard.edu/abs/2010A&A...515A..79P} {515, A79}

\bibitem[\protect\citeauthoryear{{Peters} et~al.,}{{Peters}
  et~al.}{2017}]{Peters2017}
{Peters} T.,  et~al., 2017, \mn@doi [\mnras] {10.1093/mnras/stw3216}, \href
  {https://ui.adsabs.harvard.edu/abs/2017MNRAS.466.3293P} {466, 3293}

\bibitem[\protect\citeauthoryear{{Petkova} \& {Maio}}{{Petkova} \&
  {Maio}}{2012}]{Petkova2012}
{Petkova} M.,  {Maio} U.,  2012, \mn@doi [\mnras]
  {10.1111/j.1365-2966.2012.20823.x}, \href
  {http://adsabs.harvard.edu/abs/2012MNRAS.422.3067P} {422, 3067}

\bibitem[\protect\citeauthoryear{{Quirk}}{{Quirk}}{1994}]{Quirk1994}
{Quirk} J.~J.,  1994, \mn@doi [International Journal for Numerical Methods in
  Fluids] {10.1002/fld.1650180603}, \href
  {http://adsabs.harvard.edu/abs/1994IJNMF..18..555Q} {18, 555}

\bibitem[\protect\citeauthoryear{{Raga}, {Cant{\'o}}  \&
  {Rodr{\'{\i}}guez}}{{Raga} et~al.}{2012}]{Raga2012}
{Raga} A.~C.,  {Cant{\'o}} J.,   {Rodr{\'{\i}}guez} L.~F.,  2012, \mn@doi
  [\mnras] {10.1111/j.1745-3933.2011.01173.x}, \href
  {http://adsabs.harvard.edu/abs/2012MNRAS.419L..39R} {419, L39}

\bibitem[\protect\citeauthoryear{{Raskutti}, {Ostriker}  \&
  {Skinner}}{{Raskutti} et~al.}{2017}]{Raskutti2017}
{Raskutti} S.,  {Ostriker} E.~C.,   {Skinner} M.~A.,  2017, \mn@doi [\apj]
  {10.3847/1538-4357/aa965e}, \href
  {https://ui.adsabs.harvard.edu/abs/2017ApJ...850..112R} {850, 112}

\bibitem[\protect\citeauthoryear{{Rijkhorst}, {Plewa}, {Dubey}  \&
  {Mellema}}{{Rijkhorst} et~al.}{2006}]{Rijkhorst2006}
{Rijkhorst} E.-J.,  {Plewa} T.,  {Dubey} A.,   {Mellema} G.,  2006, \mn@doi
  [\aap] {10.1051/0004-6361:20053401}, \href
  {http://adsabs.harvard.edu/abs/2006A%26A...452..907R} {452, 907}

\bibitem[\protect\citeauthoryear{{Rosdahl}, {Blaizot}, {Aubert}, {Stranex}  \&
  {Teyssier}}{{Rosdahl} et~al.}{2013}]{Rosdahl2013}
{Rosdahl} J.,  {Blaizot} J.,  {Aubert} D.,  {Stranex} T.,   {Teyssier} R.,
  2013, \mn@doi [\mnras] {10.1093/mnras/stt1722}, \href
  {https://ui.adsabs.harvard.edu/abs/2013MNRAS.436.2188R} {436, 2188}

\bibitem[\protect\citeauthoryear{{Rosen}, {Krumholz}, {Oishi}, {Lee}  \&
  {Klein}}{{Rosen} et~al.}{2017}]{Rosen2017}
{Rosen} A.~L.,  {Krumholz} M.~R.,  {Oishi} J.~S.,  {Lee} A.~T.,   {Klein}
  R.~I.,  2017, \mn@doi [Journal of Computational Physics]
  {10.1016/j.jcp.2016.10.048}, \href
  {https://ui.adsabs.harvard.edu/abs/2017JCoPh.330..924R} {330, 924}

\bibitem[\protect\citeauthoryear{{Schneider} et~al.,}{{Schneider}
  et~al.}{2012}]{Schneider2012}
{Schneider} N.,  et~al., 2012, \mn@doi [\aap] {10.1051/0004-6361/201218917},
  \href {http://adsabs.harvard.edu/abs/2012A\%26A...542L..18S} {542, L18}

\bibitem[\protect\citeauthoryear{{Skinner} \& {Ostriker}}{{Skinner} \&
  {Ostriker}}{2013}]{Skinner2013}
{Skinner} M.~A.,  {Ostriker} E.~C.,  2013, \mn@doi [\apjs]
  {10.1088/0067-0049/206/2/21}, \href
  {https://ui.adsabs.harvard.edu/abs/2013ApJS..206...21S} {206, 21}

\bibitem[\protect\citeauthoryear{{Spitzer}}{{Spitzer}}{1978}]{Spitzer1978}
{Spitzer} L.,  1978, {Physical processes in the interstellar medium}.
New York Wiley-Interscience, 1978.~333 p.

\bibitem[\protect\citeauthoryear{{Str{\"o}mgren}}{{Str{\"o}mgren}}{1939}]{Stromgren1939}
{Str{\"o}mgren} B.,  1939, \mn@doi [\apj] {10.1086/144074}, \href
  {http://adsabs.harvard.edu/abs/1939ApJ....89..526S} {89, 526}

\bibitem[\protect\citeauthoryear{{Tenorio-Tagle}}{{Tenorio-Tagle}}{1979}]{TenorioTagle1979}
{Tenorio-Tagle} G.,  1979, \aap, \href
  {http://adsabs.harvard.edu/abs/1979A%26A....71...59T} {71, 59}

\bibitem[\protect\citeauthoryear{{Walch}, {Whitworth}, {Bisbas}, {W{\"u}nsch}
  \& {Hubber}}{{Walch} et~al.}{2012}]{Walch2012a}
{Walch} S.~K.,  {Whitworth} A.~P.,  {Bisbas} T.,  {W{\"u}nsch} R.,   {Hubber}
  D.,  2012, \mn@doi [\mnras] {10.1111/j.1365-2966.2012.21767.x}, \href
  {http://adsabs.harvard.edu/abs/2012MNRAS.427..625W} {427, 625}

\bibitem[\protect\citeauthoryear{{Walch}, {Whitworth}, {Bisbas}, {W{\"u}nsch}
  \& {Hubber}}{{Walch} et~al.}{2013}]{Walch2013}
{Walch} S.,  {Whitworth} A.~P.,  {Bisbas} T.~G.,  {W{\"u}nsch} R.,   {Hubber}
  D.~A.,  2013, \mn@doi [\mnras] {10.1093/mnras/stt1115}, \href
  {http://adsabs.harvard.edu/abs/2013MNRAS.435..917W} {435, 917}

\bibitem[\protect\citeauthoryear{{Walch} et~al.,}{{Walch}
  et~al.}{2015}]{Walch2015}
{Walch} S.,  et~al., 2015, \mn@doi [\mnras] {10.1093/mnras/stv1975}, \href
  {http://adsabs.harvard.edu/abs/2015MNRAS.454..238W} {454, 238}

\bibitem[\protect\citeauthoryear{{Whitworth}}{{Whitworth}}{1979}]{Whitworth1979}
{Whitworth} A.,  1979, \mn@doi [\mnras] {10.1093/mnras/186.1.59}, \href
  {http://adsabs.harvard.edu/abs/1979MNRAS.186...59W} {186, 59}

\bibitem[\protect\citeauthoryear{{Wise} \& {Abel}}{{Wise} \&
  {Abel}}{2011}]{Wise2011}
{Wise} J.~H.,  {Abel} T.,  2011, \mn@doi [\mnras]
  {10.1111/j.1365-2966.2011.18646.x}, \href
  {http://adsabs.harvard.edu/abs/2011MNRAS.414.3458W} {414, 3458}

\bibitem[\protect\citeauthoryear{{W{\"u}nsch}, {Walch}, {Dinnbier}  \&
  {Whitworth}}{{W{\"u}nsch} et~al.}{2018}]{Wunsch2018}
{W{\"u}nsch} R.,  {Walch} S.,  {Dinnbier} F.,   {Whitworth} A.,  2018, \mn@doi
  [\mnras] {10.1093/mnras/sty015}, \href
  {https://ui.adsabs.harvard.edu/abs/2018MNRAS.475.3393W} {475, 3393}

\makeatother
\end{thebibliography}
%%%%%%%%%%%%%%%%%%%%%%%%%%
\clearpage
\appendix
\onecolumn

\section{Kernels for radial mapping of nodes onto rays}
\label{app:kernels}

In general, any kernel depends on three tree parameters: the distance from the node mass centre to the target point, $d$, the node linear size, $h_n$, and the distance from the $i$-th evaluation point on the ray to the target point, $r_i$. It is convenient to combine them into a single parameter
\begin{equation}
\label{eKern:all}
x \equiv (r_i-d)/h_n \ .
\end{equation}
Kernel evaluation is a performance-critical operation because it is carried out during the tree walk for each target point, each tree node and each intersecting ray. Therefore, it is useful to tabulate the kernels. This is easy to do, because the parameters $r_i$ and $h_n$ can have only a small number of discrete values. The possible values of $r_i$ are given by equation~(\ref{eq:etaR}); the possible values of $h_n$ are given by the AMR octal tree and their number is the binary logarithm of the ratio of the computational domain size to the size of the smallest grid cell. Parameter $d$ can have an arbitrary value between $0$ and the 3D diagonal of the computational domain, and we sample it similarly to $r_i$ but with ten times more points. The kernels are normalised so that, for each tabulated combination of $d$ and $h_n$, the sum of the values at all $r_i$ is unity .

\begin{figure*}
    \includegraphics[width=0.45\textwidth]{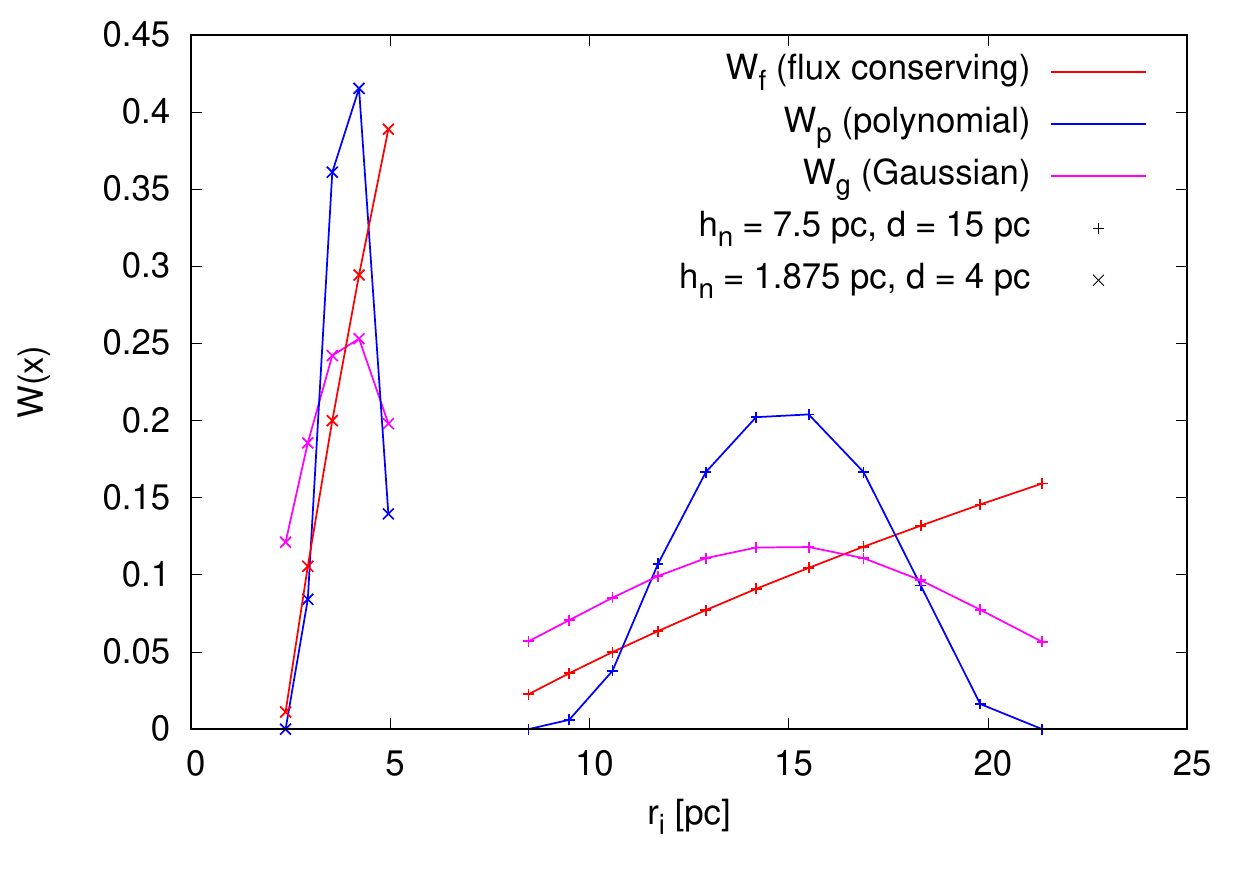}
    \includegraphics[width=0.45\textwidth]{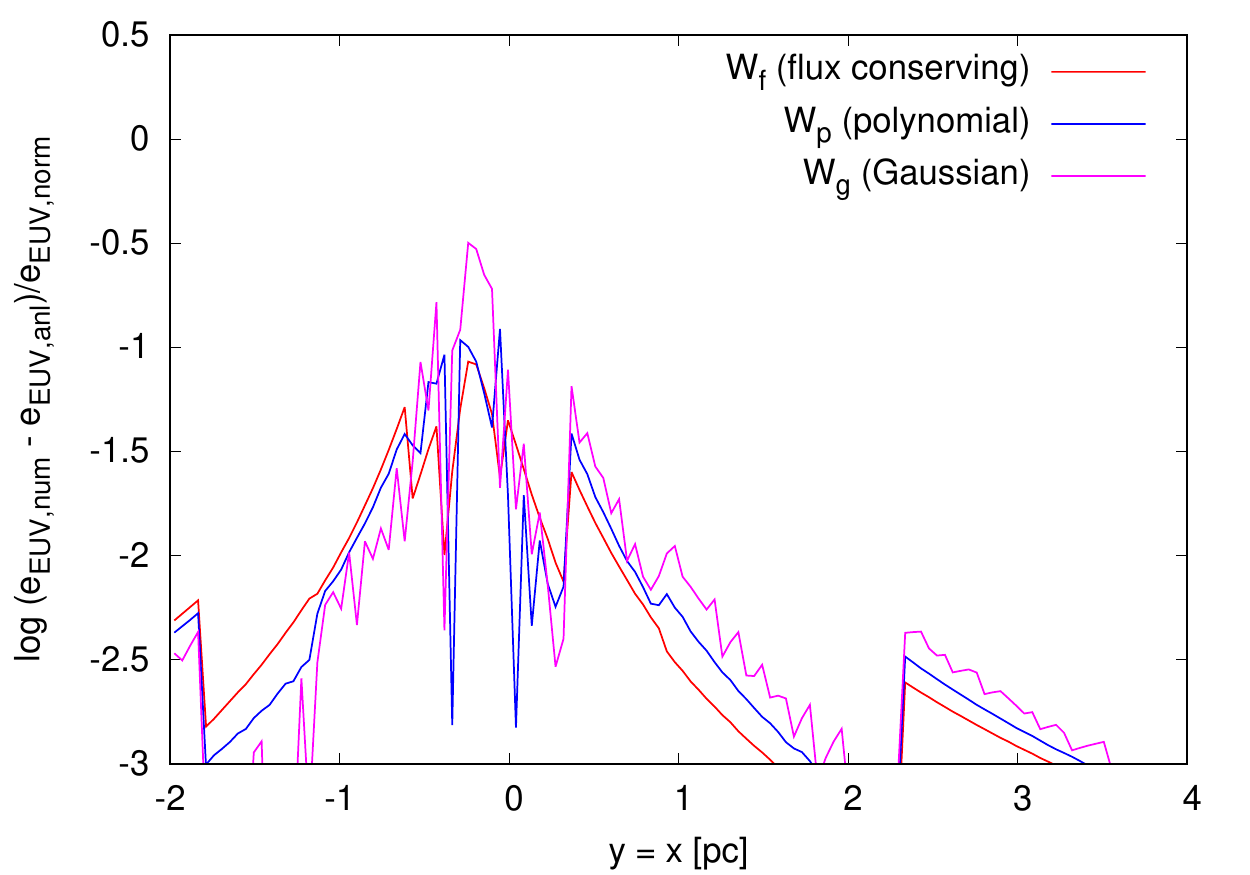}
    \caption{\textbf{Left:} the three implemented kernels $W_g$ (magenta), $W_p$ (blue), and $W_f$ (red) plotted as functions of $r_i$ for two combinations of $d$ and $h_n$: $d = 4$\,pc, $h_n = 1.875$\,pc ($\times$ symbols), and $d = 15$\,pc, $h_n = 7.5$\,pc ($+$ symbols). The positions of the evaluation points $r_i$ are calculated from equation (\ref{eq:etaR}) with $\eta_R = 2$ and $\Delta x = 0.23$\,pc corresponding to model~(a) of the Spitzer test (Section~\ref{sec:spitzer}). \textbf{Right:} the error in the radiation energy density along the line $y=x, z=0$ in the test with two radiation sources as calculated using the three kernels $W_f$, $W_p$, and $W_g$. The meaning of colour is the same as in the left panel.}
    \label{fig:kernels}
\end{figure*}

\begin{figure*}
\includegraphics[width=0.88\textwidth]{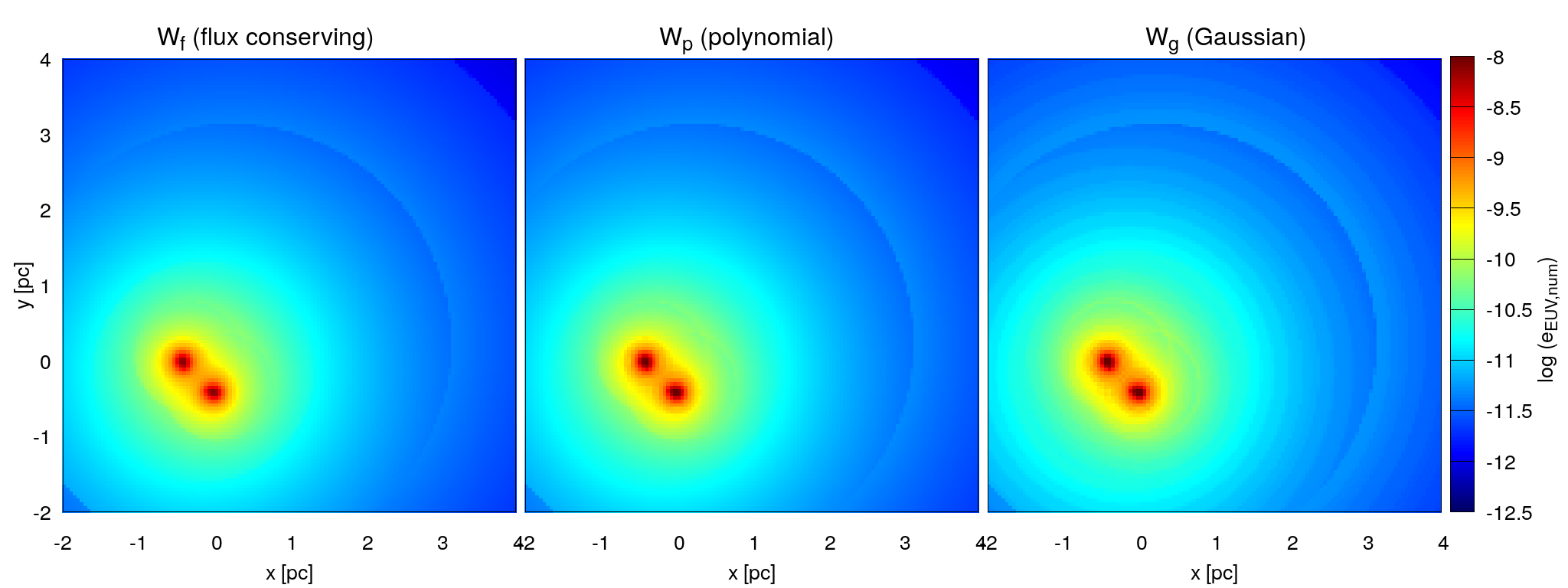}
\caption{The radiation energy density $e_{_{\rm EUV,num}}$ calculated using the three kernels $W_{f}$ (left panel), $W_{p}$ (middle panel) and $W_{g}$ (righr panel).}
\label{fig:kern:euve}
\end{figure*}

The left panel of Fig.~\ref{fig:kernels} shows the three implemented kernels, $W_g$ (Gaussian), $W_p$ (piece-wise polynomial) and $W_f$ (flux conserving), as functions of $r_i$ for two combinations of $d$ and $h_n$. The right panel of this figure displays the error in the radiation energy density when these kernels are used to calculate a model similar to the two-source model described in Section~\ref{sec:2src}. However, here the dense cloud is missing, i.e. the whole computational domain is filled with rarefied gas ($\rho = 10^{-24}$\,g\,cm$^{-3}$), and the two sources are located closer to each other (at $(x,y,z) = (-0.4,0,0)$\,pc and $(0,-0.4,0)$\,pc). The latter modification was made to increase the fraction of target cells for which the two sources are located in the same tree node. Otherwise, all the parameters are the same as for model~(b) described in Table~\ref{tab:acc:2src}. The right panel of Fig.~\ref{fig:kernels} shows the discrepancy between the numerically obtained radiation energy density, $e_{_\mathrm{EUV,num}}$, and the corresponding analytical value, $e_{_\mathrm{EUV,anl}}$, normalized by $e_{_\mathrm{EUV,anl}}$ at the point $(x,y,z) = (-0.2,-0.2,0)$\,pc, along the line of the symmetry between the sources ($y = x$, $z=0$). The corresponding radiation energy density on the $z=0$ plane for all three models is displayed in Fig.~\ref{fig:kern:euve}. We can see that the smallest error and the least pronounced numerical artifacts (visible as discontinuities in the radiation field) are obtained with the flux conserving kernel $W_{f}$. Conversely, the Gaussian kernel $W_{g}$ yields the worst results. Detailed descriptions of the kernels are given below.


\subsection{Gaussian kernel}

The Gaussian kernel is
\begin{equation}
W_g(x) = \left\{
\begin{array}{ll}
\exp(-x^2),\hspace{0.5cm} & \mbox{for }\;x < \sqrt{3}/2\,;\\
0, & \mbox{for }\;x \geq \sqrt{3}/2\,.
\end{array}
\right.
\label{eKernGaussian}
\end{equation}

\subsection{Piece-wise polynomial kernel}

Kernel $W_p$ is derived under the assumption that the emission and/or absorption coefficients -- or generally any field that the kernel represents -- are distributed uniformly within a cubic tree node. The tree node intersects with randomly oriented, randomly positioned rays (i.e. they do not have to pass through the node center) and the kernel represents the average cross-section taken over all ray positions and orientations. With this motivation, the kernel is obtained by summing up a large number of top-hat functions with extents given by the intersection of a unit cube and a randomly oriented line. The sum is then normalized and fitted with a third-order piece-wise polynomial consisting of two parts $W_{p,1}$ between $x = 0$ and $x = 1/2$ and $W_{p,2}$ between $x = 1/2$ and $x = \sqrt{3}/2$. The central value is $W_{p,1}(0) \equiv M$ and the kernel drops to zero at $\sqrt{3}/2$. We require that the kernel be continuous, and hence $W_{p,1}(1/2) = W_{p,2}(1/2) \equiv A$. Furthermore, we require that its derivative is zero at the center and at the edge, i.e. $W'_{p,1}(0) = 0$ and $W'_{p,2}(\sqrt{3}/2) = 0$. The resulting kernel has the form

\begin{equation}
W_p(x) = \left\{
\begin{array}{lcl}
a_1 x^3 - (a_1/2+4(M-A)) x^2 + M, & \mathrm{for} & x < 1/2;\\
a_2 x^3 - (a_2(1+2\sqrt{3})/2 - 2A(\sqrt{3}+2)) x^2 - &&\\
(9a_2/4-a_2(\sqrt{3}+6)+2A(2\sqrt{3}+3)) x 
+ 3\sqrt{3}a_2/4 - 3a_2(1+2\sqrt{3})/8 + 3A(\sqrt{3}+2),
& \mathrm{for} & 0.5 < x < \sqrt{3}/2;\\
0, & \mathrm{for} & x > \sqrt{3}/2
\label{eKernPolynomial}
\end{array}
\right.
\end{equation}
with
\begin{equation}
a_1 = 1.21, \quad a_2 = -6.35, \quad A = 4.08, \quad \mathrm{and}\ M = 1.184\,.
\end{equation}

\subsection{Flux conserving kernel}
\label{sec:fluxConsv}

To illustrate a limitation of the preceding kernels, 
we consider the situation in which a point source is located in a single node, as seen from a target cell, and the gas is so rarefied that the absorption is negligible. During the tree walk, individual nodes are mapped onto rays, and the location of the sources of radiation within the node is approximated by the node mass centre. The smoothing kernels have large wings spanning from $|r_i - d| - \sqrt{3}h_n/2$  to $|r_i - d| + \sqrt{3}h_n/2$, which together with the typical condition for node acceptance  $d \geq \theta_\mathrm{lim} h_n$, and with the typical value of the limiting angle $\theta_\mathrm{lim} = 0.5$, means that the source is spread out so that it contributes to evaluation points significantly closer to (and further from) the target cell than the original point source. When the emission coefficient is mapped onto a ray using a wide kernel, the source is often spread over many ray evaluation points and each of them becomes a source of radiation. It is then very unlikely that the flux at the target cell due to these multiple "sources" at the ray evaluation points is the same as the flux due to the original single source at distance $d$. Moreover, as the RTE is integrated along the ray from the source, the flux at the evaluation point $r_{i}$ is the sum of contributions from the flux coming through the evaluation point $r_{i+1}$, and from the possible source term at the evaluation point $r_{i}$. This is clearly not equal to the situation where the flux is coming from a single point source.  

To overcome this limitation, we consider a kernel which conserves the flux at the target point by construction. In the following, we neglect the ionisation rates for the sake of simplicity, so the radiation transport reduces to the situation in the absence of the absorption.  Consider a kernel which is so narrow that a single source is only mapped onto two evaluation points along a ray. Hence the emission coefficient, $\varepsilon$, of a source at distance $d$ must be divided between just two evaluation points, $r_i$ and $r_{i+1}$, such that the flux $F_0$ at $r=0$ is as close as possible to the correct value $\varepsilon/(4\pi d^2)$. We seek a numerical solution, since a general analytic solution for wide kernels with many evaluation points probably does not exist. The above condition for the emission coefficients $\varepsilon_i$ and $\varepsilon_{i+1}$, at $r_i$ and $r_{i+1}$, respectively, gives
\begin{equation}
\frac{\varepsilon}{d^2} = \frac{\varepsilon_i}{r_{i}^2} + \frac{\varepsilon_{i+1}}{r_{i+1}^2}\,.
\label{eq:kernCond0}
\end{equation}
A second condition is needed to find the particular values of $\varepsilon_i$ and $\varepsilon_{i+1}$. It can be formulated by minimizing the relative error of the flux, $e_{_{F,j}}$, at the remaining evaluation points $r_j$,
\begin{equation}
e_{_{F,j}} = \frac{(r_{j}-d)^2}{\varepsilon} \; \left| \frac{\varepsilon_i}{r_{j,i}^2} + \frac{\varepsilon_{i+1}}{r_{j,i+1}^2} - 
\frac{\varepsilon}{(r_{j}-d)^2} \right|,
\label{eKernEpsilDef}
\end{equation}
where $r_{j,i} = r_j - r_i$ and $r_{j,i+1} = r_j - r_{i+1}$. We explore three possible variants of the flux-conserving kernel.

The first variant, $W_{f,1}$, is constructed using the second condition of the form $\varepsilon_{i+1}/\varepsilon_{i} = (d - r_{i})/(r_{i+1} - d)$, where the evaluation points satisfy the constraint $r_{i} \leq d \leq r_{i+1}$. 

The second variant, $W_{f,2}$, is based on a second condition of the form $\varepsilon_{i}/(r_i - r_{i-1})^2 + \varepsilon_{i+1}/(r_{i+1} - r_{i-1})^2 = \varepsilon/(d - r_{i-1})^2$, which means that the flux is exact at the evaluation point $r_{i-1}$. Since the flux is then exact at two points along the ray (at $r = 0$ and $r = r_{i-1}$), it should be close to the correct value everywhere else on the ray between $0$ and $r_{i-1}$. 

The third variant, $W_{f,3}$, uses a simple second condition of the form $\varepsilon_{i} + \varepsilon_{i+1} = \varepsilon$. 

If the kernel spans of larger number of evaluation points, $N$, the two conditions defining the generalised kernel $W_{f,3}$ are
\begin{subequations}
\begin{align}
\frac{\varepsilon}{d^2} & = \sum_{i=j}^{j+N-1} \frac{\varepsilon_i}{r_{i}^2}, \label{eKernFC1} \\
\varepsilon & = \sum_{i=j}^{j+N-1} \varepsilon_i \label{eKernFC2}.
\end{align}
\label{eKernFC12}
\end{subequations}
A further $N-2$ conditions are then required, and these are generated by setting a linear relationship between the emission coefficient $\varepsilon_j$ and its index $j$, i.e. $\varepsilon_j = \varepsilon\times (a + b(j-i))$ for $i \le j \le i + N - 1$, where $i$ is the lowest index for which the kernel yields a non-zero value. With this prescription, equations (\ref{eKernFC1}) and (\ref{eKernFC2}) give
\begin{subequations}
\begin{align}
b & = \left(\frac{1}{d^2} - \frac{A_1}{N}\right)   \left[A_2 - \frac{(N-1)A_1}{2}\right]^{-1}, \\
a & = \frac{1}{N} - \frac{b(N-1)}{2},
\end{align}
\label{eKernCoeffsAB}
\end{subequations}
with
\begin{subequations}
\begin{align}
A_1 & = \sum_{i=j}^{j+N-1} \frac{1}{r_i^2}, \\
A_2 & = \sum_{i=j+1}^{j+N-1} \frac{i-1}{r_i^2}.
\end{align}
\label{eKernCoeffsA12}
\end{subequations}
Note that this kernel is non-local in the sense that the positions of all the evaluation points over which the kernel is non-zero have to be known to obtain the kernel coefficients at a certain point. This is straightforward in {\sc TreeRay}, because the positions of all the evaluation points on all the rays are known a priori (see equation~\ref{eq:etaR}).

\begin{figure}
\begin{center}
\includegraphics[width=0.4\columnwidth]{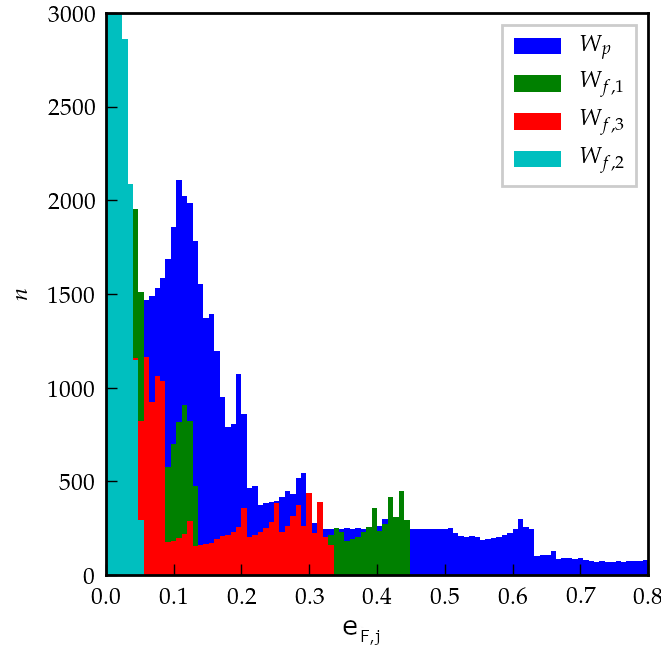}
\caption{Histograms of relative errors in the radiation flux, $e_{_{F,j}}$ (see equation \ref{eKernEpsilDef}) at evaluation points $r_j$ for the three variants of the flux-conserving kernel and the polynomial kernel (see equation \ref{eKernPolynomial}).}
\label{fig:errHist}
\end{center}
\end{figure}

Fig.~\ref{fig:errHist} compares the relative error in the radiation flux, $e_{_{F,j}}$, calculated using the three variants of the flux-conserving kernel, $W_{f,1}$ (green), $W_{f,2}$ (cyan), and $W_{f,3}$ (red). For reference, the relative error obtained using the polynomial kernel, $W_p$, is also shown (blue). Each histogram was obtained by placing a source at $10^4$ different positions on a ray with 10 evaluation points placed according to equation~\ref{eq:etaR}.  
$W_{f,2}$ leads to the most accurate results, but $W_{f,3}$ also leads to acceptable results and is much simpler. Therefore, we use $W_{f,3}$ as the default in {\sc TreeRay}.

\label{lastpage}

\end{document}